\pdfoutput=1 
\documentclass{aa}  

\usepackage{natbib}
\bibpunct{(}{)}{;}{a}{}{,} 
\usepackage{graphicx}
\usepackage{txfonts}
\usepackage{amsmath}	
\usepackage{amssymb}	
\usepackage{upgreek}
\usepackage[toc,page]{appendix}
\usepackage{longtable}
\usepackage{pifont}
\usepackage{subcaption}
\usepackage{capt-of}
\usepackage{booktabs} 
\usepackage{gensymb} 

\def\Msun{\hbox{$\rm ~M_{\odot}$}}

\def\H0{{\rm ~km~s^{-1}~Mpc^{-1}}}

\def\cm3{$\rm ~cm^{-3}$}
\def\kms{{\rm ~km~s^{-1}}}

\def\la{\mathrel{\hbox{\rlap{\hbox{\lower4pt\hbox{$\sim$}}}{\raise2pt\hbox{$<$}}}}}
\def\ga{\mathrel{\hbox{\rlap{\hbox{\lower4pt\hbox{$\sim$}}}{\raise2pt\hbox{$>$}}}}}

\def\hi{H~{\scriptsize I}$~$}
\def\hii{H~{\scriptsize II}$~$}

\def\3xmm5{3XMM-DR5\thinspace}

\newcommand{\cmark}{\ding{51}}%
\newcommand{\xmark}{\ding{55}}%

\begin{document} 

   \title{The Coordinated Radio and Infrared Survey for High-Mass Star Formation}
   \subtitle{III. A catalogue of northern ultra-compact H {\large II} regions}

   \author{I. E. Kalcheva
          \inst{1}\fnmsep\thanks{E-mail: pyiek@leeds.ac.uk}
          \and
          M. G. Hoare\inst{1}
          \and
 		  J. S. Urquhart\inst{2} 
          \and
          S. Kurtz\inst{3}
          \and
          S. L. Lumsden\inst{1}     
          \and                         \\
          C. R. Purcell\inst{4,5}     
          \and                     
          A. A. Zijlstra \inst{6} 
          }

  \institute{School of Physics and Astronomy, University of Leeds, Leeds LS2 9JT, UK
  \and
  Centre for Astrophysics and Planetary Science, University of Kent, Canterbury CT2 7NH, UK
  \and
  Instituto de Radioastronom{\'i}a y Astrof{\'i}sica, Universidad Nacional Aut{\'o}noma de M{\'e}xico, 58089 Morelia, Michoac{\'a}n, M{\'e}xico 
    \and
Research Centre for Astronomy, Astrophysics, and Astrophotonics, Macquarie University, NSW 2109, Australia
 \and
 Sydney Institute for Astronomy, School of Physics, The University of Sydney, NSW 2006, Australia
  \and
  Jodrell Bank Centre for Astrophysics, Alan Turing Building, School of Physics and Astronomy, The University of Manchester, Oxford Road, Manchester M13 9PL, UK
   }

   \date{Received Month Day, 2018; accepted Month Day, 2018}

 
 \abstract{ 
 A catalogue of 239 ultra-compact \hii regions (UCHIIs) found in the CORNISH survey at 5 GHz and 1.5$''$ resolution in the region $10\degree<l<65\degree, ~|b|<1\degree$ is presented. This is the largest complete and well-selected sample of UCHIIs to date and provides the opportunity to explore the global and individual properties of this key state in massive star formation at multiple wavelengths. The nature of the candidates was validated, based on observational properties and calculated spectral indices, and the analysis is presented in this work. The physical sizes, luminosities and other physical properties were computed by utilising literature distances or calculating the distances whenever a value was not available. The near- and mid-infrared extended source fluxes were measured and the extinctions towards the UCHIIs were computed. The new results were combined with available data at longer wavelengths and the spectral energy distributions (SEDs) were reconstructed for 177 UCHIIs. The bolometric luminosities obtained from SED fitting are presented. By comparing the radio flux densities to previous observational epochs, we find about 5\% of the sources appear to be time variable.   
 This first high-resolution area survey of the Galactic plane shows that the total number of UCHIIs in the Galaxy is $\sim$ 750 -- a factor of 3-4 fewer than found in previous large area radio surveys.
 It will form the basis for future tests of models of massive star formation.
}


\titlerunning{A catalogue of northern ultra-compact H {\scriptsize II} regions}
\authorrunning{Ivayla E. Kalcheva et al.}

   \maketitle
%

\section{Introduction}

  The puzzle of the birth and early life of stars exceeding 8\Msun$ $  is not yet fully assembled. Some of the obstacles towards building a complete evolutionary sequence for these massive stars include their rarity due to their brief lifetime and the rapid evolution of each observable stage. The main sequence is reached while the young star is still embedded within a dense core and as a result the early phases of its development are hidden behind a heavy veil of dust. A well-founded distinction between global and individual properties of sources in each evolutionary stage is hampered by the strong influence of other objects within the multiple systems where massive stars typically form.
  
  It is a vital task for modern astronomy to overcome these challenges. Massive stars affect not only their immediate surroundings, but also shape their parent galaxy. Their formation controls phase changes in the interstellar medium (ISM) via the profuse emission of ionising UV photons \citep{molinari:2014}. Processes associated with their evolution, such as winds, outflows, expanding \hii regions and supernovae, stir the ISM and enrich it with heavy elements \citep{yorke:2007}. This makes their understanding a stepping stone towards a more detailed picture of the Milky Way, as well as the extent to which galaxy formation and evolution in general is driven by stellar populations. 
  
  After a massive star has formed, it ionises a pocket of hydrogen gas which remains confined in its vicinity while expanding -- thus forming an \hii region. \hii regions are highly convenient tracers of massive star formation, as they are clearly visible across the Galactic plane in the cm-regime \citep{bania:2009}. Distinguished as a separate observational 
  class by \cite{churchwell:1989a}, ultra-compact \hii (UCHII) regions link the accretion phase when a massive protostar is formed, and the development of a more diffuse and less obscured \hii region. UCHII regions are defined as embedded photoionised regions $\lesssim$ 0.1~pc in diameter, with emission measures $\gtrsim$ 10$^{7}$ $\rm ~pc ~cm^{-6}$ and electron densities $n_{\rm e}$ $\gtrsim$ 10$^{4}$ $\rm ~cm^{-3}$ \citep{churchwell:1989a}.
  They are the most luminous objects in the Milky Way in the far-IR, and are observable in the radio part of the spectrum if their luminosities are equivalent or higher than a B0.5 main-sequence star. The Lyman continuum ionising flux corresponding to zero-age main-sequence stars with spectral class from B2 to O5 is in the range $10^{44}$ -- $10^{49}$ $\rm ~photons ~s^{-1}$.
  Estimating the distances to UCHII regions, together with their density distributions, luminosities, morphologies, kinematics, and relationship to the parent molecular clouds is essential. These properties can be used to help understand not only the effect of UCHII regions on their environment, but also test the existing evolutionary models of massive star formation and the structure of the Milky Way \citep{hoare:2007}. 
  
  The bounds of current understanding of massive star formation are widened by the modern family of Galactic plane surveys, covering the dust (from hot to cold), the molecular and the ionised gas. These include the GLIMPSE programme  \citep{churchwell:2009, benjamin:2003} and its companion MIPSGAL survey (mid-IR) \citep{carey:2009}, the UKIDSS GPS survey (near-IR) \citep{lucas:2008}, the BU-FCRAO Galactic Ring Survey (CO) \citep{jackson:2006}, the ATLASGAL survey (sub-mm) \citep{schuller:2009}, the VGPS survey (\hi) \citep{stil:2006}, the CORNISH survey (radio) \citep{hoare:2012, purcell:2013}. These legacy surveys provide resolution and sensitivity apposite to the detection and discerning of sources occupying angular scales down to $\sim$ 1$''$. At the same time, they cover wide areas on the sky and overcome the high extinction of the plane. 
  In this way, a multi-wavelength treasure trove of unbiased, high-resolution and statistically representative data are available to aid the studies of the earliest phases of massive star formation.

  The CORNISH survey\footnote{\url{http://cornish.leeds.ac.uk/public/index.php}}, the first Galactic plane survey that is comparable in resolution and coverage with the GLIMPSE data, maps the compact ionised gas within the ISM. At present, the CORNISH catalogue of the northern Galactic plane, imaged with the VLA, is the most uniformly sensitive, homogeneous and complete list of northern compact radio sources at 5 GHz. 
  The CORNISH team identified 240 ultra-compact \hii region candidates. 
  The sample provides the largest unbiased and uniform collection of these objects to date.
  
  Previous radio UCHII samples comprise predominantly IR-targeted surveys based on IRAS point sources with far-IR colours similar to well-known UCHIIs. \cite{churchwell:1989a, churchwell:1989b} selected a sample of 75 UCHIIs (out of $\sim$ 1600 candidates in the Galaxy) to observe at 6~cm and 2~cm with the VLA using this method and classified them morphologically. 
  Similarly, \cite{miralles:1994} selected and observed 12 sources at 6 and 2~cm with the VLA. \cite{garay:1993} also based their selection on strong IRAS point sources associated with compact \hii regions and produced multi-frequency observations with the VLA (resolved and morphologically classified). \cite{kurtz:1994} performed radio-continuum observations on 59 UCHIIs, again IRAS-selected.
  \cite{depree:2005} located and resolved a hundred objects  within the massive star forming regions W49A and Sgr B2 from VLA radio continuum and radio recombination line emission observations (and revisited the \cite{churchwell:1989a} morphological classification).
    
  The RMS survey \citep{urquhart:2007, urquhart:2009, lumsden:2013} marked a new era of massive star formation studies. Colour-selected sources from MSX \citep{price:2001} and 2MASS \citep{skrutskie:2006} were followed up by arcsecond-resolution IR, as well as mm and radio observations. These, together with archival data, were used  to identify for the first time a Galaxy-wide sample of $\sim$ 2000 candidate massive young stellar objects (MYSOs) and \hii regions in approximately equal numbers \citep{urquhart:2012}.
  
  The biggest disadvantage of IR-selection in UCHII studies is the discrimination against the most deeply-embedded sources. 
  The issue is resolved by unbiased radio surveys.  
  The first larger-scale unbiased survey at 1.4 GHz (inner Galaxy, VLA B and A/B configuration) was conducted by \cite{zoo:1990} and was followed by (VLA C configuration) 5 GHz observations  \citep{becker:1994} covering about a fourth of the GLIMPSE region.
  The survey \citep[now contained within a larger collection of re-reduced archival radio data known as MAGPIS, see][]{helfand:2006} is useful for the study of extended thermal sources such as evolved \hii regions, bubbles, etc. However, UCHII regions are unresolved or marginally resolved, and even in some instances missed altogether due to insufficient sensitivity. The CORNISH survey covers the entire GLIMPSE region and its noise level of 0.4~mJy  ensures the detection of virtually all UCHIIs  around a B0.5V star or earlier within the covered area \citep{hoare:2012}. 
    
  This work explores the sample of northern ultra-compact \hii regions from the CORNISH survey, the majority of which are also conveniently available within the related surveys, 
  to study the properties of this deeply embedded phase. 
  The sample selection procedure is presented (\S \ref{id}) and the nature of the identified sources is  verified  through their observational properties (\S \ref{validation}) and spectral indices (\S \ref{spind}). Candidate short-timescale variable sources are presented in \S \ref{variability}. The methodology of obtaining the distance information and the computed distances are presented in \S \ref{distances}. The derived physical properties are discussed in \S \ref{physprop}. 
  Results from performing automated polygon-based aperture photometry on UKIDSS and GLIMPSE infrared associations are discussed in \S \ref{IR} and presented in an extended source catalogue table.  
  The spectral energy distributions of the sample sources from near-IR to sub-mm wavelengths 
  were explored and utilized via SED fitting to obtain the UCHII bolometric luminosities (\S \ref{lbol}). In \S \ref{comparison_methods}, different UCHII search methods in blind surveys are compared.
  The present work is summarised in \S \ref{summary}.


\section{Identification of the CORNISH UCHII sample}\label{id}

\begin{figure}
\centering
\includegraphics[width = \columnwidth]{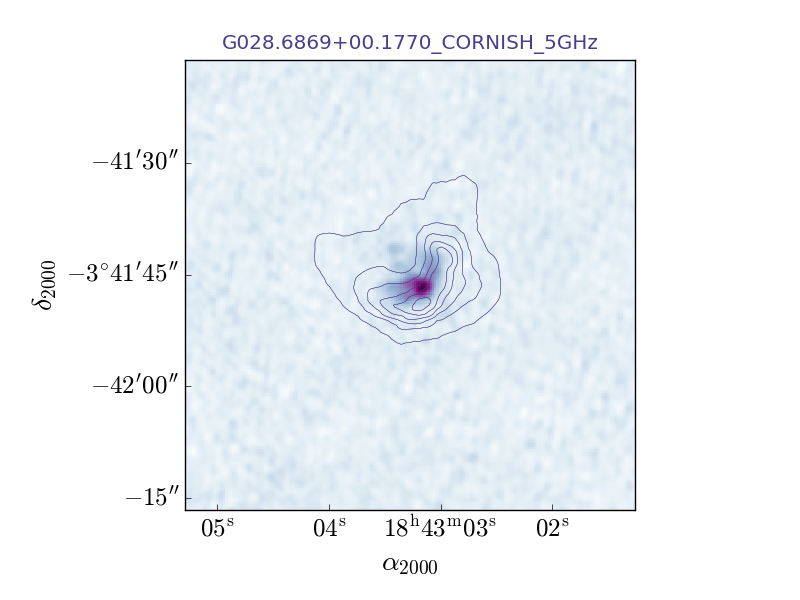} 
\includegraphics[width = \columnwidth]{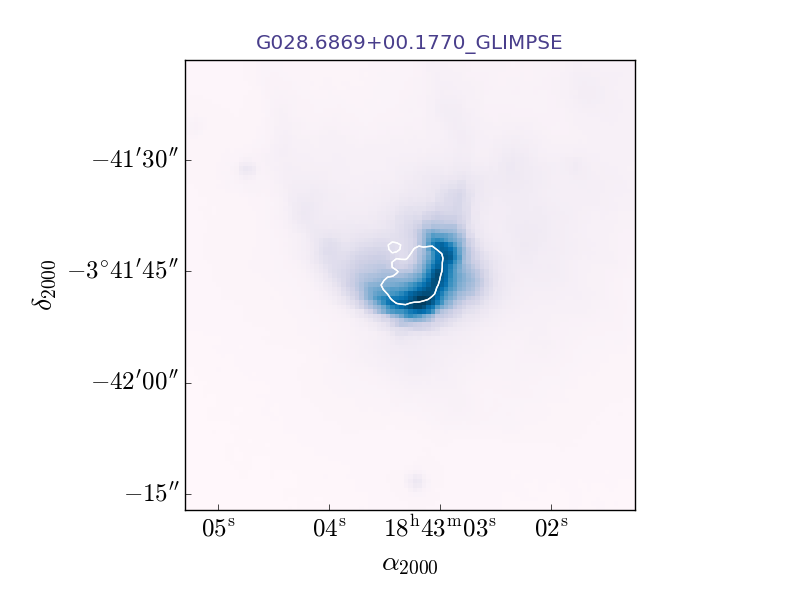}
\includegraphics[width = \columnwidth]{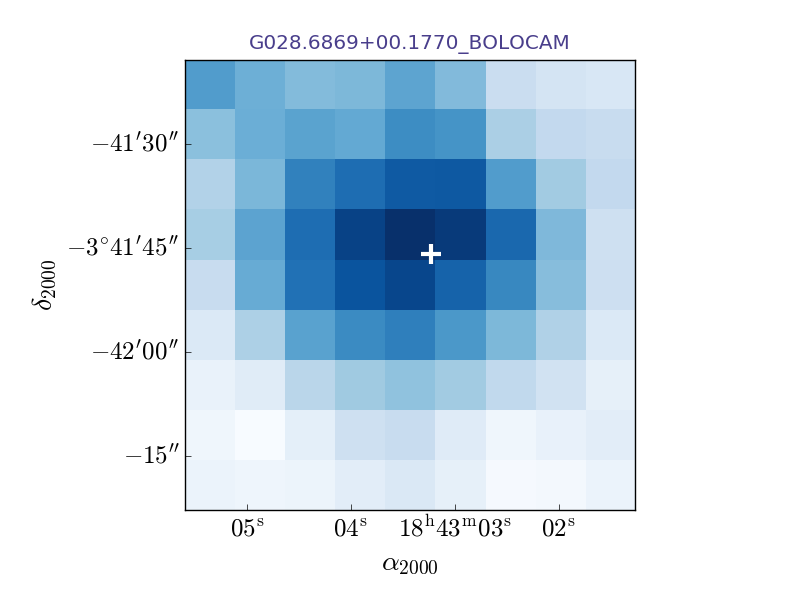}
\caption{Images of the cometary UCHII G028.6869+00.1770. Top: CORNISH 6~cm image with overplotted GLIMPSE 8~micron contours, with the corresponding GLIMPSE image shown below, with overplotted radio contours. The morphology at both wavelengths is in very good agreement, and there is excellent positional coincidence between the two catalogues, allowing reliable source identification. The BOLOCAM image (bottom panel) shows a bright unresolved source at 1.1 mm coincident with G028.6869+00.1770 (the source position is marked by a white cross).}
\label{fig:iding}
\end{figure}

The CORNISH catalogue comprises 3062 sources above a 7$\sigma$ detection limit. Above this limit, less than one spurious source is expected \citep{purcell:2013}. The 240 UCHII regions were selected from this high-reliability catalogue. 
All UCHIIs were visually identified, following criteria similar to the RMS survey, where millimetre, infrared and radio data were used for source classification \citep[see][]{lumsden:2013}. It should be noted that the CORNISH team also identified 48 diffuse \hii regions (as judged by comparison to the MAGPIS and GLIMPSE surveys), which are a part of the larger sample of CORNISH \hii regions. 

\begin{figure}
\includegraphics[width = \columnwidth]{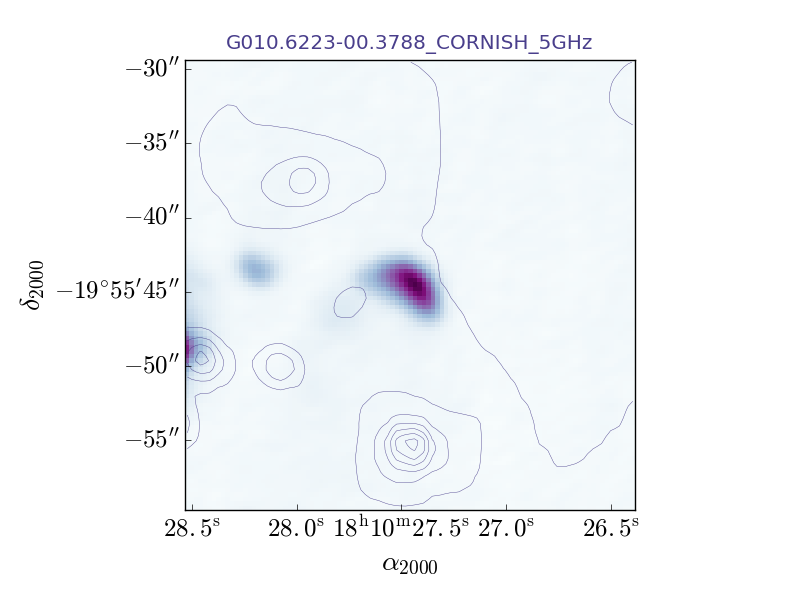}
\includegraphics[width = \columnwidth]{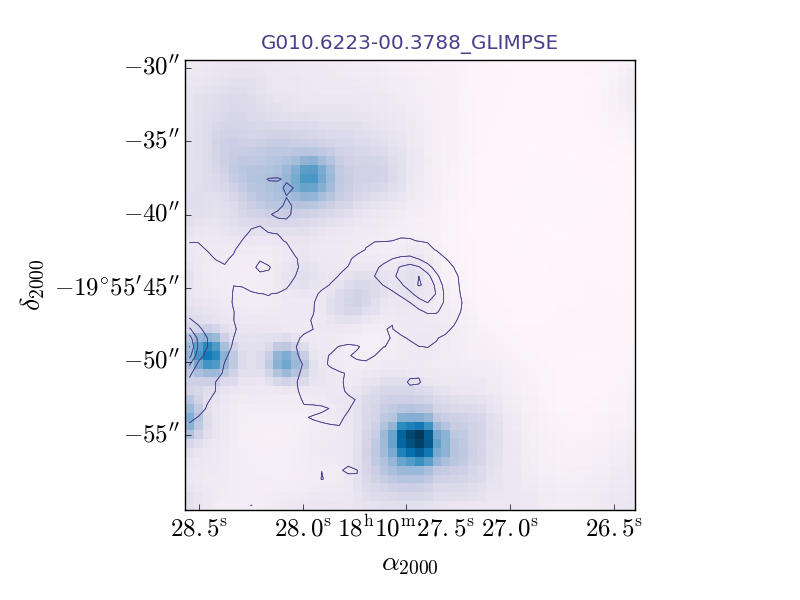} 
\caption{CORNISH (top panel) and GLIMPSE (bottom panel) images of G010.6223\allowbreak$-$00.3788, presented as in Fig. \ref{fig:iding}, with overplotted 8~micron and 6~cm contours, respectively. A dust lane is hiding the source in the 8~micron image.} 
\label{fig:big_difference1}
\end{figure}

The full UCHII radio sample has counterparts in GLIMPSE, in all four bands (namely, IRAC 3.6, 4.5, 5.8, and 8.0~$\upmu$m), with excellent positional accuracy in both surveys. This was utilised for the source identification. In the case of UCHIIs, there is overall a good agreement between the mid-IR and the 5 GHz source morphology, which ensures that the same source was captured by both surveys. A particularly good check for this are the 8~$\upmu$m images. They show the morphology produced by a combination of warm Lyman-$\alpha$ heated dust inside the ionised zone \citep{hoare:1991} and polyaromatic hydrocarbon (PAH) emission from just outside the ionisation front \citep{watson:2008}.
This can be seen in Fig. \ref{fig:iding}. Comparison between both wavelengths is therefore useful for the distinction of adjacent unrelated sources and over-resolved emission \citep[see][]{purcell:2013}. It can also reveal the most heavily obscured objects (those deeply embedded in infrared-dark clouds (IRDCs) or hidden behind dust lanes), as shown in Fig. \ref{fig:big_difference1}. 

MYSOs, unlike UCHIIs, do not have strong 8.0~$\upmu$m PAH emission, which is consistent with the lack of a strong UV continuum \citep[e.g.][]{gibb:2004}. 
 They are also generally undetected at 5 GHz, even though there are a few known MYSOs observed at radio wavelengths, with radio luminosities ($S_{\nu}D^{2}$) always below $\sim$ 30~mJy~$\rm kpc^{2}$ (discussed in  \citealt{hoare:2007}, \citealt{lumsden:2013}, and seen from the recent sample by \citealt{purser:2016}). Sources above this limit are thus \hii regions or planetary nebulae (PNe). 

This leaves PNe as possible contaminants of the selected sample. 
Unlike PNe, UCHIIs are found within molecular clouds, often in close proximity to IR clusters and dust lanes, which aids the visual classification. 
A lower but significant fraction of sources are found near other radio sources. About 33\% of the CORNISH UCHIIs are situated in a radio cluster (within 12$''$ of another source), with 30\% in a sky region containing more than seven detections of 7$\sigma$ sources. The outlines of 24\% of the UCHIIs overlap one or more 7$\sigma$ sources (see \S \ref{validation}). 
\hii regions are expected to be strong sources in 1 mm continuum (which maps the cool dust), whereas planetary nebulae are not. 
BOLOCAM 1.1 mm images \cite[see][]{rosolowsky:2010} centred at the radio source position were visually inspected in conjunction with the IR images to verify that the UCHII sample is not contaminated by PNe. 

It is easier to sift out other classes of sources such as radio stars and radio galaxies. Radio stars can be distinguished by their lack of mid- and far-IR emission, whereas radio galaxies have no infrared counterparts.

\section{Radio properties of the CORNISH UCHIIs}\label{validation}

\begin{figure}
\centering
\includegraphics[width=0.95\columnwidth]{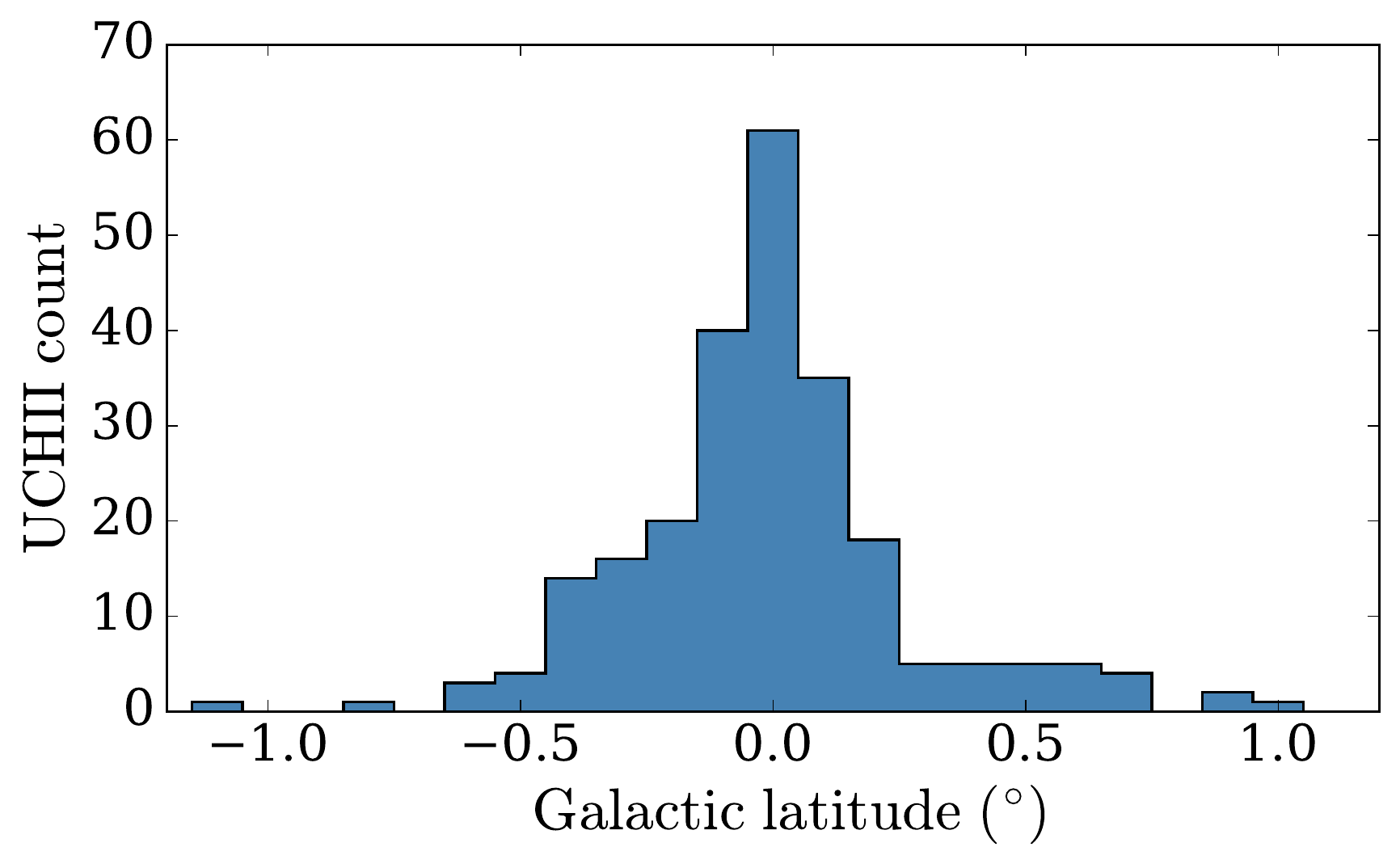} \\ 
\includegraphics[width=\columnwidth]{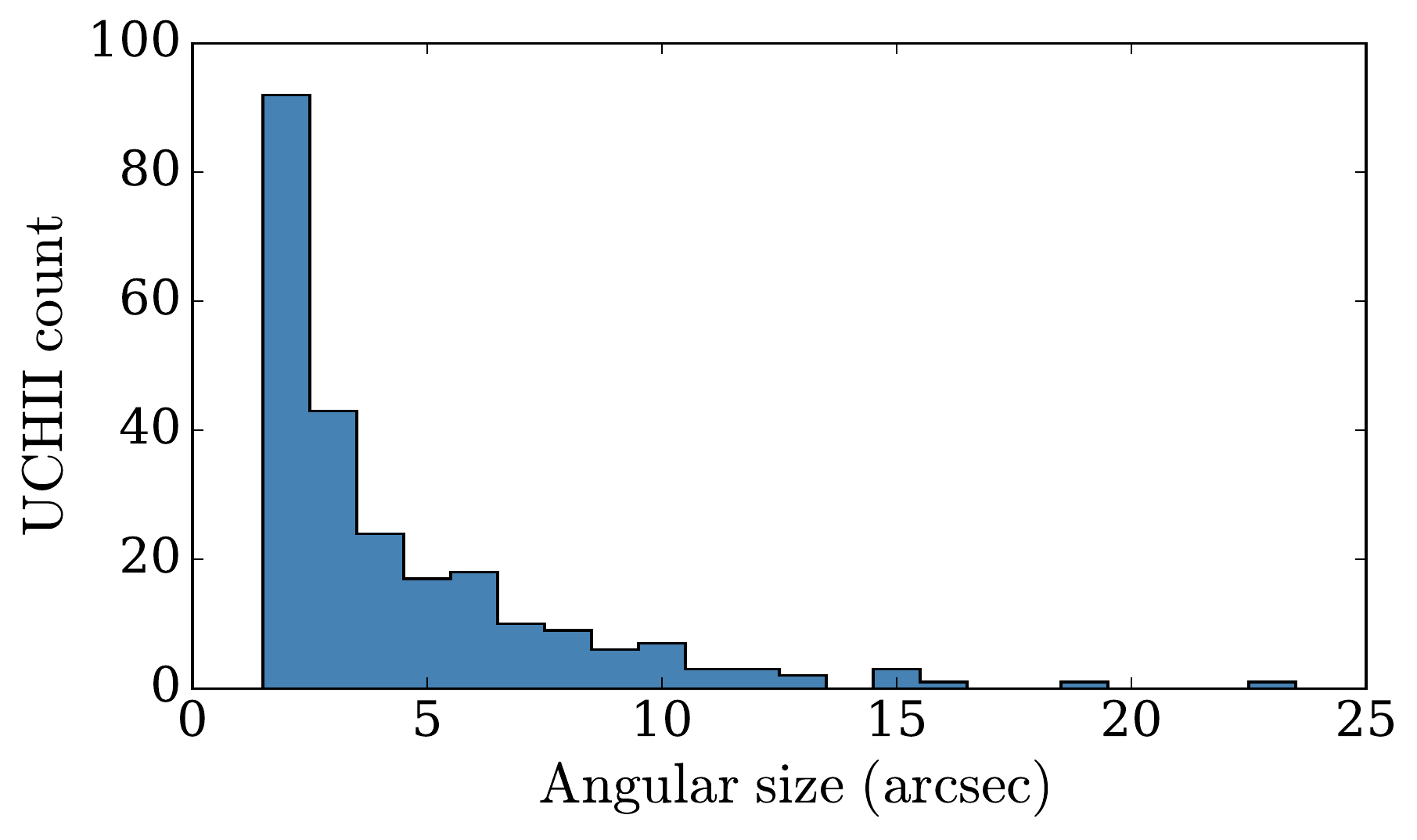} \\  
\includegraphics[width=\columnwidth]{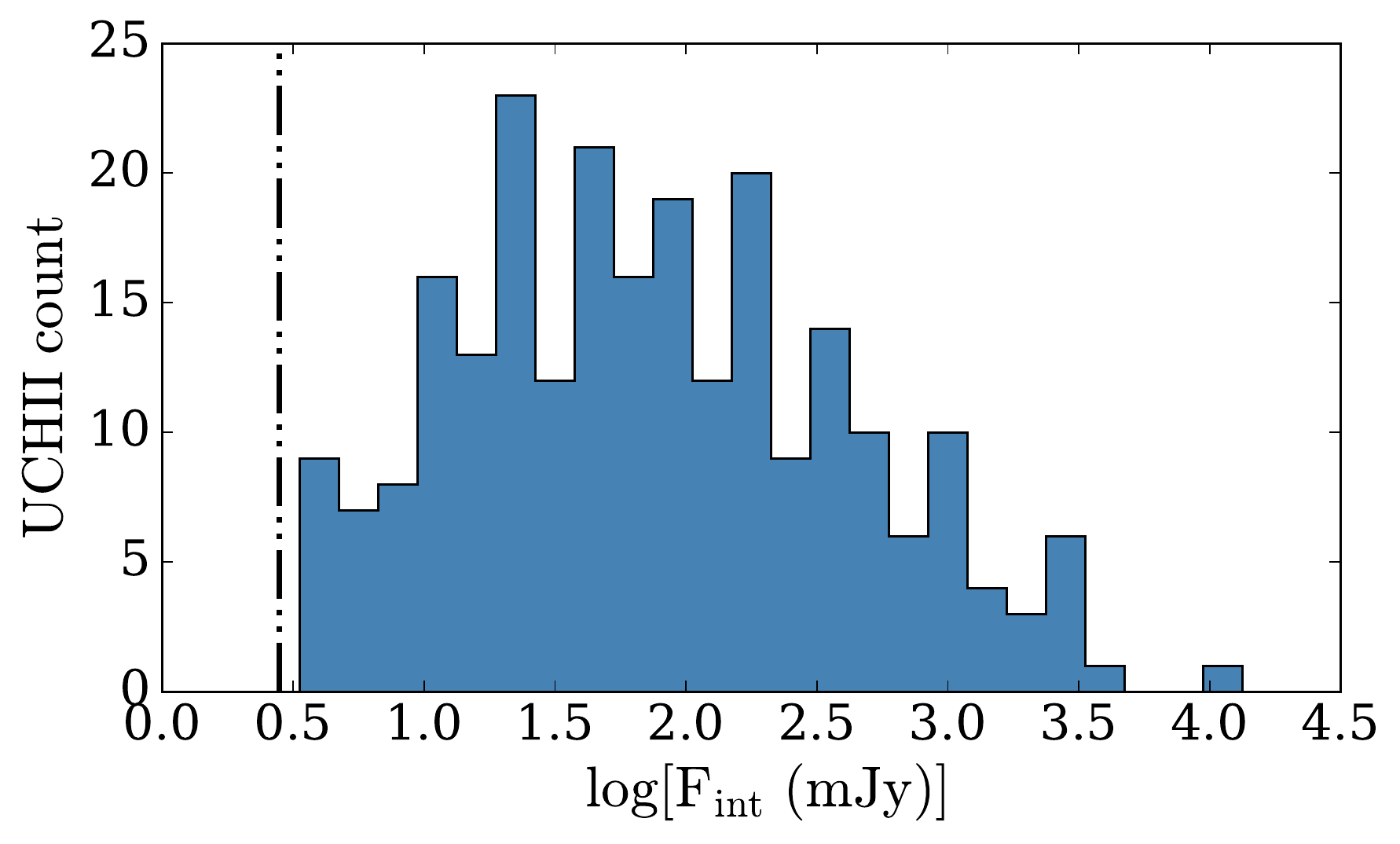}
\caption{Observational properties of the candidate sample -- the confinement to the Galactic plane (top panel), small angular sizes (middle panel), and total radio fluxes (bottom panel) are consistent with UCHII regions. The dot-dashed line in the bottom panel marks the 7$\sigma$ (2.8~mJy) sensitivity limit of CORNISH.} 
\label{fig:lat}
\end{figure}

The distribution of the Galactic latitudes, angular sizes and integrated fluxes  of the sample of candidate UCHIIs are shown in Fig. \ref{fig:lat}. 
The sources are closely confined to the Galactic plane, as expected for very young massive star forming regions. 
Using the CORNISH survey, \cite{urquhart:2013} fit a scale-height of 20.7 $\pm$ 1.7~pc for compact and ultra-compact \hii regions. 

The majority of the ultra-compact \hii regions have angular sizes below 5$''$, with the histogram peaking towards unresolved sources.  
As discussed in \cite{purcell:2013}, sources begin to suffer from over-resolution above angular sizes of 14$''$.
 Most sources have integrated fluxes above 10~mJy, and the brightest source has a flux density of 12.6~Jy. The integrated flux histogram shows a clear downturn towards the lowest values, that is, the UCHII sample is close to complete. 
 
\begin{figure}
\centering
\includegraphics[width=\columnwidth]{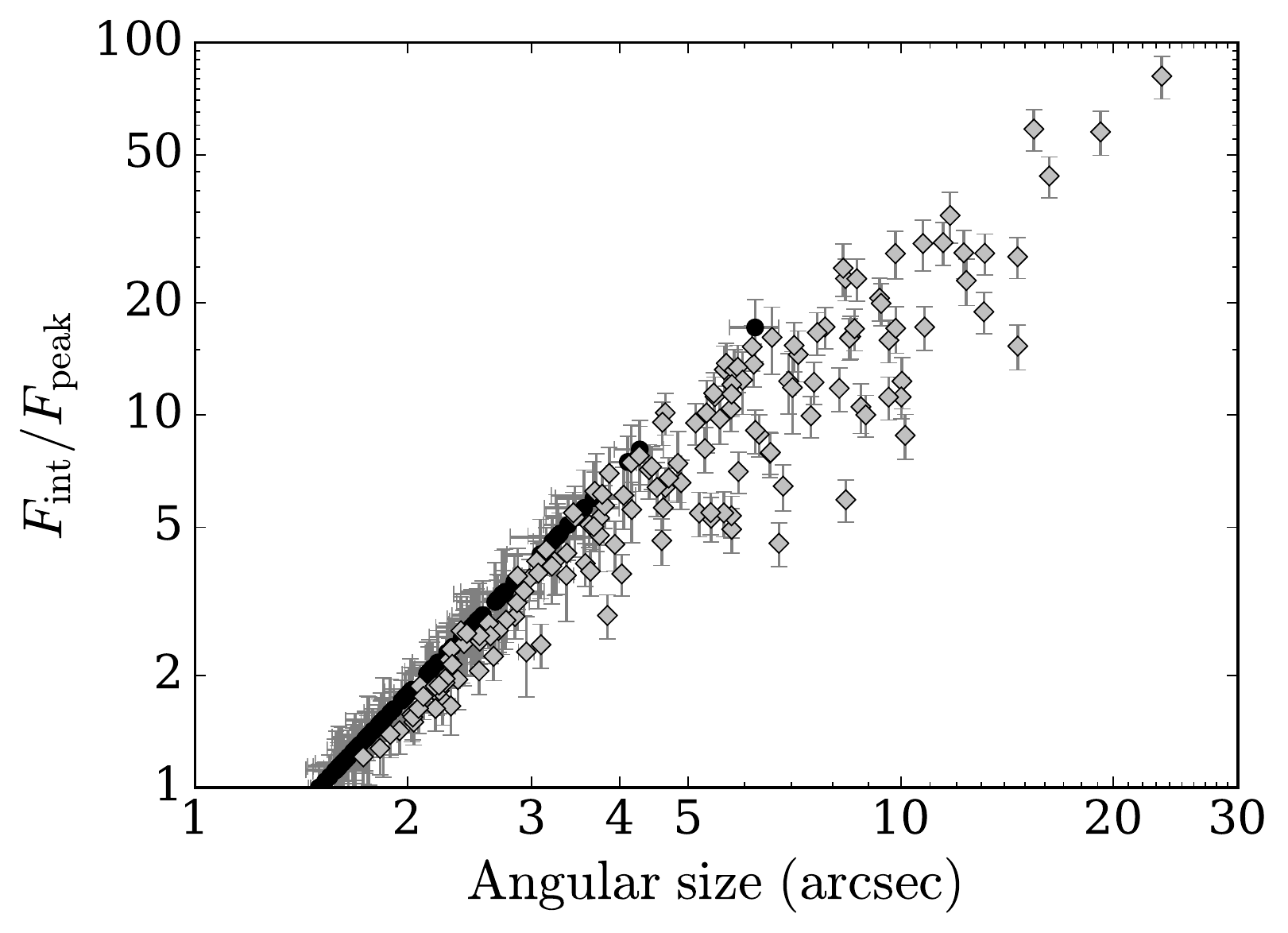}
\caption{Ratio of integrated to peak flux vs. angular size of the 239 CORNISH UCHIIs. Fluxes estimated from Gaussian fits and from polygon apertures are shown as black circles and grey diamonds, respectively.}
\label{fig:resunres}
\end{figure}

The CORNISH beam size is 1.5$''$ and all sources with $\theta <$ 1.8$''$ are marked as unresolved in the catalogue table (detailed checks for the entire CORNISH catalogue are discussed in  \citealt{purcell:2013}). 
Within the CORNISH UCHII candidate sample, the flux was measured by fitting a Gaussian in 90 out of the 239 cases (angular size range 1.5$''$ to 6.2$''$), and for the remaining 149 sources, a hand-drawn polygon was used instead (angular size range from 1.8$''$ to 23.4$''$). 
The integrated and peak fluxes were compared as a function of angular size of each source, as shown in Fig. \ref{fig:resunres}. 
 Naturally, those sources whose fluxes were measured from a Gaussian fit show a clear trend of the flux ratio with increasing angular size, whereas the remaining sources with manually drawn contours show more variation. 

\subsection{Lower-resolution radio counterparts}\label{lowres}
\begin{figure}
\centering
\includegraphics[width=\columnwidth]{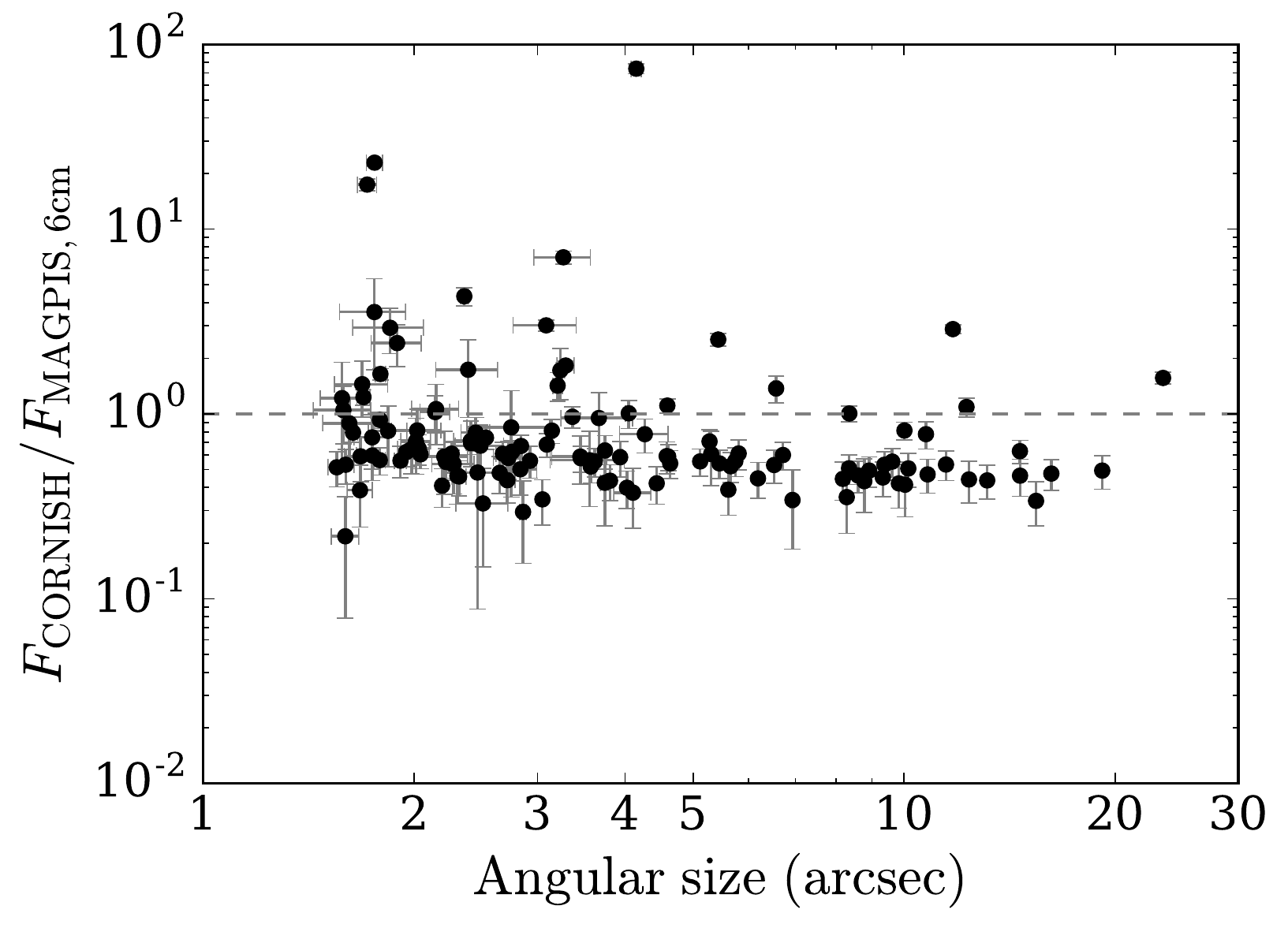}
\caption{Comparison of the MAGPIS 6~cm and CORNISH integrated fluxes (black circles) with plotted line of equality, plotted against the corresponding CORNISH angular size.} 
\label{fig:fluxes}
\end{figure}

The Multi-Array Galactic Plane Imaging Survey (MAGPIS) \citep{helfand:2006} is useful for the study of evolved \hii regions and other extended, optically thin thermal emitters. However, it is not  well-suited to explore dense, thermal sources, as those are unresolved or only marginally resolved. 
Catalogues at 20~cm (VLA B,C,D configuration) and at 6~cm (VLA C configuration) are available \citep{white:2005}. The 6~cm catalogue covers $\sim$ 23\% of the northern-GLIMPSE region, and at the survey resolution most of the detected CORNISH counterparts are unresolved -- no morphological information is available. 

The benefits of a comparison between CORNISH and MAGPIS are explained in detail in \cite{purcell:2013} (see Figs. 20, 21). 
In brief, for extended sources in the CORNISH sample, the measured flux density could be a lower limit in cases where some of the extended emission was filtered out due to gaps in $\textit{uv}$ coverage, and the measured angular sizes could be underestimated by as much as 50\% for non-Gaussian sources in interferometric measurements \citep[see e.g.][]{panagia:1978}. To test for such instances, a comparison to the MAGPIS 6~cm data is useful, as the survey configuration allowed the recovery of more diffuse emission at the cost of lower resolution.

It is important to note that there are instances of repeats in the combined 6cm-20cm MAGPIS catalogue. This is due to the fact that sources matching more than once are listed multiple times, for example a 6~cm source listed with each 20~cm counterpart and vice versa \citep{white:2005}.  Since the structure of the MAGPIS catalogue leads to repeats even within a small cross-matching radius, to obtain a sufficiently reliable cross-match with the MAGPIS catalogue table, the matched sample had to be limited to only 47 associations. Visual identification of the associations was therefore preferred.  The CORNISH team visually identified 216 20~cm matches (out of which 162 have both a 6~cm and 20~cm MAGPIS detection). 

The fluxes of the 20~cm and 6~cm MAGPIS UCHII associations were measured independently of the catalogue table values via automated aperture photometry scripts. 
Radio outlines from the CORNISH database $-$ hand-drawn polygons in the case of extended radio sources,  and Gaussian outlines in the case of compact radio sources, were utilised. In order to use these outlines as apertures for the lower-resolution MAPGIS data, they were expanded accordingly. The necessary `padding' value (i.e. the required radial expansion) was determined after multiple runs with different aperture sizes. Based on the curve of growth, a padding value of 4$''$ (i.e. total expansion of 8$''$) was chosen for the flux measurement.

Fig. \ref{fig:fluxes} shows that the majority of the flux ratio values occupy the range between $\sim$ 0.2 and 1.1, which indicates that some flux was not recovered at the higher resolution (around 65\% of the flux was detected for these sources). The slightly negative slope of the flux ratio is due to worsening over-resolution with increasing angular size. 
Sources overlapping with one or more 5$\sigma$ or 7$\sigma$ CORNISH neighbours were excluded, as these sources are unresolved and merged in the MAGPIS 6~cm images and their flux measurements are unreliable.
The outliers above the equality line, that is, with CORNISH fluxes significantly higher than their MAGPIS 6~cm counterpart, are investigated for potential short-timescale variability in \S \ref{variability}. 

The reliability of the measured 216 20-cm and 162 6-cm flux values was judged on the basis of their median brightness level of the sky. 
For all sources with abnormal (i.e. outside the range of the majority of sources) median level in the sky-annuli, upper flux limits are included in the presented flux table (Table \ref{table:fluxes_M} in  Appendix \ref{appendixD}). 
Sources with consistent sky values that are also not overlapping with a 5$\sigma$ or 7$\sigma$ source are here considered to be the highly-reliable MAGPIS flux measurement subset. These were used to compute the spectral indices of the CORNISH UCHIIs -- see \S \ref{spind}.

A few MAGPIS UCHII associations were picked at random out of the highly-reliable subset and their fluxes were also measured with \texttt{CASA}, by Gaussian fits in unresolved cases or computing the flux for the extended source region otherwise\footnote{The same method was tested on the CORNISH images, in good agreement with the catalogued values.}. 
Table~\ref{table:casa} shows a comparison of the flux values in the published MAGPIS catalogue table with our remeasured fluxes for these sources. 
Clearly, neither the photometry performed with the automated script or with \texttt{CASA} reproduced the catalogued 6~cm MAGPIS values to a reasonable degree, with a much better agreement between our two photometric measurements. This also appears to be true when comparing the results for extended 20~cm sources. This is why the fluxes of all available lower-resolution counterparts were remeasured in this work.
Our results obtained for unresolved 20~cm sources appear to be overall in better agreement with the MAGPIS catalogue.  

\begin{table*}
\centering
\caption{Comparison of a few MAGPIS flux values in the published 6cm-20cm catalogue table (5th and 8th columns) with the measured flux values presented in this work -- as from the automated script (3rd and 6th columns) and individual measurements with \texttt{CASA} (4th and 6th columns). Only highly-reliable values (based on median sky values and no overlapping sources) are compared in the table. All values are in mJy. The second column includes the corresponding CORNISH fluxes. The source denoted by $\ddagger$ could be intrinsically variable at 6~cm.}
\begin{tabular}{lr|rrr|rrr}
\hline
\hline
CORNISH name&$F_{\rm C}$ &$F_{\rm M6cm}$&$F_{\rm M6cm}$& $F_{\rm M6cm}$ &$F_{\rm M20cm}$&$F_{\rm M20cm}$& $F_{\rm M20cm}$ \\
 & (cat.) & (ap.) & ($\texttt{CASA}$) & (cat.)  & (ap.) & ($\texttt{CASA}$) & (cat.) \\
\hline
  G011.1104$-$00.3985 & 305.37 $\pm$ 28.55& 303.63 $\pm$ 9.12 & 320.3 & 112.37 & 326.73 $\pm$ 3.86 & 370.91 & 187.21\\
  G012.1988$-$00.0345 & 62.71 $\pm$ 5.92 & 102.95 $\pm$ 1.47 & 110.18 & 59.82 & 42.52 $\pm$ 1.34 & 48.00 & 62.05\\
  G016.3913$-$00.1383$^{\ddagger}$ & 124.27 $\pm$ 15.43 & 43.18 $\pm$ 3.61 & 42.89 & 20.34 & 57.52 $\pm$ 5.18 & 66.71 & 58.23\\
  G018.7106+00.0002 & 107.46 $\pm$ 10.62 & 177.74 $\pm$ 1.71 & 188.78 & 102.96 & 36.60 $\pm$ 0.94 & 41.97 & 35.39\\
  G021.8751+00.0075 & 566.73 $\pm$ 54.14 & 1064.73 $\pm$ 8.85 & 1073.00 & 264.32 & 685.90 $\pm$ 7.84 & 719.53 & 599.63\\
  G037.8731$-$00.3996 & 2561.21 $\pm$ 234.04 & 5189.35 $\pm$23.38 & 5219.50 & 1518.5 & 1769.56 $\pm$ 6.57 & 1800.10 & 1279.40\\
\hline
\end{tabular}
\label{table:casa}
\end{table*}

\subsection{5GHz--1.4GHz spectral indices}\label{spind}
\begin{figure}
\includegraphics[width=\columnwidth]{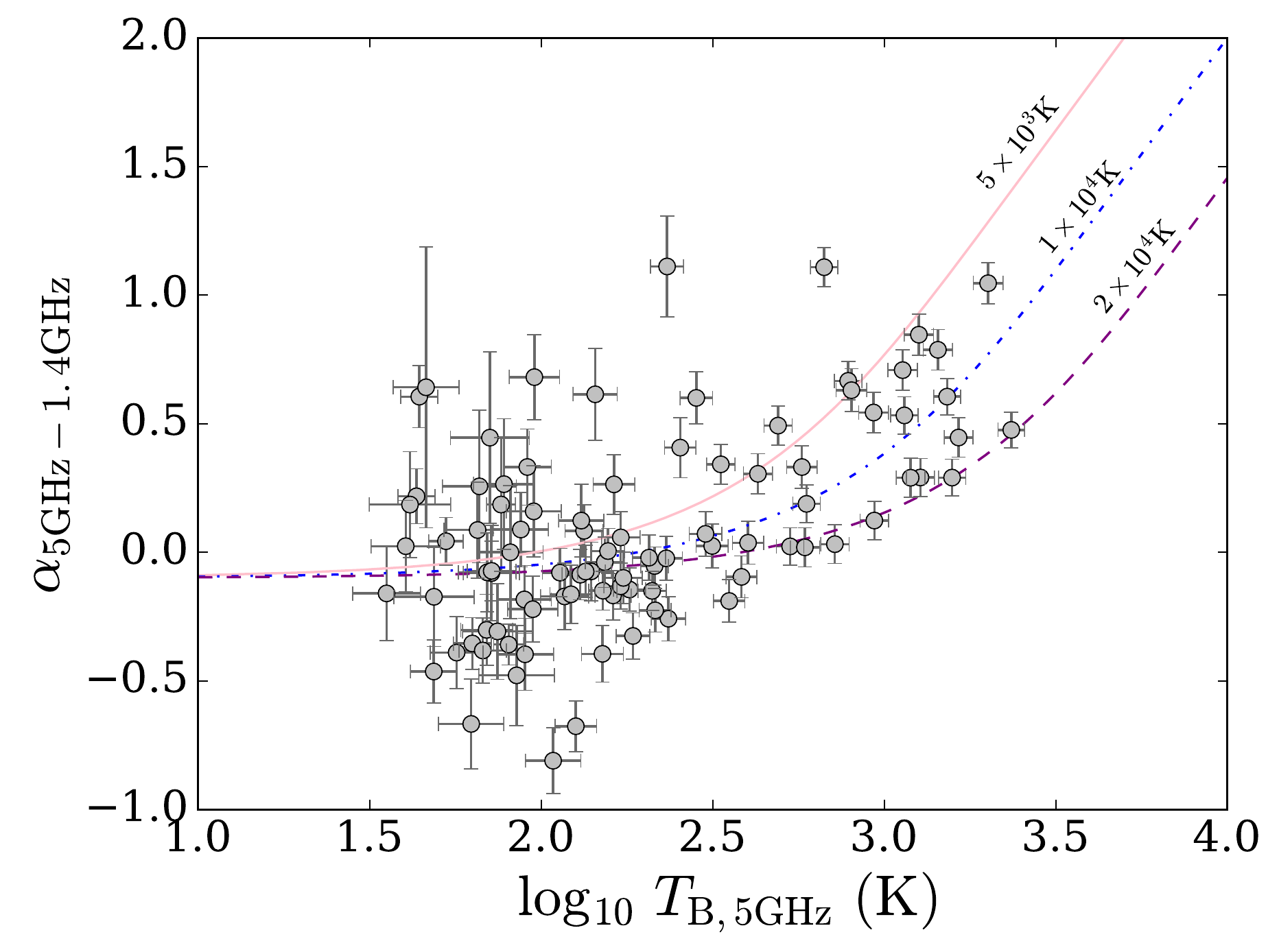}
\caption{Spectral indices between 1.4 GHz and 5 GHz against brightness temperature of the corresponding CORNISH source at 5 GHz. The model lines signify different ~\protect $T_{\rm e}$. All CORNISH sources with an overlapping ~\protect 5$\sigma$ or 7$\sigma$ source were excluded, as they are not resolved in MAGPIS.} 
 \label{fig:spind_tb}
\end{figure}

The spectral indices ($\alpha$) were computed from the relation
\begin{equation}
\alpha =\frac{\mathrm{ln}\left({S}_{\rm 5GHz}/{S}_{\rm 1.4GHz}\right)}{\mathrm{ln}\left(5/1.4\right)} ~,
\label{eq:sp}
\end{equation}
where ${S}_{\rm 5GHz}$ and ${S}_{\rm 1.4GHz}$ are the integrated fluxes at 6~cm and 20~cm, respectively. The uncertainty was found from
\begin{equation}
\Delta \alpha =\frac{\sqrt{{\left({\sigma }_{\rm 5GHz}/{{S}_{\rm 5GHz}}\right)}^{2}+{\left({\sigma }_{\rm 1.4GHz}/{{S}_{\rm 1.4GHz}}\right)}^{2}}}{\mathrm{ln}\left({5}/{1.4}\right)}.
\label{eq:sperr}
\end{equation}

The indices were calculated for the CORNISH 6~cm flux and using the newly measured 20~cm fluxes. As discussed in the previous section, for the purpose of  obtaining reliable spectral indices, all overlapping sources were excluded, thus making it possible to compute 93 spectral indices. 
This should be sufficiently representative of the sample as a whole, as the selection is in this way based on image quality alone.

The brightness temperatures of the sources in the UCHII sample at 6~cm were also calculated. This was done for all CORNISH UCHII regions (see Fig. \ref{fig:tb}), using the equation 
\begin{equation}
{T}_{\rm b}=\frac{1.36{\lambda }^{2}{S}_{\nu}}{{\theta }^{2}} ~,
\label{eq:Tb}
\end{equation}
where the flux density ${S}_{\nu}$ is in mJy, the brightness temperature ${T}_{\rm b}$ is in K, the wavelength $\lambda$ is in cm and the HPBW (half-power beam width) $\theta$ for a Gaussian beam is in arcseconds. For all unresolved sources, this provides a lower limit. 
${T}_{\rm b}$ serves as a measure of optical thickness -- increasing with larger optical depth and reaching the electron temperature ($T_{\rm e}$) of the ionised region in the optically thick limit \citep{siodmiak:2001}. 
The optical depths $\tau_{\nu}$ at 5 GHz were found from
\begin{equation}
{T}_{\rm b}={T}_{\rm e}\left(1-{e}^{-\tau_{\nu}}\right) ~,
\label{eq:tau}
\end{equation}
for ${T}_{\rm e}$ = $10^{4}$~K (Eqn. 4 in \citealt{siodmiak:2001}).
We note that the optical depth results (Fig. \ref{fig:optdepth}) are lower limits in the case of unresolved sources (due to the angular size dependence of ${T}_{\rm b}$), as well as for the most extended sources (due to the possible underestimation of the 5 GHz flux, see Fig. \ref{fig:fluxes}). 

The spectral index analysis was performed by plotting the UCHII spectral indices (Fig. \ref{fig:spind_tb}) as a function of their corresponding 6~cm brightness temperatures and comparing the result to the theoretical model developed by \cite{bojicic:2011} (see Eqn. 6 in their paper), who followed up on the work by \cite{siodmiak:2001}. 
The focus of these works were planetary nebulae, however the model is general and should be applicable to all ionised nebulae. It assumes a uniform nebula characterized by its electron temperature and optical thickness at a corresponding reference frequency.
The individual spectral index values (along with the measured fluxes at 6~cm and 20~cm) can be found in Table \ref{table:fluxes_M}.

The majority of UCHII candidates were found to be within the expected theoretical limits for free-free emission ($-0.1 \lesssim \alpha \lesssim 2$) $-$ that is, between sources with optically thin or optically thick radio continuum emission at both frequencies.
None of the sources have a spectral index above 1.5. However, $\sim$ 18\% of the sources are found below the lower limit (taking errors into account), which is inconsistent with thermal emission. The reason for the apparently non-thermal spectral indices is likely to be the difference in the VLA configuration at 1.4 GHz and 5 GHz (i.e. the $uv$ coverage at higher resolution resulting in filtering out of some of the flux, as seen in Fig. \ref{fig:fluxes}). 
If the 6~cm fluxes were $\sim$ 1.6 times higher (as high as the measured 6~cm MAPGIS fluxes, Fig. \ref{fig:fluxes}), all but two sources (G010.6297$-$00.3380 and G023.8985+00.0647) would be shifted within the thermal bounds. Despite this, we choose to use the CORNISH 6~cm data over the MAGPIS 6~cm images, as the latter show significantly greater image-to-image variations in median sky levels and thus provide fewer 6~cm frames that are viable for the spectral index calculation (only 71/162 MAGPIS 6~cm sources when overlapping sources are excluded). The 20~cm images appear overall of higher quality than the MAGPIS 6~cm data. Therefore, by utilising the CORNISH 6~cm images and the (reduced sample of 93) MAGPIS 20~cm images, we have limited our spectral index analysis to the best available data.
Due to the time difference between the 20~cm observations and (both sets of) the 6~cm observations, source variability, as discussed in \S \ref{variability}, is another possible explanation for some instances of non-thermal (appearing) UCHIIs. Unfortunately, it is not possible at this time to quantify this effect without more 20~cm observations. 

The location of each UCHII in Fig. \ref{fig:spind_tb}  is determined by both its optical thickness and its electron temperature. The model lines in the diagram correspond to applying Eqn. 6 of \cite{bojicic:2011} for electron temperatures $T_{\rm e}$ = 5~$\times \ 10^{3}$~K,  $T_{\rm e}$ = $\rm 1 \ \times \ 10^{4}$~K and $T_{\rm e}$ = 2~$\times \ 10^{4}$~K. For reference, the mean value for the electron temperature of Galactic \hii regions has been estimated to be 8000~K \citep{quireza:2006}.
At lower brightness temperatures, the UCHIIs scatter around the optically thin limit, where we find the majority of the sample.  
 The much larger associated uncertainties of the sources with lower $T_{\rm b}$ (i.e. of lower opacity at 6~cm) should be noted in this case. With increasing $T_{\rm b}$, $\alpha$ increases in agreement with the model, corresponding to $5 \times 10^{3} \rm \ K \lesssim $ $T_{\rm e}$  $\lesssim 2 \times 10^{4} \rm \ K$, revealing sources that appear optically thick. 
Lower electron temperatures would be a better fit to the sample if we take into account the possible under-estimation of the 6~cm flux.  

The presented spectral index results are a validation of the UCHII region nature of our sample.

\subsection{Evidence for short-timescale UCHII variability?}\label{variability}

All sources with available MAGPIS and CORNISH 6~cm images were revisited to look for significant flux changes between the two epochs that cannot be attributed to image quality or other individual reasons. 
Such sources could be variable over short timescales comparable with the difference in time between the MAGPIS 6~cm  survey (obs. 1989--1991) and the CORNISH survey (obs. 2006--2008). 
  The flux ratio lower limits were taken into account to quantify this. 
  The higher flux at the earlier epoch (due to the detection of more extended emission) prevents a reliable investigation of instances of intrinsic flux decrease, particularly for sources larger than 5$''$. 
A further hindrance is that any hypothetical UCHIIs that are completely invisible at the later epoch due to a significant flux decrease over time cannot be reliably differentiated from more extended \hii region phases due to the lower resolution at the earlier epoch. 

 Due to these limitations imposed by the different VLA configurations used at the two epochs, in the context of this work, we use the term variability to refer to increase in flux in the $\sim$ 15~years separating the two surveys. 
The sources which appear to be intrinsically variable all have a flux increase greater than $\sim$ 50\%. 
These are listed in Table \ref{table:variables}. 

\begin{table}
\centering
\caption{Ratio of CORNISH to MAGPIS 5 GHz fluxes for candidate variable UCHII regions (observed 15~years apart). We note that in some cases, the source was not detected in MAGPIS, so the upper flux limit was used instead for the comparison.}
\begin{tabular}{lr}
\hline
\hline
CORNISH name& $F_{\rm C}/F_{\rm M,6cm}$\\
\hline
G011.0328+00.0274 & 2.42 $\pm$ 0.61 \\
G011.9786$-$00.0973 & 3.56 $\pm$ 1.83 \\
G014.5988+00.0198 & 2.92 $\pm$ 0.80 \\
G016.3913$-$00.1383 & 2.88  $\pm$ 0.15 \\
G023.4553$-$00.2010 & 17.42 $\pm$ 1.28 \\
G025.7157+00.0487 & 4.33 $\pm$ 0.49 \\
G030.7579+00.2042 & 73.85 $\pm$ 4.49 \\
G030.7661$-$00.0348 & 7.04 $\pm$ 0.56 \\
G037.7347$-$00.1128  & 22.90 $\pm$ 1.09 \\
\hline
\end{tabular}
\label{table:variables}
\end{table}

\paragraph*{}
The 6-cm variables appear to have several properties in common: 
\begin{enumerate}
\item They exist in relative isolation (no overlapping sources, no other radio sources within $\gtrsim$ 1$'$, no busy complexes). An exception to this is G030.7661\allowbreak$-$00.0348, which is likely experiencing substantially different effects than the rest in the bustling environment of W43; 
\item They are all near-IR dark (apart from G030.7661\allowbreak$-$00.0348); 
\item Most are particularly compact ($\lesssim$ 0.1~pc), with the exception of G016.3913\allowbreak$-$00.1383. 
 \item Their 5GHz--1.4GHz spectral indices are not anomalous with respect to the rest of the sample.

\end{enumerate}

The potentially variable sources comprise $\sim$ 5\% of the CORNISH UCHII sample. 
 In four cases (G011.9786\allowbreak$-$00.0973, G014.5988\allowbreak+00.0198, G023.4553\allowbreak$-$00.2010, and G030.7661\allowbreak$-$00.0348), the CORNISH source was not detected at both 6 and 20~cm at the earlier observational epoch. 
 Three sources (G025.7157$+$00.0487, G030.7579\allowbreak+00.2042, and G037.7347\allowbreak$-$00.1128) all have only a 20~cm detection at the earlier epoch (and there is a dim 20~cm counterpart for G011.0328\allowbreak+00.0274). The extended G016.3913\allowbreak$-$00.1383 has 6 and 20~cm counterparts, but these appear smaller and dimmer at the earlier epoch. 
 Although there are some suggestive correlations, this group of candidate variable sources is clearly statistically insufficient to establish any common pattern linking the (presence or lack of)  emission at 6 and 20~cm at the same epoch. 
   Unfortunately, the 6~cm variability cannot be linked in any way to variability at 20~cm, due to the lack of other available 20~cm data of comparable quality for the 6-cm variables.   

Time-variable radio flux densities of several ultra-compact and hyper-compact (HC) \hii regions have been reported previously (\citealt{acord:1998}, \citealt{hernandez:2004}, \citealt{rodriguez:2007}, \citealt{gomez:2008}, \citealt{madrid:2008}). The flux changes have been associated with morphological changes across observational epochs, on timescales of a few years. 
The UC and HC \hii regions have been caught expanding \citep{acord:1998}. \cite{madrid:2008} discuss contracting UC and HC \hii regions; however their sources are unresolved. Variability over observable timescales could be caused by factors that are either external or internal to the forming star. The former could be the result of chaotic motions of the material surrounding the ionising star. In this scenario, optically thick gas (e.g. clumps in the stellar wind) would occasionally block the outgoing radiation, shielding the outer ionised gas layers and thus neutralising them \citep{peters:2010a}.
In the latter scenario, the forming star itself is undergoing changes \citep{hernandez:2004} $-$ surface temperature fluctuations affect the UV flux and thus the \hii region size.  

\parskip 0pt 

Theoretical studies have reproduced this behaviour, referred to as flickering. 
The three-dimensional collapse simulations of massive star formation of \cite{peters:2010a, peters:2010b, peters:2010c} and \cite{klassen:2012a} include feedback by ionising radiation and show time variability leading to changes in \hii region appearance and flux comparable to observations. 
The \cite{peters:2010a} model produces flickering on scales of $\sim$ 10~years. In this model, accretion has not ceased prior to the UCHII stage. The infalling neutral flow becomes ionised when in close proximity to the star. The \hii region is gravitationally trapped early on, which is followed by a fluctuation between trapped and extended states, and thus changes in flux, size and morphology are seen over time. 
\cite{madrid:2011} performed statistical analysis of simulated radio-continuum observations separated by 10 year steps, using the \cite{peters:2010a, peters:2010b, peters:2010c} models to form \hii regions. They found that 7\% of the simulated HC and UC \hii regions have a detectable flux increase (larger than 10\%) and 3\% have a detectable flux decrease, but expect only $\sim$ 0.3\% of their \hii regions to have a flux increase of over 50\%.
The observations discussed in this work show that $\sim$ 5\% of the CORNISH UCHII regions have become brighter by 50\% or more over a comparable time scale ($\sim$ 15~years), based on the 6~cm data. 
In practice, any similar statistic of observed sources with flux increase $\lesssim$ 10\% would be unreliable, given the associated flux uncertainties. In any case, assuming that all of our candidate variable sources truly undergo intrinsic changes, variable HC and UC \hii regions could be significantly more common and their brightness could fluctuate more than predicted. If this is the case, invoking ongoing accretion cannot account for the observed dramatic change in flux.
The \cite{peters:2010a} model does not include radiation pressure, magnetic fields, winds and outflows, all of which are components of the star formation process and might be related to variability.

\section{Distances}\label{distances}\label{analysis}

A crucial step towards characterising UCHII regions is to determine the distance to each source. 
One can then convert measured parameters (e.g. fluxes and angular sizes) into physical quantities  (e.g. luminosities and physical sizes). Accurately derived distances to UCHII regions are used to test the current models of the face-on Galactic structure \citep[see e.g.][]{urquhart:2013}. 
As the heavy obscuration hinders the use of any optical distance determination techniques, the distances to most UCHII regions are kinematically derived.  The kinematic distance is found by fitting the radial velocity of the source to a Galactic rotation curve (e.g. \citealt{brand:1993, reid:2009}), using radio or mm spectral line data. Errors in the calculated distance arise when the source radial velocity differs from the one assumed by the model (e.g. velocity errors  of about 10$\kms$ due to the velocities departing from circular rotation as a result of streaming motions) \citep{bania:2009}.

Distance estimates for the outer Galaxy are relatively straightforward. However, a major obstacle arises when one seeks the kinematic distance for objects within the Solar circle. At Galactic radii smaller than that of the Sun, two possible distances exist for each radial velocity. These distances, known as near and far, are situated at equal intervals from the tangent point distance. The kinematic distance ambiguity (KDA) is not present only for the tangent point velocity, which is the maximum radial velocity. Different methods exist to assign the correct kinematic distance to the sources of interest -- for example, \hi Emission/Absorption (\hi E/A), \hi Self-Absorption (\hi SA), or using absorption lines from other molecules, for example formaldehyde ($\rm H_{2}CO$) (as discussed by \citealt{bania:2009}). 

These methods are particularly effective for bright UCHII regions, whose free-free continuum emission is substantially stronger than the Galactic \hi emission, thus resulting in unambiguous absorption spectra.  
 As the maximum radial velocity along the line of sight is always the tangent velocity, lack of absorption between the source and tangent velocity reveals that the source is located at the near distance. Otherwise, the far distance is assigned (or the tangent distance, in the cases when the source velocity equals the tangent velocity).
In this work, the \hi E/A method was adopted \citep[see e.g.][]{urquhart:2012}. This method makes use of CO emission line data and \hi absorption to obtain the near and far kinematic distances and to attempt to resolve the KDA. 

\subsection{Distances from ATLASGAL and RMS}

The work by \cite{urquhart:2013} presents an unbiased and complete sample of 170 molecular clumps with 213 embedded compact and ultra-compact \hii regions over the common GLIMPSE, ATLASGAL and CORNISH survey region.
Table 3 from \cite{urquhart:2013} contains distances to all CHII- and UCHII-hosting clumps and Table 4 contains Lyman continuum fluxes (${F}_{\rm Ly}$) and source physical sizes. Out of the associated 213 \hii regions, 203 also belong to the CORNISH UCHII sample (ten were classified as more extended).   
Of the remaining 36 UCHII candidates without ATLASGAL distances, eight were found to have a distance estimate in the Red MSX source survey (RMS) database\footnote{\url{http://rms.leeds.ac.uk/cgi-bin/public/RMS_DATABASE.cgi}} \citep{lumsden:2013}. These are the distances for G010.6297\allowbreak$-$00.3380, G030.6881\allowbreak$-$00.0718, G032.0297\allowbreak+00.0491, G035.0524\allowbreak$-$00.5177, G038.5493\allowbreak+00.1646, G048.6099\allowbreak+00.0270, G060.8842\allowbreak$-$00.1286, and G061.7207\allowbreak+00.8630.

\subsection{Distance estimates for remaining sources using CO data}

\cite{anderson:2009} describe a large-scale study of the molecular properties of \hii regions of different sizes and morphologies using fully sampled CO maps. This is the BU-FCRAO Galactic Ring Survey \citep{jackson:2006}, which uses $^{13}$CO $J=1\longrightarrow0$ emission. 
This has advantages over the commonly used $^{12}$CO isotopologue, as $^{13}$CO is $\sim$ 50 times less abundant and thus provides a smaller optical depth, and consequently smaller line widths and better separation of velocity components along the line of sight. 

CO data-cubes from the Galactic Ring Survey\footnote{\url{http://www.bu.edu/galacticring/}} were used to obtain radial velocities for the remaining sources without ATLASGAL or RMS distances. 
The positional accuracy of the GRS is $\sim$ 2.3$''$, which is equivalent to 1/10 of the spacing between grid points on the map. 
Data cubes for the available sources were obtained, and the radial velocities were measured for the sources coinciding with the CORNISH coordinates. The emission line structure of the data is very complex, often with multiple emission lines. This reflects the complexity of the ($l, b, v_{\rm LSR}$) structure of the molecular gas in the line of sight of the \hii region.
The sources were located within the 15$'$ data cubes by going manually through the cube channels and then mapping the cube once the source was found at or close to the precise CORNISH coordinates. A source's velocity was taken to be equal to the velocity of the most prominent emission line in the map at the exact source position. The final CO source velocity results were compared to Table 3 in \cite{anderson:2009} for the sources in common. 
The code by \cite{reid:2009} was used to obtain near and far kinematic distances corresponding to each radial velocity estimate (see \S \ref{kda}). 

\begin{table}
\centering
\caption{Comparison between velocities (in $\kms$) measured in this work and  by \protect\cite{anderson:2009}, Table 3 for the sources in common. As the peak channel was used to calculate the velocities, the associated errors are given by the channel width. This equals 0.21~$\rm km~s^{-1}$ for each BU-FCRAO Galactic Ring Survey cube. In practice this error is outweighed by the error due to peculiar motions, which is $\sim$ 10 $\kms$.} 
\begin{tabular}{lr|lr}
\hline
\hline
\multicolumn{1}{l}{CORNISH} & \multicolumn{1}{c}{$v$} & \multicolumn{1}{|l}{Anderson}&  \multicolumn{1}{c}{$v$} \\
\hline
G024.4698+00.4954&103&C24.47+0.49 &102.67\\
G024.4721+00.4877&102.8&C24.47+0.49 &102.67\\
G024.4736+00.4950&102.6&C24.47+0.49 &102.67\\
G024.8497+00.0881&109.3&C24.81+0.10 &108.31\\
G030.7661$-$00.0348&96.1&C30.78$-$0.03 &94.76\\
G030.7661$-$00.0348&96.1&U30.84$-$0.11b&96.89\\
G031.2420$-$00.1106&21.1&U31.24$-$0.11a&21.07\\
G034.2544+00.1460&57.7&U34.26+0.15 &57.1\\
G034.2571+00.1466&57.7&U34.26+0.15 &57.1\\
G037.9723$-$00.0965&54.7&C38.05$-$0.04 &54.1\\
G049.4640$-$00.3511&59.5&U49.49$-$0.37 &60.08\\
G049.4891$-$00.3763&60.9&U49.49$-$0.37 &60.08\\
G050.3157+00.6747&26.5&U50.32+0.68 &26.31\\
\hline
\end{tabular}
\label{table:comparison}
\end{table}

\subsection{Resolving the KDA} \label{kda}

The \hi Emission/Absorption (\hi E/A) method was implemented as a standard way to choose between the calculated near and far distances \citep[see e.g.][]{bania:2009,urquhart:2012}. The method has proven to be very successful for KDA resolution \citep{bania:2009}. 

\begin{table}
\centering
\caption{KDA-resolved UCHII distances. The distances and errors were derived with the  ~\protect\cite{reid:2009} code and the \hi E/A method was used to resolve the KDA. The sources that are found at the near, far, and tangent distance, are labelled with n, f, and t, correspondingly. The third column indicates whether the \hi spectra used for the KDA resolution were of good quality (i.e. mostly in absorption, marked with \cmark) or poor (i.e. mostly in emission, indicated with \xmark).} 
\begin{tabular}{lccr}
\hline
\hline
CORNISH name&KDA & Strong \hi abs.&$d$ (kpc)\\
\hline
G024.4698+00.4954&n & \cmark & 5.5 $\pm$ 0.3 \\ 
G024.4721+00.4877&n&\cmark & 5.5 $\pm$ 0.3  \\ 
G024.4736+00.4950&n &\cmark & 5.5 $\pm$ 0.3 \\
G024.8497+00.0881&n& \cmark & 5.8 $\pm$ 0.3\\	 
G026.0083+00.1369&f&\xmark & 13.8 $\pm$ 0.5 \\ 
G026.8304$-$00.2067&f&\xmark & 11.9 $\pm$ 0.3\\
G029.7704+00.2189&f&\xmark & 9.8 $\pm$ 0.3 \\ 
G030.7579+00.2042&t &\xmark & 7.2 $\pm$ 0.6\\		
G030.7661$-$00.0348&t& \cmark & 7.2 $\pm$ 0.6\\ 
G031.2420$-$00.1106&f& \xmark & 12.7 $\pm$ 0.4 \\ 
G034.2544+00.1460&n& \cmark & 3.6 $\pm$ 0.4 \\ 
G034.2571+00.1466&n& \cmark &3.6 $\pm$ 0.4 \\ 
G035.4570$-$00.1791&f& \xmark & 9.7 $\pm$ 0.4 \\ 
G037.7562+00.5605&f&\xmark & 12.2 $\pm$ 0.5 \\  
G037.9723$-$00.0965&f& \xmark & 9.7 $\pm$ 0.4\\ 
G045.5431$-$00.0073&t& \xmark & 5.9 $\pm$ 0.9 \\ 	
G049.4640$-$00.3511&t&\cmark & 5.5 $\pm$ 2.2 \\
G049.4891$-$00.3763&t& \cmark &5.5 $\pm$ 2.2\\
G050.3157+00.6747&f& \xmark & 8.6 $\pm$ 0.5 \\ 
G061.4763+00.0892&t& \cmark & 4.0 $\pm$ 1.5\\ 
G061.4770+00.0891&t&\cmark &4.0 $\pm$ 1.5\\
\hline
\end{tabular}
\label{table:kda_dist}
\end{table}
Out of the 28 CORNISH UCHIIs with missing CO distances, CO and VGPS \hi data were found for 21 (CO data were not available within the range 10$^{\circ}\leq l \leq 17\degree)$. 
In the VGPS spectra, the \hi brightness temperature at the source location  is overall $\lesssim$ 100 K, as expected for optically thin \hi gas.  
Therefore, a background source of higher temperature should be seen in \hi absorption. 
The presence of a \hii region would provide a sufficiently strong continuum to detect a line in absorption. 
It should be noted, however, that  the VGPS synthesized beam size is $\sim$ 45$''$ (FWHM) at 21 cm (see \citealt{stil:2006} for the VGPS survey paper). The \hii region may not be detected at all if the source size is not comparable to the survey beam size \citep{urquhart:2012}. The extent of the effects of this on our distance estimates are hard to quantify. 
Moreover, neighbouring CORNISH UCHIIs found within the beam cannot be distinguished. This is the case, for instance, for G024.4721\allowbreak+00.4877, G024.4736\allowbreak+00.4950, and G024.4698\allowbreak+00.4954, whose \hi spectra are practically identical.  The latter issue is alleviated by the fact that spatially close UCHIIs are likely to belong to the same star-forming region, and hence to be situated at the same distance. 

To assign a near or a far distance, the \hi data were plotted, together with a line marking the assigned CO radial velocity and its associated error ($\pm$ 10$\kms$ due to peculiar motions) and the calculated tangent velocity was included as well (see Appendix \ref{appendixB}). 
 All sources with CO velocity within 10$\kms$ of the tangent point velocity were assigned the tangent point distance, in order to limit wrong assignments to the near distance (as suggested by \citealt{bania:2009}). Additional uncertainty arises due to half of the \hi spectra being dominated by emission instead of absorption at the UCHII source position. The presence or lack of convincingly strong absorption is indicated in Table \ref{table:kda_dist} for each \hii region. An example of a reliable \hi spectrum is that coinciding with G049.4891$-$00.3763, as it clearly shows very strong absorption lines. Emission-dominated spectra appear similar to the one coincident with G029.7704\allowbreak+00.2189 (Appendix \ref{appendixB}).
  
All 21 UCHIIs with available GRS CO and VGPS \hi data were assigned a distance, and four more  distances were adopted from the paper by \cite{cesaroni:2015}. These are the distances for G010.3204\allowbreak$-$00.2328, G011.1712\allowbreak$-$00.0662, G014.1741\allowbreak+00.0245, and G016.3913\allowbreak$-$00.1383. 
It should be noted that for these four sources, as for all sources without data to resolve the ambiguity, \cite{cesaroni:2015} assigned the far distance. Placing sources without a distance solution at the near distance is preferable.  
Only three sources (G010.8519$-$00.4407, G011.9786$-$00.0973, and G014.1046+00.0918) with no known distance and no available emission and absorption data remain in our sample. From the group of 21 UCHIIs, six were assigned the near distance, eight -- the far, and the remaining seven sources were assigned the tangent point distance. 

\section{Derived physical properties} \label{physprop}

\begin{figure*}
\begin{subfigure}[b]{0.5\textwidth}
\caption{}
\vspace{-2.5mm}
   \includegraphics[width=0.95\columnwidth]{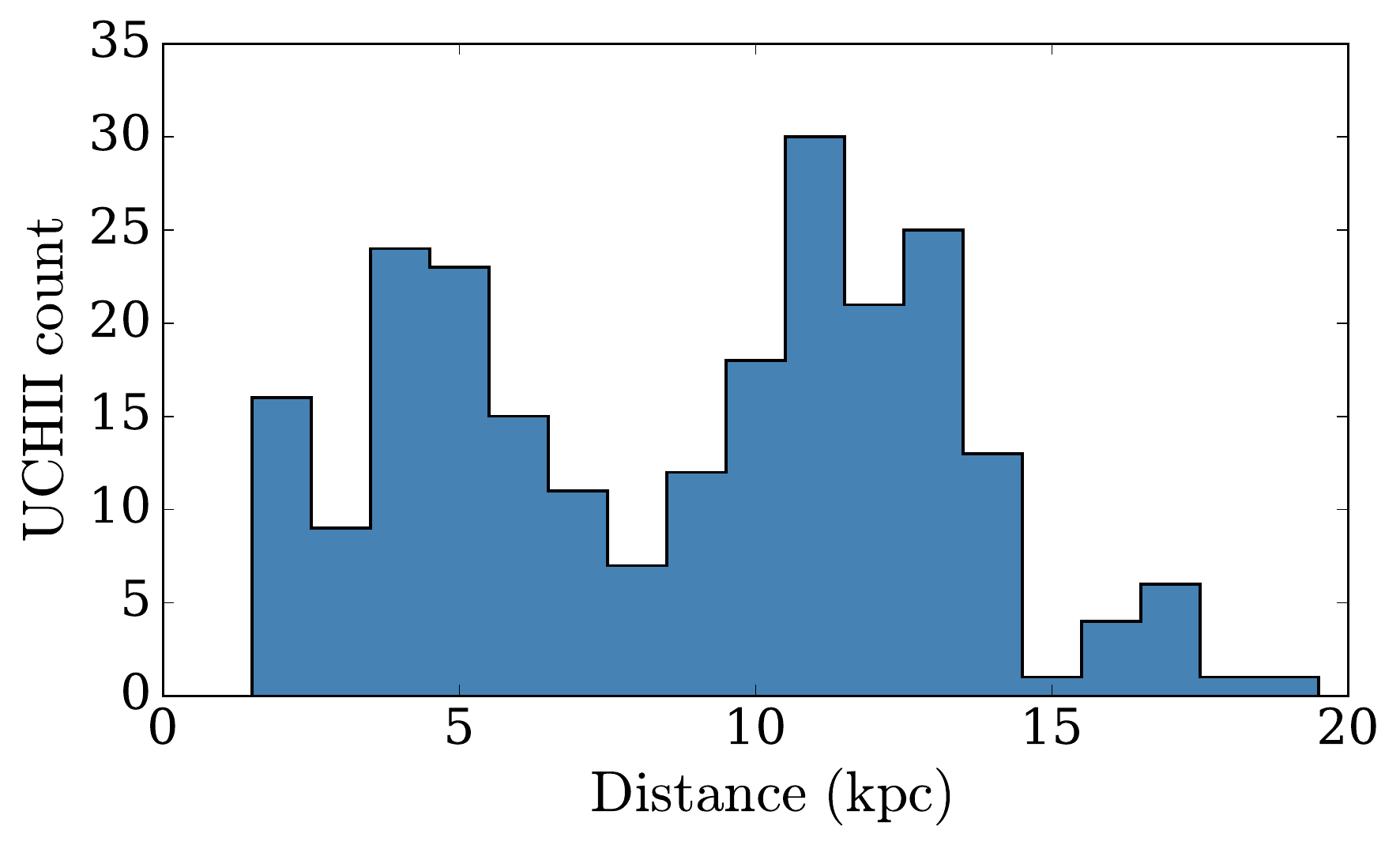}
      \label{fig:dist} 
\end{subfigure}
\begin{subfigure}[b]{0.5\textwidth}
\caption{}
\vspace{-2.5mm}
   \includegraphics[width=0.95\columnwidth]{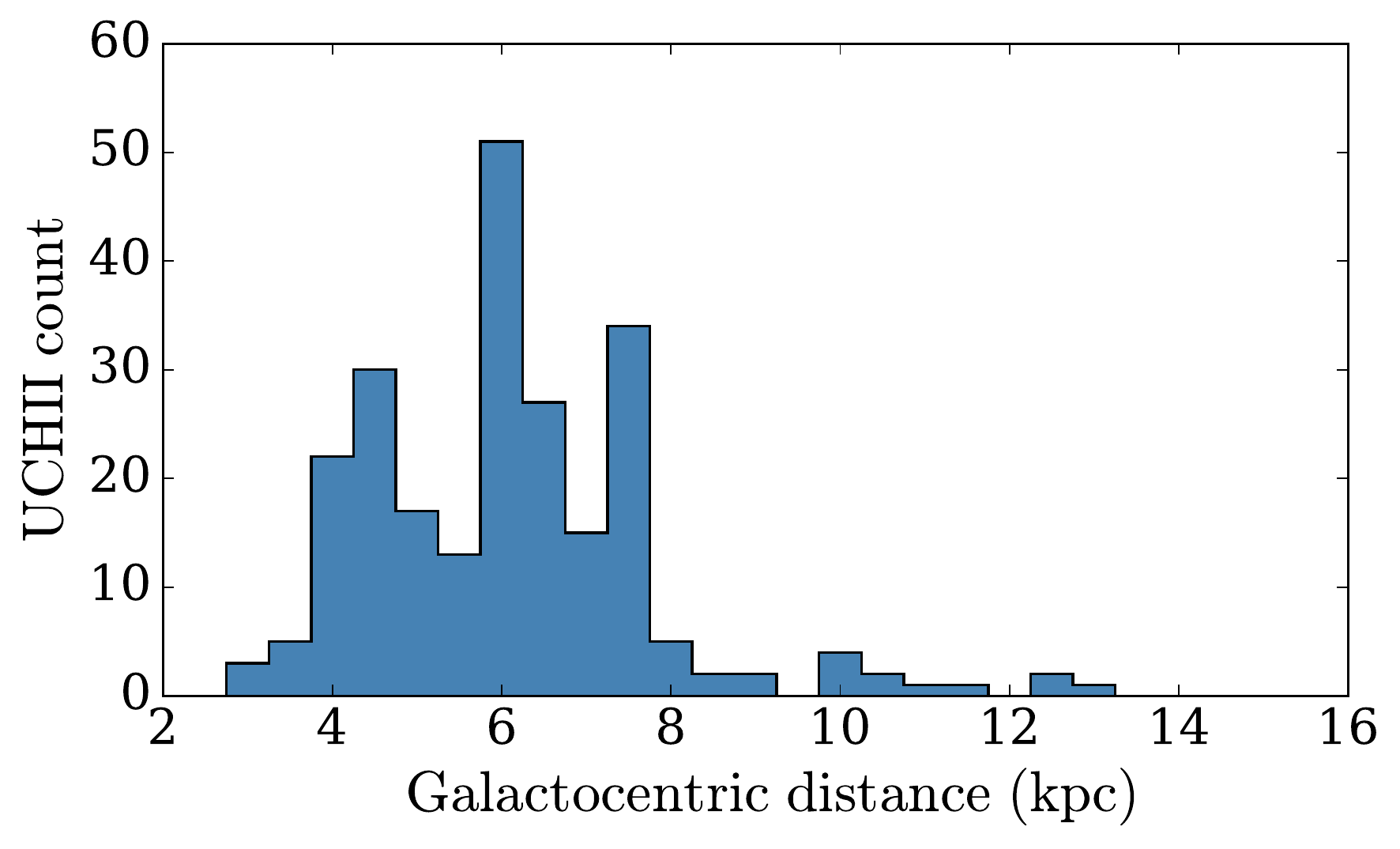}
      \label{fig:galdist} 
\end{subfigure}
\newline
\vspace{1mm}
\begin{subfigure}[b]{0.5\textwidth}
\caption{}
\vspace{-2.5mm}
   \includegraphics[width=0.95\columnwidth]{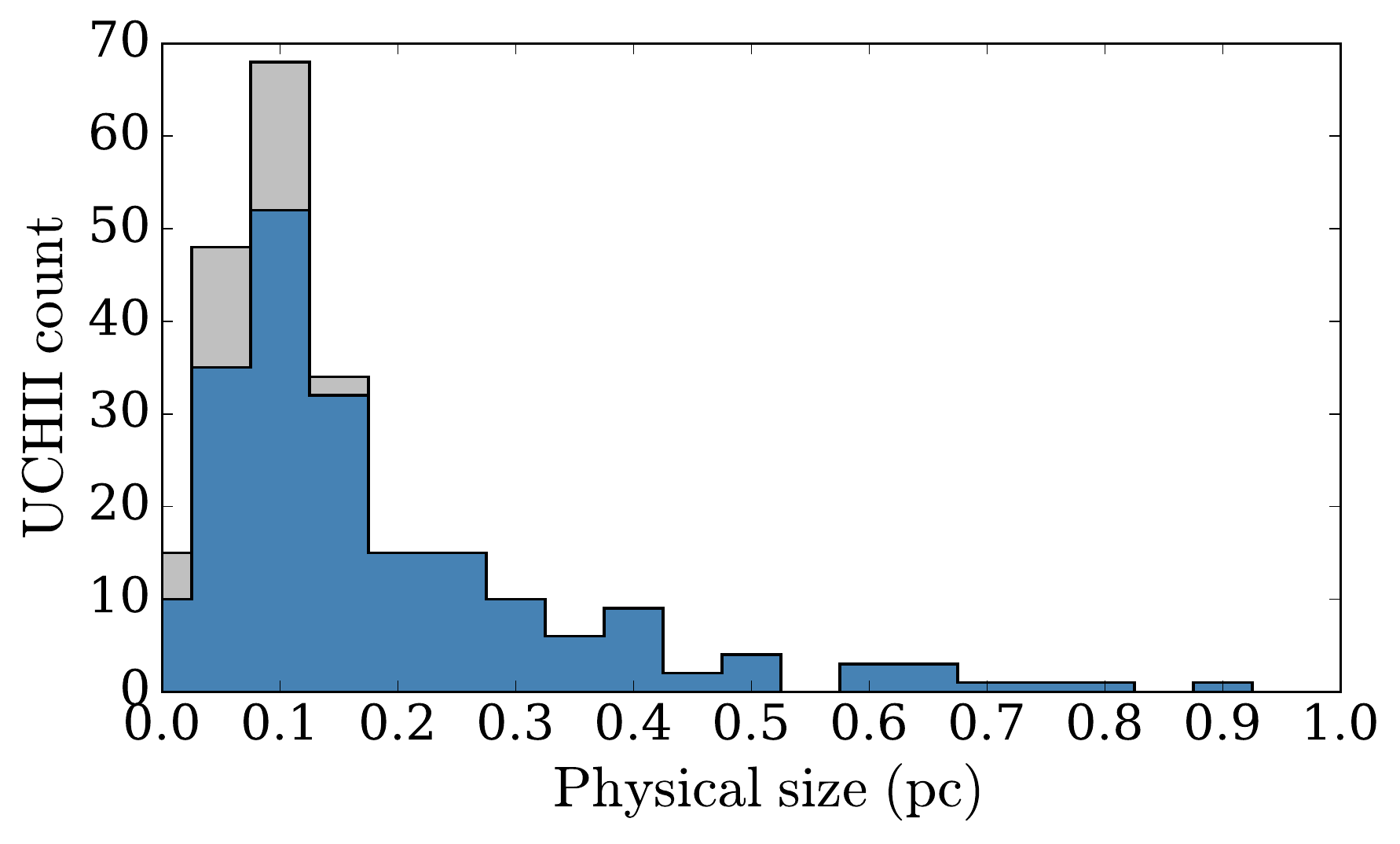}
      \label{fig:phsize} 
\end{subfigure}
\begin{subfigure}[b]{0.5\textwidth}
\caption{}
\vspace{-2.5mm}
   \includegraphics[width=0.95\columnwidth]{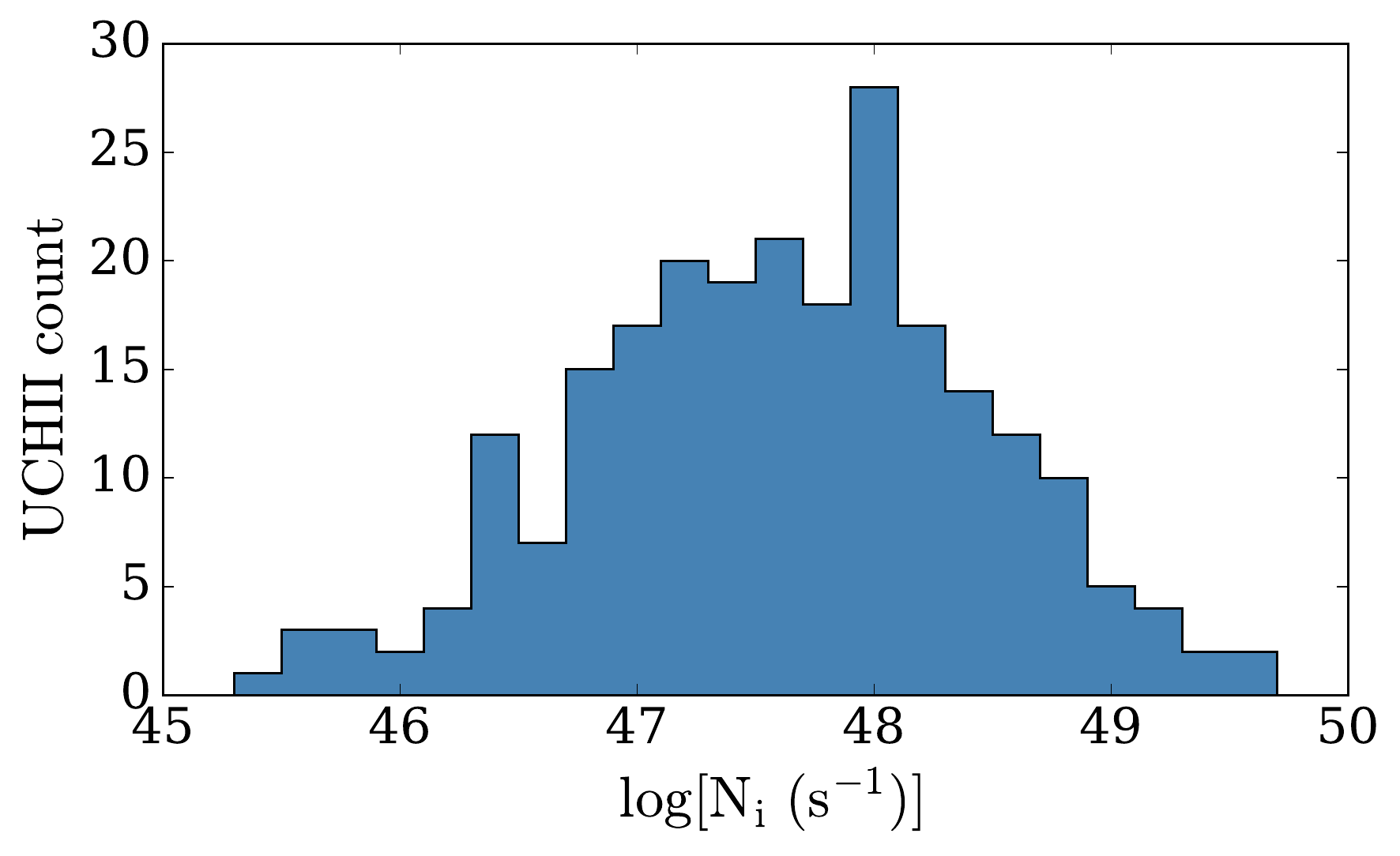}
      \label{fig:ly} 
\end{subfigure}
\newline
\vspace{1mm}
\begin{subfigure}[b]{0.5\textwidth}
\caption{}
\vspace{-2.5mm}
   \includegraphics[width=0.95\columnwidth]{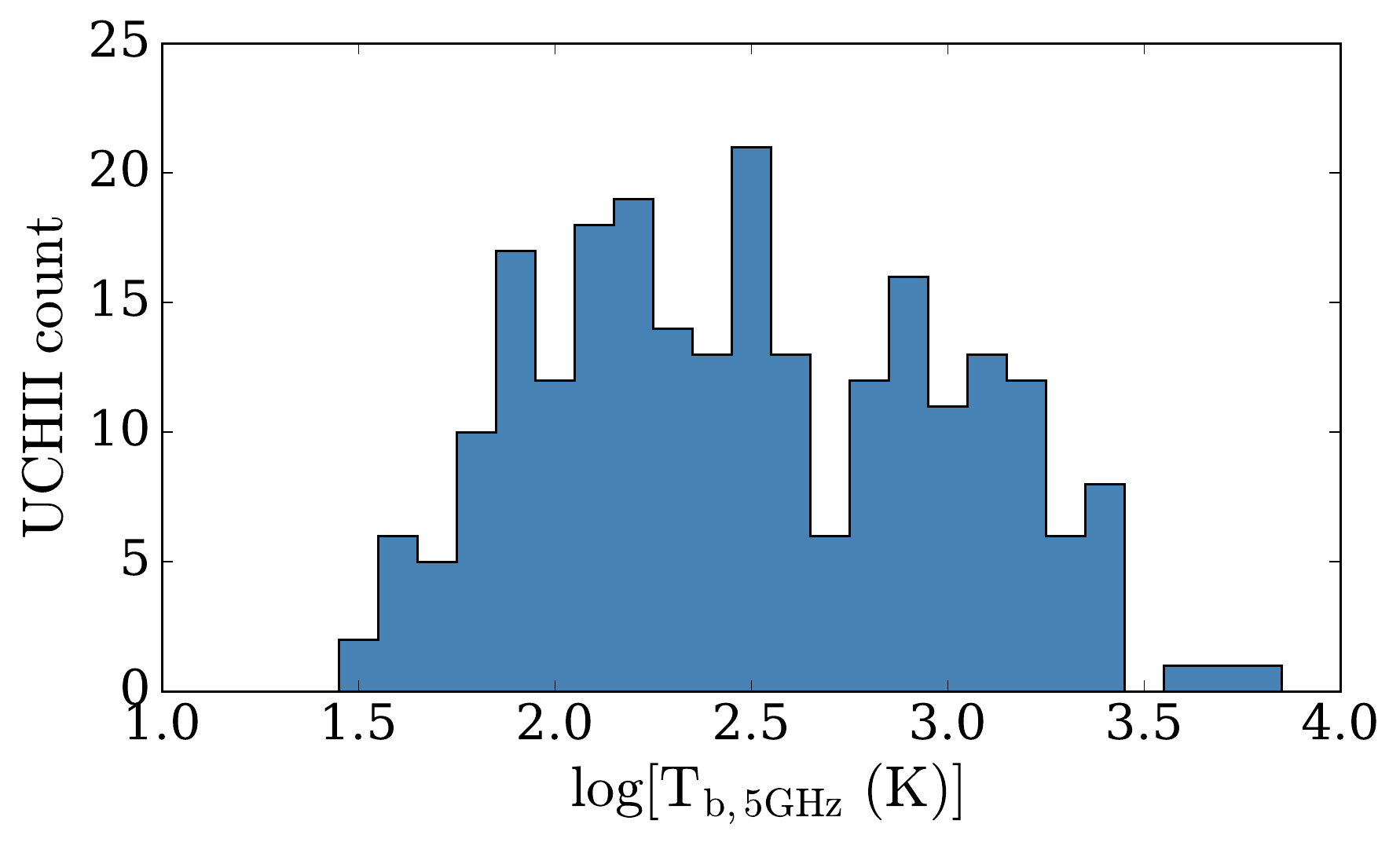}
      \label{fig:tb} 
\end{subfigure}
\begin{subfigure}[b]{0.5\textwidth}
\caption{}
\vspace{-2.5mm}
   \includegraphics[width=0.95\columnwidth]{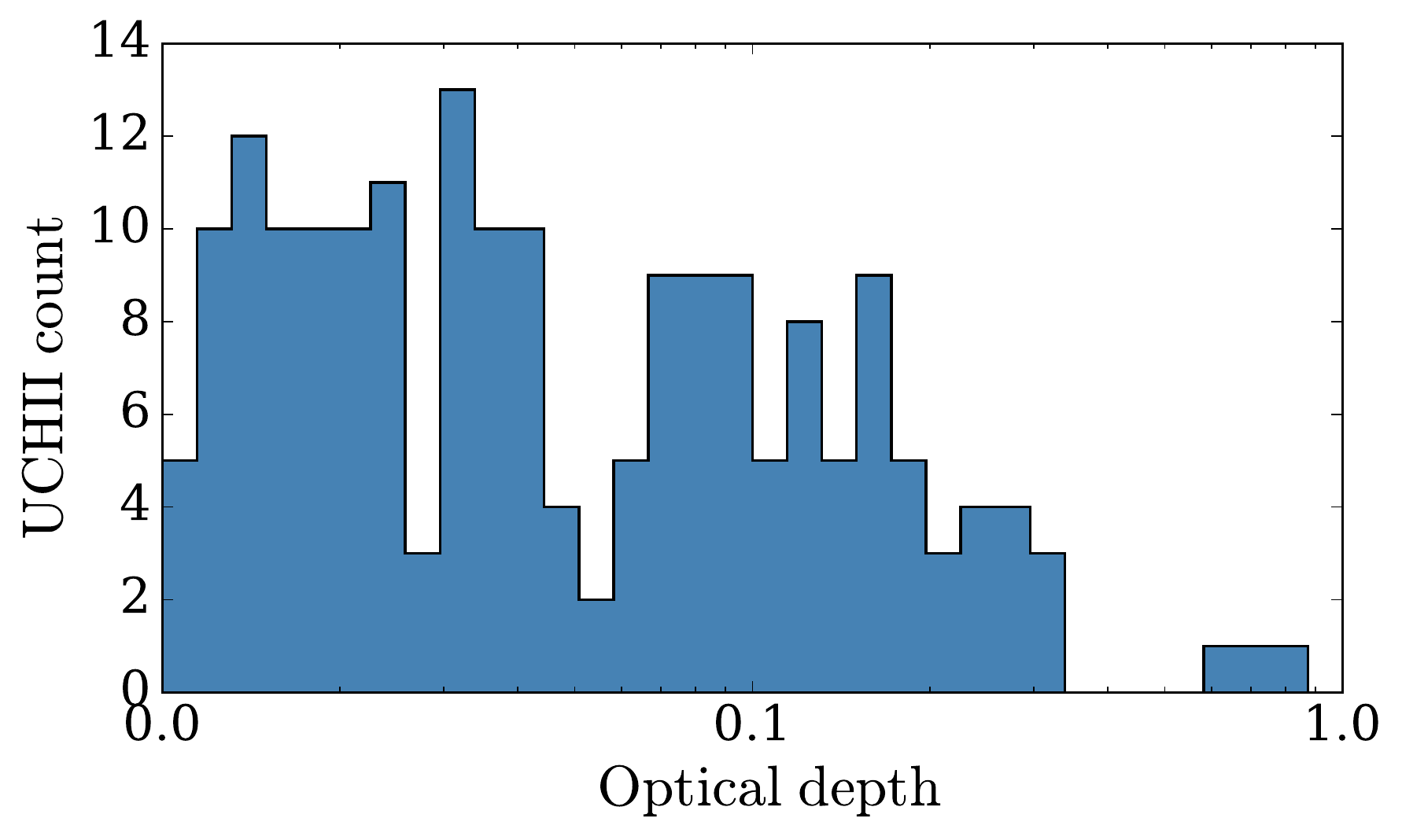}
      \label{fig:optdepth} 
\end{subfigure}
\newline
\vspace{1mm}
\begin{subfigure}[b]{0.5\textwidth}
\caption{}
\vspace{-2.5mm}
   \includegraphics[width=0.95\columnwidth]{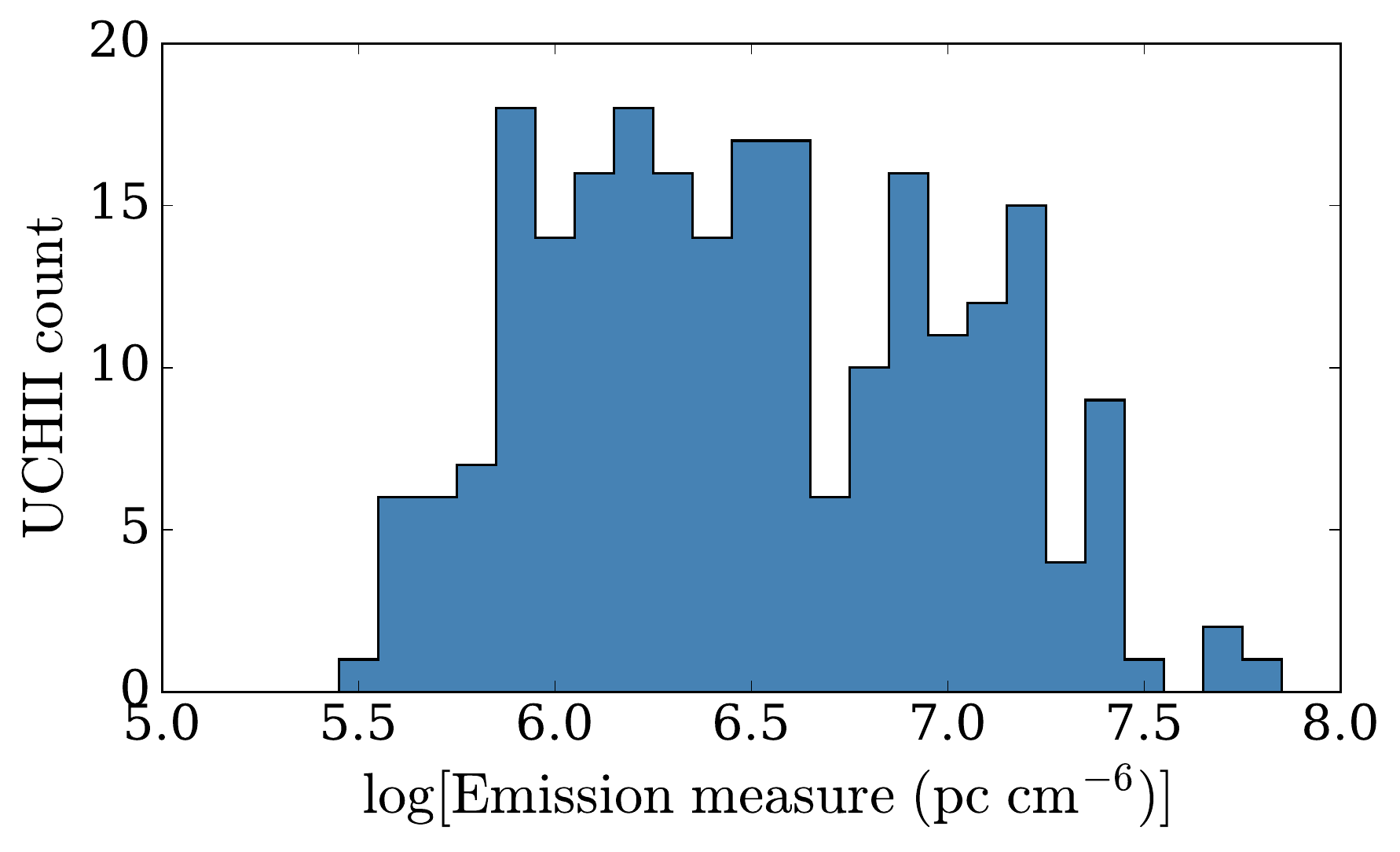}
      \label{fig:em} 
\end{subfigure}
\begin{subfigure}[b]{0.5\textwidth}
\caption{}
\vspace{-2mm}
   \includegraphics[width=0.95\columnwidth]{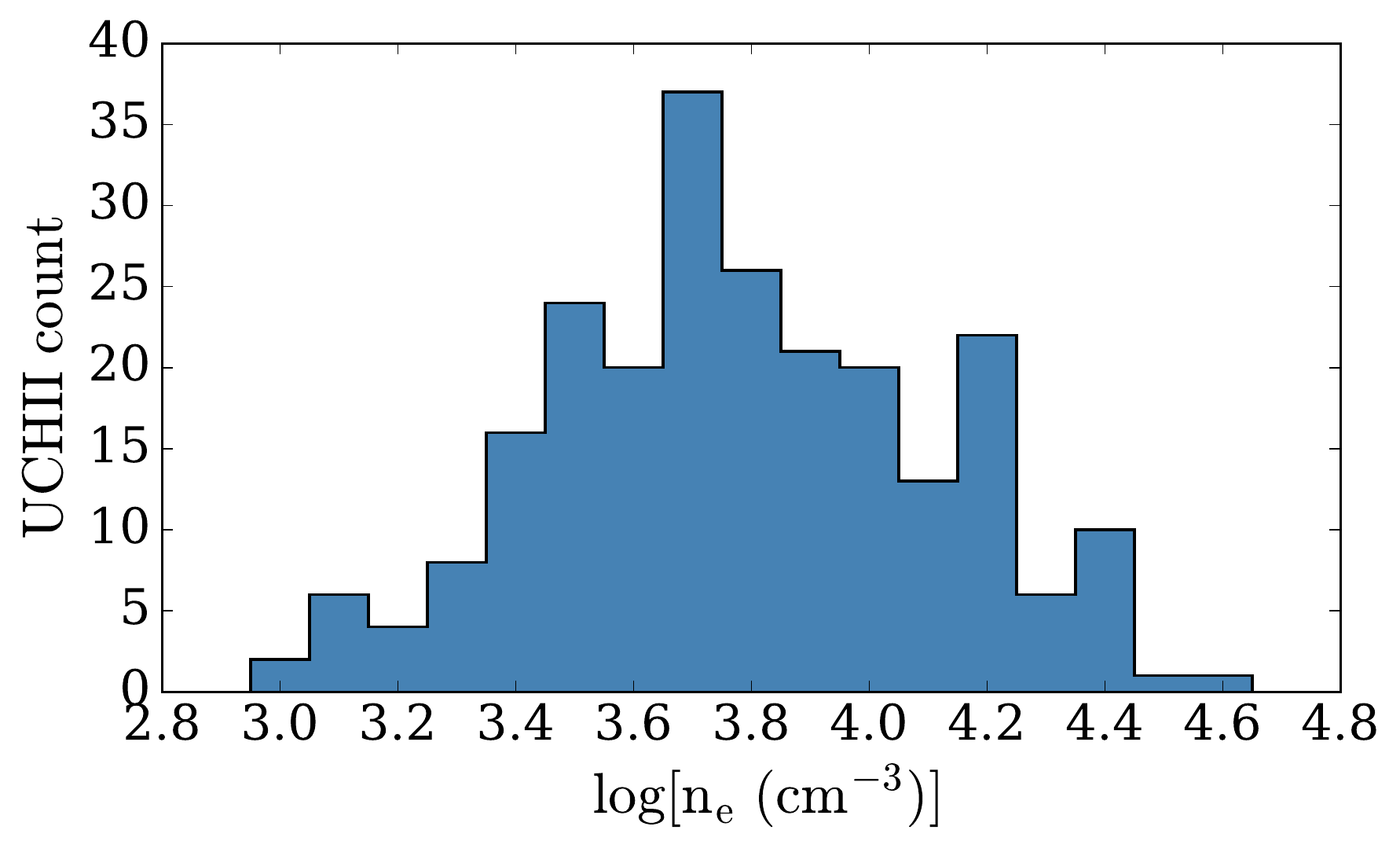}
      \label{fig:ne} 
\end{subfigure}
\caption{Parameter distributions of the CORNISH UCHIIs. The distance histogram (a) includes values from ~\protect \cite{urquhart:2012} and \protect \cite{urquhart:2013}, as well as the distances computed in this work. The grey area in panel (c) shows upper limits for sizes of unresolved sources. The Lyman continuum flux histogram  (d) includes results from \protect \cite{urquhart:2013}, as well as results from this work, with $ \pm $ 10\% associated error. 
There are $ \pm $ 30\% errors on the emission measures (g) and  $ \pm $ 20\% errors on the electron densities (h).} 
\label{fig:allparameters}
\end{figure*}

A summary of the physical properties of the sample is presented in Fig. \ref{fig:allparameters}. 
 The distributions of heliocentric and galactocentric distances, physical sizes, optical depths, brightness temperatures, Lyman continuum fluxes, emission measures, and electron densities are shown. The computed values corresponding to each CORNISH UCHII can be found in Appendix \ref{appendixE}.
 
Fig. \ref{fig:dist} shows the distribution of the UCHII regions with heliocentric distance.  The five peaks are at 2, 4, 11, 13, and 17~kpc. As noted by \cite{urquhart:2013} (who also find five similarly situated peaks at 2, 5, 10, 12, and 16~kpc), they likely correspond, in turn, to the near side of the Sagittarius arm, the end of the bar and Scutum-Centaurus arm, the far sides of the Sagittarius and Perseus arms, and the Norma arm. \cite{urquhart:2013} studied the associated clumps to the CORNISH compact \hii regions and found that the most prominent peak in the heliocentric distance histogram at $\sim$ 11~kpc  corresponds to the W49A complex (as this peak is not seen for clumps). 

The galactocentric distances were also estimated (Fig. \ref{fig:galdist}). The galactocentric distance distribution depends only on the choice of Galactic rotation curve \citep{urquhart:2011}. The peaks are located at $\sim$ 4.5, 6, and 7.5~kpc, similarly to the findings of \cite{urquhart:2011} (peaks at $\sim$ 4, 6, and 8~kpc) for the young massive star sample in the RMS survey. \cite{urquhart:2011} identify the peak at approximately 4~kpc to be at the intersection of the Long Bar and the Scutum-Centaurus arm, also coinciding with the W43 complex. The 6~kpc peak is coincident with the Sagittarius arm, and the 7.5~kpc peak corresponds to the Perseus arm (the bin is dominated by the W49A complex -- the most active star-forming region in the Galaxy). 

The distribution of the physical sizes is shown in Fig. \ref{fig:phsize}. All sources with available deconvolved angular sizes \citep[see][]{purcell:2013} are presented in the histogram in blue. The physical sizes of 36 UCHIIs could not be obtained because the corresponding sources were unresolved and hence their true sizes are unknown (and three more sources do not have a computed distance). Upper limits on the physical sizes were computed for these unresolved sources with available distances (grey region in the histogram). 
Most sources ($\sim$ 66\%) are larger than 0.05~pc and smaller than 0.2~pc in diameter, and the distribution peaks at 0.1~pc, which is consistent with the typical sizes of UCHII regions. About 12.5\% are with sizes between 0.01~pc and 0.05~pc, and the remaining $\sim$ 21.5\% sources are between 0.2~pc and 0.9~pc. The investigated properties of these sources are in accord with the rest of the sample and therefore likely exhibit the natural variation in size one would expect for a continuum of spectral subtypes and ages, and for different ambient densities. 
 
Lyman continuum fluxes from \cite{urquhart:2013} were available for 203 out of the 239 sources in the CORNISH sample. 
The Lyman continuum fluxes ${N}_{i}$ for 33 of the remaining sources were computed from 
\begin{equation}
\left[\frac{{N}_{i}}{\rm ~photons~ {s}^{-1}}\right]=9\times{10}^{43}\left[\frac{{S}_{\nu }}{\rm ~mJy}\right]\left[\frac{{d}^{2}}{\rm ~kpc}\right]\left[\frac{{\nu }^{0.1}}{5\rm ~GHz}\right] ~,
\label{eq:Ni}
\end{equation} 
where ${S}_{\nu}$ is the flux at a frequency $\nu$ and $d$ is the distance (Eqn. 6  from \citealt{urquhart:2013}).
The associated error on ${N}_{i}$ is 10\%, which is dominated by the distance error. As for the emission measures and electron densities, a dust-free, optically thin emitter is assumed here. The Lyman continuum flux distribution (Fig. \ref{fig:ly}) of the CORNISH UCHII sample peaks between 47.2 < log($N_{\rm~ i}$) < 48.5~$\rm~photons~s^{-1}$, which is also consistent with the Lyman continuum ionising flux from zero-age main-sequence stars with spectral class from B0 to O7. The computed Lyman continuum fluxes should be treated as lower limits.  

The brightness temperature distribution at 6~cm and the corresponding optical depth distribution, as calculated in \S \ref{spind} (Eqns. \ref{eq:Tb} and \ref{eq:tau}), are shown in Figs. \ref{fig:tb} and \ref{fig:optdepth}, respectively. The optical depth histogram, as well as Fig. \ref{fig:spind_tb}, confirm that assuming optically thin sources at 5 GHz is justified for the majority of UCHIIs in our sample.

The emission measures ($EM$, Fig. \ref{fig:em}) and the electron densities ($n_{\rm e}$, Fig. \ref{fig:ne}) of the UCHII regions were computed from the equations

\begin{equation}
\left[\frac{EM}{\rm~ pc \ {cm}^{-6}}\right]=1.7\times{10}^{7}\left[\frac{{S}_{\nu }}{\rm~Jy}\right]{\left[\frac{\nu}{\rm~ GHz}\right]}^{0.1}{\left[\frac{{T}_{\rm e}}{\rm~ K}\right]^{0.35}}{\left[\frac{\theta_{S}}{''}\right]^{-2}}  ~
\label{eq:EM}
\end{equation}
and
\begin{equation}
\begin{split}
\left[\frac{n_{\rm e}}{\rm~ {cm}^{-3}}\right]=2.3\times{10}^{6}\left[\frac{{S}_{\nu }}{\rm~Jy}\right]^{0.5}{\left[\frac{\nu}{\rm~ GHz}\right]}^{0.05}{\left[\frac{{T}_{\rm e}}{\rm~ K}\right]^{0.175}}{\left[\frac{\theta_{S}}{''}\right]}^{-1.5}
\\
\times {\left[\frac{d}{\rm~ pc}\right]}^{-0.5} ~,
\label{eq:ne}
\end{split}
\end{equation}
where ${T}_{\rm e}$ = $10^{4}$~K is the electron temperature and $\theta_{S}$ is the source angular size. 
Both equations were adopted from \cite{sanchez:2013}, who followed the formalism of \cite{mezger:1967} and \cite{rubin:1968}. It is assumed that the cm continuum flux is emitted from homogeneous optically thin \hii regions. As can be seen from the optical depth results shown in Fig. \ref{fig:optdepth}, this is a good description for the sample. 
The typical uncertainties on the flux density and angular diameter imply an uncertainty on the emission measure of 30\% and uncertainty on the electron density of 20\%. Taking the 5$\sigma$ flux sensitivity of CORNISH (2~mJy, see \citealt{hoare:2012}) and the 1.5$''$ resolution, we estimate the CORNISH sensitivity to log(\textit{EM}) to be $\sim$ 5.5~$\rm pc \ cm^{-6}$.  

The computed UCHII emission measures and electron densities are generally consistent with the results by \cite{kurtz:1994}. No sources in the sample have computed electron densities and emission measures that would exceed $10^{5} \ \rm cm^{-3}$ and $10^{8} \ \rm pc \ cm^{-6}$, respectively, even when taking into account the associated uncertainties. Hyper-compact regions have electron densities in excess of $10^{6} \ \rm cm^{-3}$ and emission measures in excess of $10^{10} \ \rm pc \ cm^{-6}$ \citep{hoare:2007}. 
Thus, the CORNISH survey has identified only UCHIIs; no HCHIIs are reported. This is not surprising, because the high density of HCHIIs implies high turnover frequencies, $\sim$ 30 GHz.
For an optically thick free-free spectrum with $S_\nu \propto \nu^2$, the flux density at the CORNISH observing frequency of 5 GHz will be of the order of $40\times$ lower than the flux density near the turnover frequency. 

The computed optical depths at 5 GHz were used to quantify how much of the 5 GHz flux density would be missed in (the few) potential cases of optically thick \hii regions (as well as how much this effect varies from source to source). In 80\% of all the \hii regions observed at 5 GHz, the difference between the measured and the theoretical unattenuated flux is below 10\%. The bright UCHII G049.4905$-$00.3688 has the highest computed difference ( $\sim$ 56.4\%). 
 The distribution of the flux difference due to attenuation tapers off above differences greater than  $\sim$ 20\%, indicating that there is most likely no significant fraction of sources that have been missed altogether. 
The same should be true even if the electron temperature varies from region to region (within the expected physical bounds). 
It should also be noted that the computed Lyman continuum flux in this case is not significantly underestimated due to optically thick free-free emission, but could still be affected by loss of ionising photons (e.g. via dust absorption), or for radio flux that was not recovered. 

\section{Infrared properties} \label{IR}

\subsection{Associations with mid-infrared data}\label{glimpse} 

\begin{figure}
\includegraphics[width=\columnwidth]{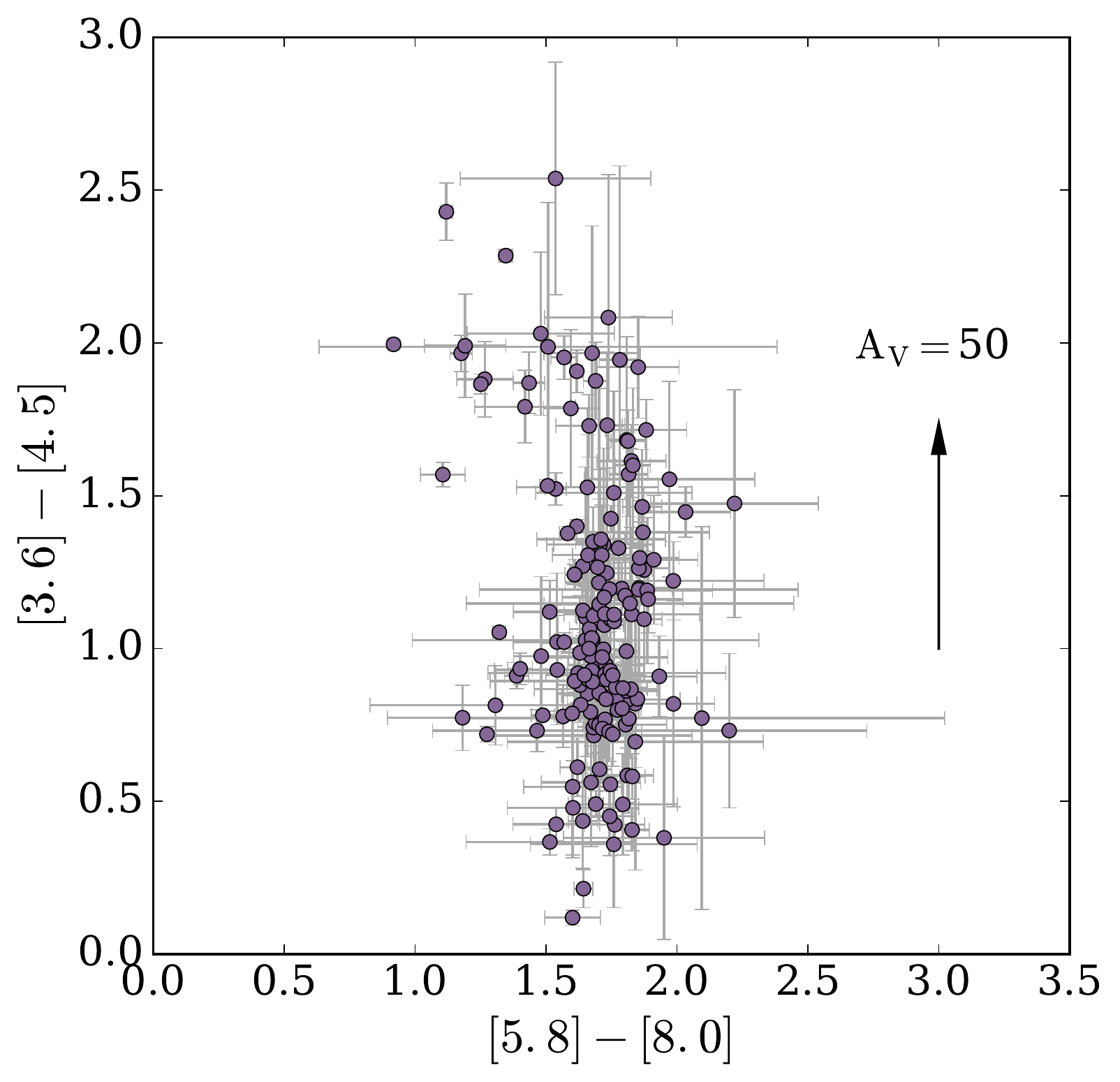}
\caption{Colour-colour diagram  of  the CORNISH UCHIIs (excluding all sources with one or more saturated GLIMPSE images or otherwise unreliable GLIMPSE fluxes). The arrow shows the reddening vector, based on the extinction law of \protect \cite{indebetouw:2005}. The plot is in the Vega magnitude system.}
\label{fig:colour}
\end{figure}

As discussed by \cite{watson:2008}, the emission detected in the vicinity of a hot star  is dominated by a different emission process in each of the four mid-IR GLIMPSE-IRAC bands. In particular, the presence of PAH emission is based on 8 $\upmu$m, 5.8~$\upmu$m, and lack of 4.5 $\upmu$m emission. Similar to larger IR bubbles, the mid-IR images of UCHII regions also show the presence of 8 $\upmu$m shells, dominated by strong PAH features in the IRAC  bands at 3.6, 5.8, and 8 $\upmu$m. The inner surface of each shell is located at a distance from the ionising star that is equal to the PAH destruction radius. 
The 3.6 $\upmu$m band is dominated by stars, with contributions from the diffuse PAH feature at 3.3 $\upmu$m and perhaps from scattered starlight. The 4.5 $\upmu$m band exhibits no PAH features; the brightest contributors are stars and the diffuse emission is due to the shared contribution of lines from \hii regions and from shocked molecular gas.    
The diffuse emission in the 5.8~$\upmu$m band is dominated by the 6.2~$\upmu$m PAH feature, apart from the immediate vicinity of O stars (where PAHs are destroyed). The diffuse emission in the 8~$\upmu$m band is dominated by the 7.7 and 8.6~$\upmu$m PAH features, or by thermal emission of dust heated by the hot stars and  Lyman-$\alpha$ photons.

The CORNISH survey was designed to cover the GLIMPSE region of the Galactic plane  \citep{hoare:2012}, ensuring that all sources have mid-infrared counterparts to the radio continuum sources.
Photometry of all GLIMPSE UCHII sources in the four IRAC bands was performed. 
Selecting the correct size of the IR source would not have been straightforward without knowledge of the position and size of the ultra-compact radio source, as the IR environment is more complex than in the radio view due to the different contributions to the emission. It is also difficult to disentangle individual sources in busy neighbourhoods. 

In order to use the hand-drawn polygons (in the case of extended radio sources), and Gaussian outlines (in the case of compact radio sources) as apertures for the IR data, they were expanded accordingly. The exact padding value necessary for each of the four GLIMPSE bands was chosen after measuring  (for each band) the counts at different aperture sizes and examining where the curve of growth begins to plateau before starting to increase again with the inclusion of unrelated sources. 
This was done for a small sub-sample of sources with a range of sizes representative of the UCHII sample, in all four bands. The padding radii used for the 3.6, 4.5, 5.8, and 8~$\upmu$m images were 2$''$, 2.4$''$, 3.4$''$, and 4.3$''$, respectively. The measured fluxes are included in Table \ref{glimpse_long} in Appendix \ref{appendixC}. Median absolute deviation from the median (MADFM) background estimation was utilised (for all photometry in this work), as it is insensitive to the presence of outliers and is a reliable estimate of the noise \citep{purcell:2013}\footnote{All sources were detected at at least 9 times the median background value. Errors were computed following \protect \cite{masci:2009} and include prior (noise-model) and derived uncertainties.}. 
From all 956 GLIMPSE images (all four bands), 36 8.0~$\upmu$m images were found to be saturated at the source location after visual inspection, as well as two 5.8~$\upmu$m images and one 4.5~$\upmu$m image. These, together with non-detections, were excluded from the final results.  
The photometric results for G031.2801\allowbreak+00.0632 were also excluded, as diagnostic diagrams showed them to be dominated by a neighbouring YSO (seen in all GLIMPSE bands) rather than the UCHII region.
  This left 180 sources with 3.6~$\upmu$m fluxes, 191 -- with 4.5~$\upmu$m fluxes, 190 -- with 5.8~$\upmu$m fluxes, and 184 -- with 8.0~$\upmu$m fluxes.

A colour-colour plot is shown in Fig. \ref{fig:colour}. Only sources for which it was possible to compute both the [3.6]$-$[4.5] and [5.8]$-$[8.0] colours were included (and no upper or lower limits for the remaining sources), to avoid overcrowding the plot. This was not found to affect the exhibited trends for the mid-IR colours. From the 174 sources with reliable flux values in all bands, $\sim$ 85\% occupy the zone 1.5 $<$ [5.8]$-$[8.0] $<$ 2. The [3.6]$-$[4.5] colour ranges between 0.1 and 2.1.
This is similar to the results reported by \cite{fuente:2009} for 19 ultra-compact \hii regions. They find that about 75\% of the UCHIIs are grouped around [5.8]$-$[8.0] $\simeq$ 1.7 and 0.5 $\lesssim$ [3.6]$-$[4.5] $\lesssim$  2.0. 

\subsection{Associations with near-infrared data and extinctions}\label{ukidss}
\begin{figure}
\includegraphics[width=\columnwidth]{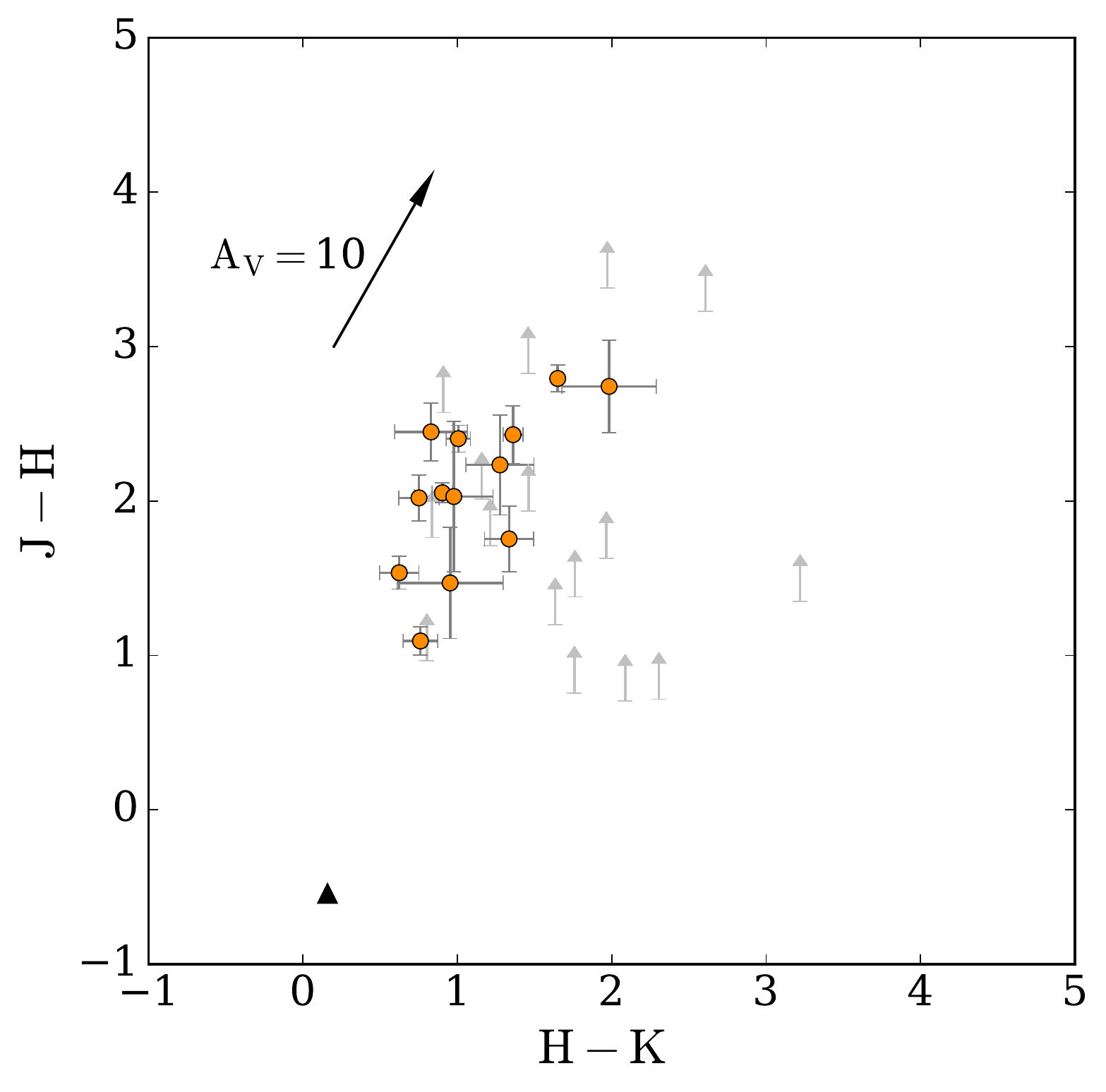}
\caption{Near-IR colour-colour diagram of near-IR nebulae associated with the CORNISH UCHIIs. The stellar contamination has been removed. Lower limits for sources visible only in H and K are also shown. The arrow shows the extinction vector, calculated using the extinction law from \protect \cite{stead:2009}. The predicted intrinsic colours of an ionised nebula are shown with a black triangle. The observed data and the intrinsic colours are shown in the AB magnitude system.} 
\label{fig:colour_nearir}
\end{figure}

\begin{figure}
\includegraphics[width=\columnwidth]{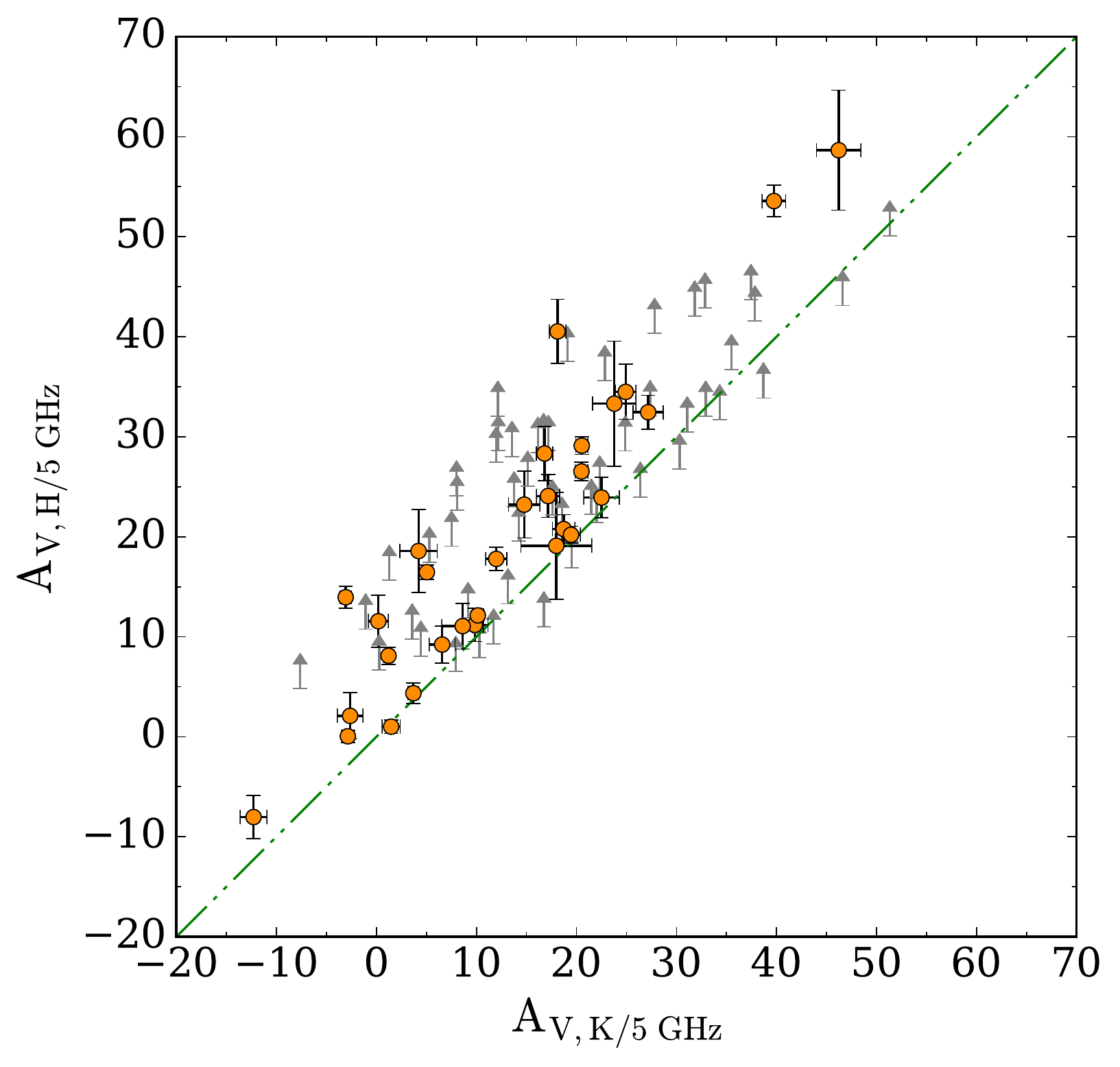}
\includegraphics[width=\columnwidth]{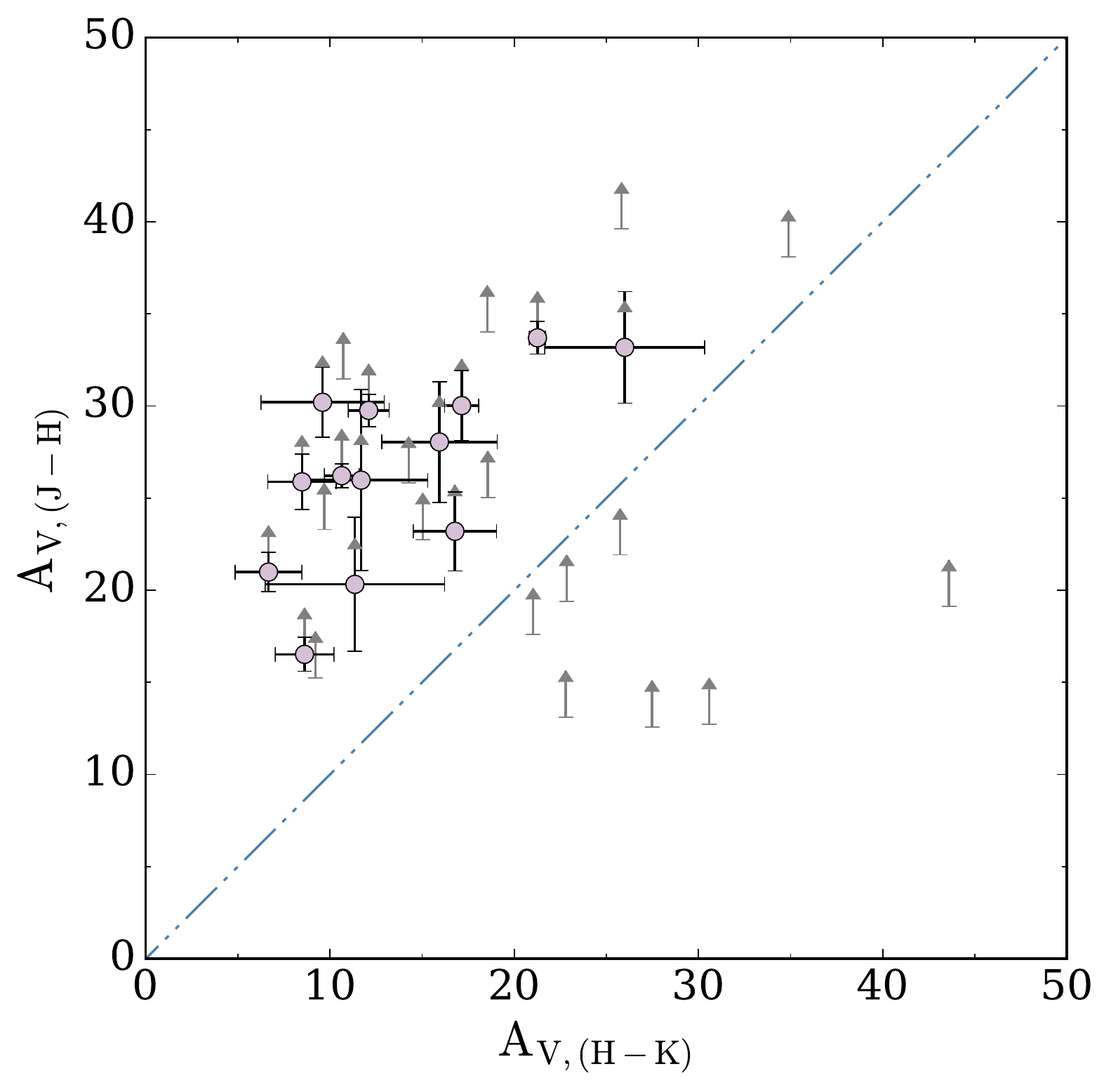}
\caption{Visual extinctions of the UCHIIs, computed from their near-IR fluxes from four methods -- $F_{\rm H}/F_{5 \rm GHz}$, $F_{\rm K}/F_{5 \rm GHz}$, $J-H$, and $H-K$. The top panel compares the H- and K-band derived $A_{\rm V}$. In the bottom panel, the comparison is shown for the $J-H$ against the  $H-K$ derived $A_{\rm V}$. Lower limits on the extinction are shown in the cases where the UCHII is seen only in H and K (grey arrows). Lines of equality are also plotted.} 
\label{fig:Av}
\end{figure}

\begin{table*}
\centering
\caption{Comparison between some extinction values in this work and in the literature, derived from comparison to 6~cm, Brackett-$\gamma$ line (`Br$\gamma$'), or multi-configuration radio observations (`radio'). 
Our results and the referenced literature values are in good agreement with extinction-law fits to \hii recombination-line data \citep{moore:2005}.} 
\begin{tabular}{lcr|lcrr}
\hline
\hline
\multicolumn{1}{l}{CORNISH name} &  \multicolumn{1}{c}{ $x$} & \multicolumn{1}{c}{$A_{x}$ \ (mag)} & \multicolumn{1}{|l}{Literature name}&  \multicolumn{1}{c}{ $x$} &\multicolumn{1}{c}{$A_{x}$ \ (mag)} & \multicolumn{1}{c}{Reference}  \\
\hline
G029.9559$-$00.0168& K/6cm &  2.33 $\pm$ 0.07 & G29.96$-$0.02 & K/radio &2.14 $\pm$ 0.08 & \cite{watson:1997}\\  
 &  &  &  & Br$\gamma$ & 2.20 $\pm$ 0.25 &  \\ 
 &  &   & &  Br$\gamma$  &2.16 $\pm$ 0.07 & \cite{moore:2005} \\

G043.8894$-$00.7840& K/6cm &  2.83 $\pm$ 0.12 & G43.89$-$0.78 & Br$\gamma$&  3.32 $\pm$ 0.21 & \cite{moore:2005}\\ 

G045.4545+00.0591 & K/6cm &  2.21 $\pm$ 0.11 & G45.45+0.06 & K/6cm & 2.5 & \cite{feldt:1998} \\ 

 G049.4905$-$00.3688   & K/6cm  & 2.33 $\pm$ 0.07 & W51d &  K/6cm & 2.6 $\pm$ 0.3 & \cite{goldader:1994}\\
&  &   &  & Br$\gamma$&  1.59 $\pm$ 0.07 & \cite{moore:2005}\\ 
\hline
\end{tabular}
\label{table:av_comparison}
\end{table*}

The visual extinction in the line of sight to UCHIIs can be estimated from their J, H, and K fluxes, assuming the diffuse emission is due to purely nebular gas emission. Several methods can be utilised to achieve this, such as the use of near-IR colours, or ratios of the near-IR flux to the radio flux \citep{willner:1972}. 
When using the colour-dependent methods, provided sufficiently reliable magnitude measurements of the embedded sources, one's choice comes down to a compromise between scattering effects and infrared excess. Dust excess (typically pronounced in the K-band) causes the $H-K$ colours to appear redder, and scattered light (due to dust grains) at shorter wavelengths results in bluer  $J-H$ colours \citep{porter:1998}. 
It is useful to compare the results from the different methods and in this way weigh the severity of systematic issues while providing an extinction range for the studied sources. 

The UKIDSS Galactic plane survey \citep{lucas:2008} covered the northern and equatorial Galactic plane at $|b|<5\degree $ in the J (1.17$-$1.33~$\upmu$m), H (1.49$-$1.78~$\upmu$m), K (2.03$-$2.37~$\upmu$m) bands and provides an opportunity to investigate the near-IR properties of the CORNISH UCHII sample, such as fluxes and detection statistics.
Only a point-source UKIDSS catalogue is available at present. Therefore automated photometry was performed in the same manner as described in \S \ref{glimpse} on all CORNISH sources with available UKIDSS data. In total, 230 sources had available UKIDSS images in the J band, 228 -- in H, 227 -- in K.  
The visual inspection revealed that out of all sources, 83 have a visible nebula in K, out of which 31 also have an H-band nebula. Out of these, 14 nebulae are visible in J\footnote{G023.9564\allowbreak+00.1493 has a visible nebula in the near-IR, but photometric issues due to image quality lead to unreliable flux values, and the source is not included in the final results table.}. 
Aperture photometry (with median background subtraction) of all contaminant bright stars found within the expanded polygon aperture used for the automated J, H, and K flux measurements was also performed. This was done for all images with a visible near-IR nebula coinciding with an UCHII. The measured stellar fluxes were subtracted from the total photometric fluxes in order to obtain the nebular fluxes. The near-IR UCHII fluxes and corresponding AB magnitudes are presented in Table \ref{ukidss_long} in Appendix \ref{appendixC}. Figure \ref{fig:colour_nearir} shows a diagram of the $J-H$ against $H-K$ nebular colours. The mean colour of the nebulae visible in all three bands is 2.1 for $J-H$, and 1.1 for $H-K$.

Figure \ref{fig:Av} presents a comparison between the computed visual extinctions for the UCHII sample obtained from four methods:  $F_{\rm H}/F_{5 \rm GHz}$ and $F_{\rm K}/F_{5 \rm GHz}$ (top panel); $J-H$ and $H-K$ (bottom panel). The empirically-derived $R_{V}$-dependent extinction law $A_{\lambda}/A_{V}$ from \cite{cardelli:1989} (Eqns. 1-3b) was used to convert from near-IR to visual extinction. 
The (standard for the ISM) optical total-to-selective extinction ratio $R_{V}$ = 3.1 was assumed. This has been found to reach values of 5-6 towards dense clouds \citep[e.g.][]{cardelli:1989} but using $R_{V}$ = 5 did not affect the results within error. 

 The near-IR extinction was computed from the difference between the measured and expected near-IR magnitudes.
For the flux-ratio methods, the expected H- and K-band fluxes were obtained by utilising the intrinsic ratios between IR and radio flux found by \cite{willner:1972}, $F_{\rm H}/F_{\rm 5GHz}$ = 0.26 and $F_{\rm K}/F_{\rm 5GHz}$ = 0.3.  Using the Willner ratios, a value of 0.68 for $H-K$ was computed in this work.
 In order to obtain the $J-H$ extinction, $F_{\rm J}/F_{\rm 5GHz}$  = 0.43 was computed, using Equation 1 and Table 2 in \cite{brussaard:1962}, taking into account the significant  Paschen-$\beta$  line contribution to the J band \citep{hummer:1987}.  
This resulted in a value of $-$0.1 for $J-H$. 
These predicted intrinsic $J-H$ and $H-K$ colours of ionised nebulae agree well with those from  near-IR photometry of planetary nebulae (with 2MASS data) studied by \cite{larios:2005}: $H-K$ = 0.65 and  $J-H$ $\sim$ $-$0.1.  
\cite{weidmann:2013} also found comparable values, using PNe in the VVV survey\footnote{\url{https://vvvsurvey.org/}}: $H-K$ = 0.62 and  $J-H$ = 0.0. This comparison is in the Vega system, taking into account the 2MASS (Vega system) and UKIDSS (AB system) magnitude offsets for each band. Our computed intrinsic AB colours, $J-H$ = $-$0.54 and $H-K$ = 0.15, were converted to the Vega system using Table 7 in \cite{hewett:2006}: the AB offsets for the $J$, $H$, and $K$ bands are 0.938, 1.379, and 1.9, respectively. 

The presented results from the four different extinction methods are consistent with previous estimates of visual extinctions towards UCHIIs, which are $\sim$ 0--50 mags \citep[see e.g.][]{hanson:2002}. 
\cite{moore:2005} calculated the extinction for a number of compact and ultra-compact \hii regions, using observed hydrogen recombination lines. 
A comparison between extinction results from this work and from literature is presented in Table \ref{table:av_comparison}, showing good agreement. 
 
As can be seen in Fig. \ref{fig:Av}, the offsets between the extinctions obtained from the different methods are clearly systematic.  
Values derived using the K band are $\sim$ 10 magnitudes lower (i.e. brighter) than those using H (and J). An addition of $\sim$ 10 magnitudes to the K band would bring the $F_{\rm H}/F_{5 \rm GHz}$ and $F_{\rm K}/F_{5 \rm GHz}$ methods to agreement and eliminate most of the unphysical negative values. Such an addition would mimic eliminating the expected boost to the K band from the contribution of very hot dust in the vicinity of the ionising star. However, a $\sim$ 10-magnitude addition to $K$ in the $A_{\rm V,(J-H)}$ vs. $A_{\rm V,(H-K)}$ diagram to exclude potential hot dust contribution would actually result in a $\sim$ 20-magnitude systematic offset between these two methods. 
Such a discrepancy likely stems from the general nature of each pair of utilised methods and highlights the need for in-depth investigation, preferably on a case-to-case basis (beyond the scope of this work). 

Two of the UCHIIs with computed extinction have coinciding $XMM-Newton$ hard x-ray sources. The extinction was computed independently from x-ray spectral fitting (which also revealed that the sources have $k_{\rm B}T \ > \ 2$~keV, i.e. $T \ > \ 10^{7}$~K). The $H-K$ $-$ derived visual extinction for G030.7661$-$00.0348 is $\sim$ 35 mags, whereas the x-ray$-$fitted hydrogen column density translates to $A_{\rm V}$  $\approx$ 67 mags (towards the W43 star cluster as a whole). G025.3824$-$00.1812 has  $H-K$ $-$ derived $A_{\rm V}$  $\approx$ 12, and the value derived from the x-ray spectral fit is $\sim$ 11 mags\footnote{The x-ray --  derived value is the same also for G025.3809\allowbreak$-$00.1815, as the two neighbouring sources are not resolved by $XMM-Newton$. }. 

\section{Spectral energy distributions and bolometric luminosities}
\label{lbol}

\begin{figure}
    \centering
    \includegraphics[width=\columnwidth]{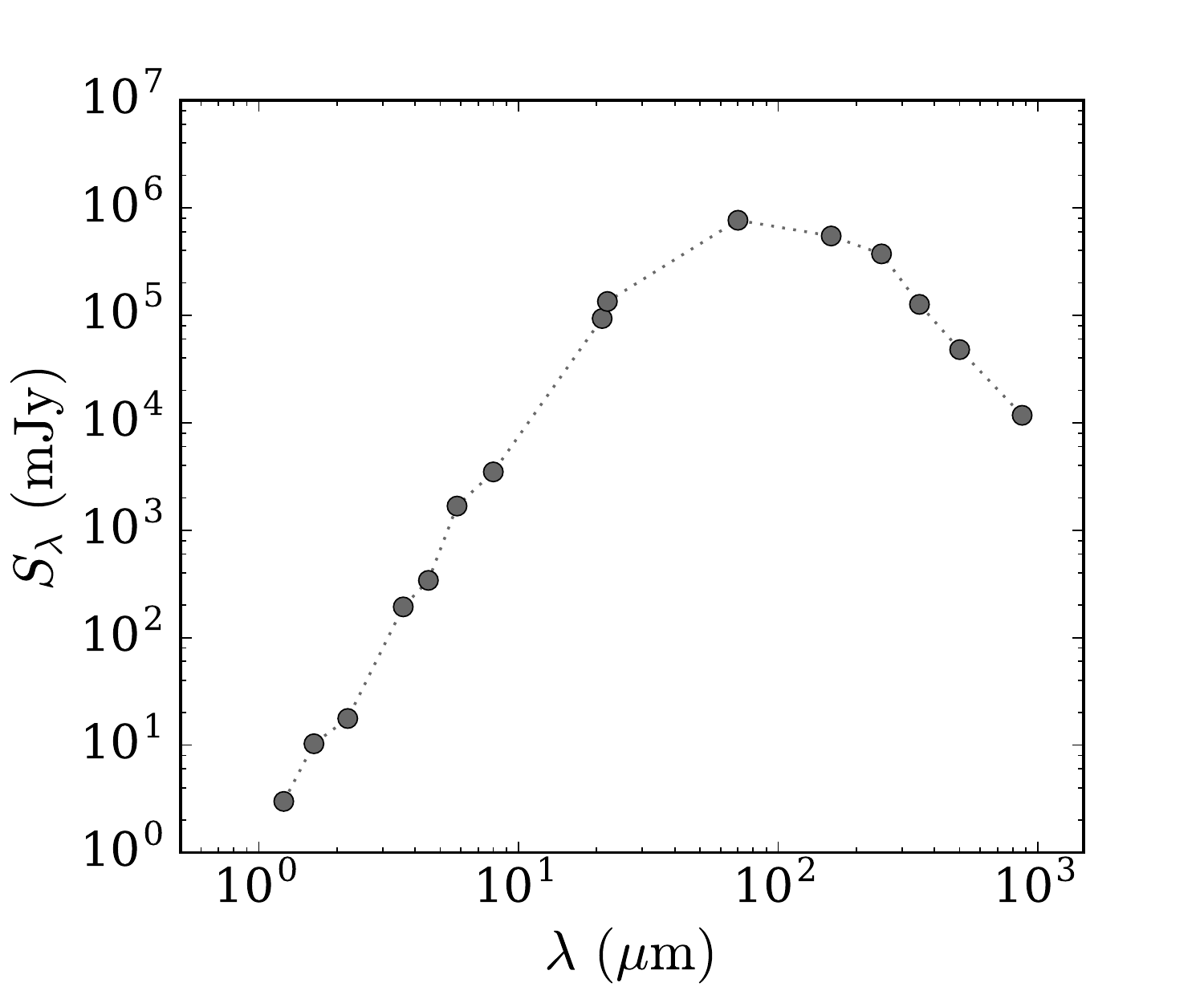}
    \caption{Average SED for all UCHIIs with available multi-wavelength data, normalised to the sample median distance of 9.8~kpc. The plot includes the $J$, $H$, and $K$ UKIDSS fluxes and the 3.6, 4.5, 5.8, and 8.0~$\upmu$m GLIMPSE fluxes from this work, together with fluxes from MSX (21~$\upmu$m), WISE (22~$\upmu$m), HiGAL (70, 160, 250, 350, and 500~$\upmu$m), and ATLASGAL (870~$\upmu$m) (see table A.1 in ~\protect\citealt{cesaroni:2015}).} 
    \label{fig:seds}
\end{figure}

The computed UKIDSS and GLIMPSE fluxes were combined with  multi-wavelength data from MSX (21~$\upmu$m), WISE (22~$\upmu$m), HiGAL (70, 160, 250, 350, and 500~$\upmu$m), and ATLASGAL (870~$\upmu$m). These data were available for 177 UCHIIs and the SEDs were reconstructed for these sources. 
The majority of the SEDs have very reasonable shapes and exhibit the same average shape.  
There are a few SEDs with irregularities, typically the flux at 350~$\upmu$m and 22~$\upmu$m. There are many SEDs (90/177) where the 4.5~$\upmu$m flux is low (comparable to the 3.6~$\upmu$m flux), which is most likely caused by the gap in PAH emission at this wavelength. The average SED for the sample (when all sources are placed at the median distance of 9.8~kpc) is presented in Fig. \ref{fig:seds}. 
\begin{figure}
\includegraphics[width=\columnwidth]{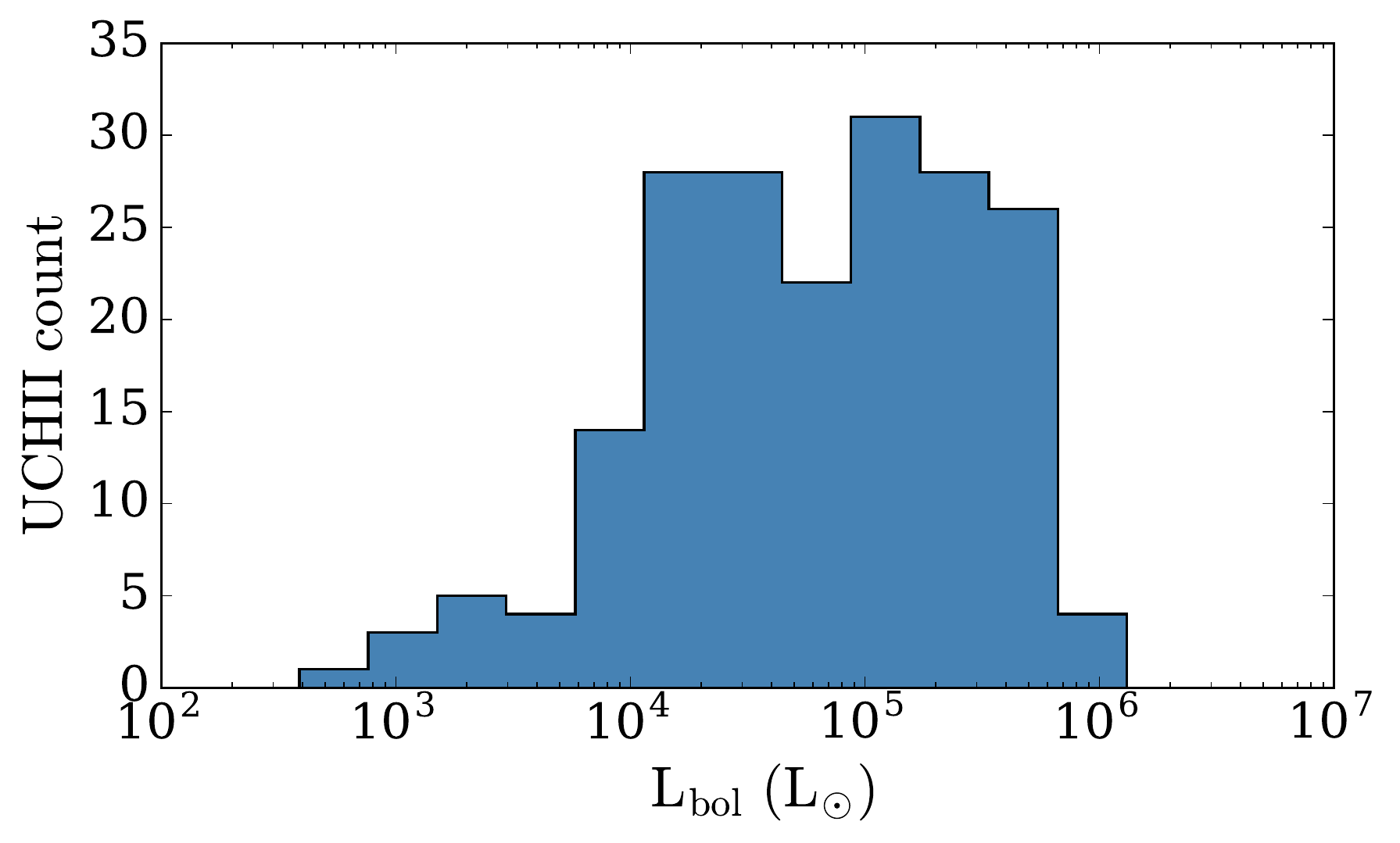}
\caption{Bolometric luminosity distribution for the UCHIIs with fitted SEDs.} 
\label{fig:lbol_hist}
\end{figure}

\begin{figure}
\includegraphics[width=\columnwidth]{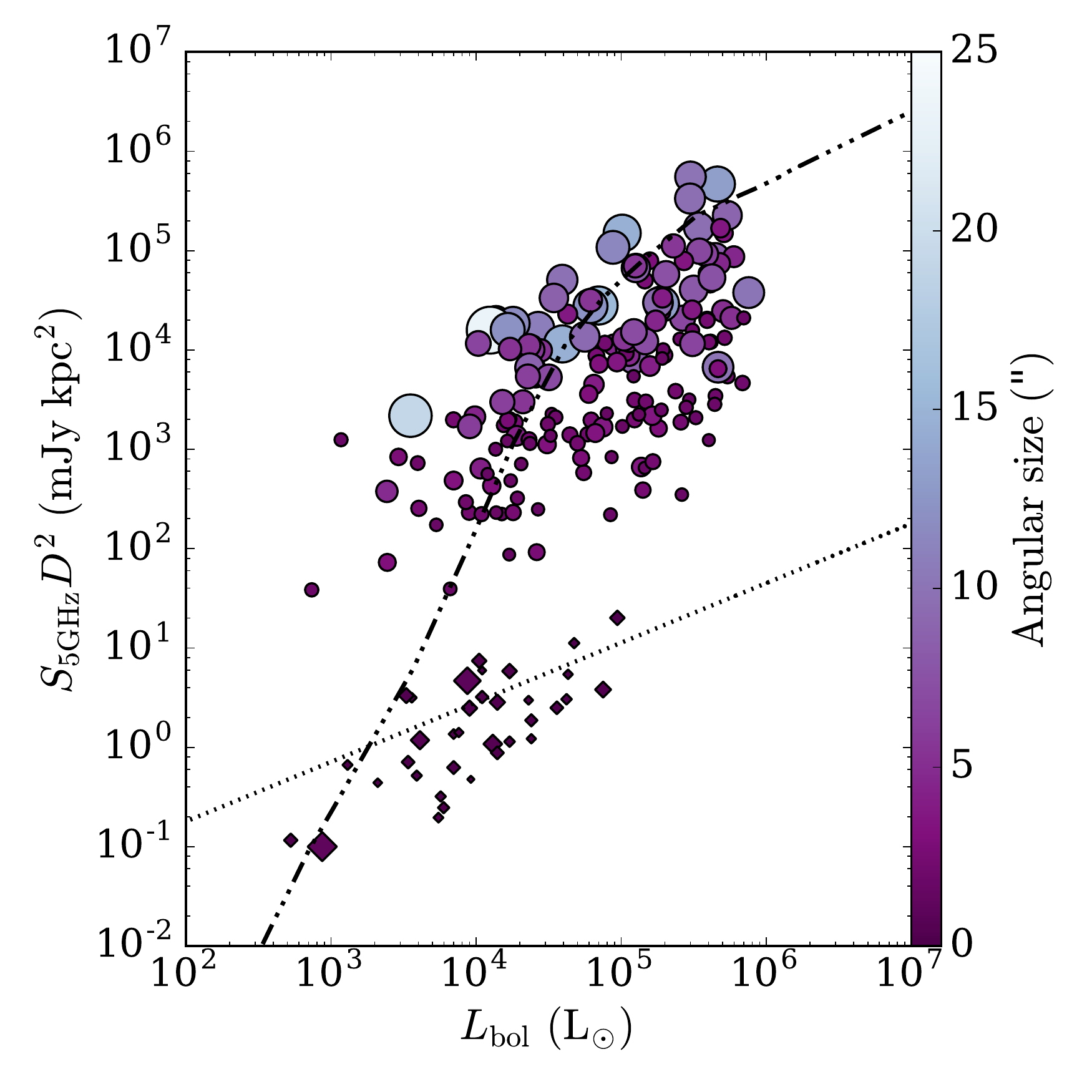}
\caption{5 GHz radio luminosity against bolometric luminosity for the UCHIIs with fitted SEDs, shown with circles. 
A sample of confirmed ionised MYSO jets by \protect\cite{purser:2017} is also included (indicated by diamonds). The dash-dotted line marks the expected radio luminosity from Lyman continuum emission. The dotted line shows the empirical relation extrapolated for low-mass YSO jets \protect \citep{anglada:1995}. The angular size of all sources is colour-coded and emphasized through the marker sizes (the jet sample marker sizes are exaggerated, as their resolution is higher).} 
\label{fig:lbol}
\end{figure}

The bolometric luminosities of the CORNISH UCHIIs were calculated from fitting the SEDs of the sources with available data across multiple wavelengths. The SED Fitter from \cite{robitaille:2007} was utilised, following the procedure prescribed by \cite{mottram:2011}. 
 The complete sample was split into subsamples according to their distance (as computed in this work) to limit the distance range when running the fits. The fitted bolometric luminosities were then converted to the final bolometric luminosity values by replacing the automatically fitted distances with the distances presented in this work. 
The results are shown in Fig. \ref{fig:lbol_hist}. These are in good agreement with Fig. 19 in \cite{urquhart:2013}, which summarises the RMS bolometric luminosities of 135 associated clumps. We note that the models used in the SED fitting are tailored to YSOs and do not take into account additional sources of dust heating in the ionised region, such as Lyman-$\alpha$ \citep{hoare:1991}. However, the SEDs of UCHIIs and YSOs are sufficiently similar up to radio wavelengths (which are not included in the models) and the fits are useful for computing the bolometric luminosities. 

Possible contamination by MYSO jet emission was investigated with a plot of radio luminosity at 5 GHz against bolometric luminosity of the UCHII regions and comparing to a sample of confirmed MYSO jets by \cite{purser:2017}. This is presented in Fig. \ref{fig:lbol}. The UCHII regions are brighter, as expected, and the two populations clearly separate.
The predicted optically-thin 5 GHz radio luminosities corresponding to a range of bolometric luminosities, from the stellar models by \cite{thompson:1984} for $L_{\rm bol} \ \leq \ 10^{3} \ \rm L_{\odot}$ and from the models summarised in Table 1 of \citealt{davies:2011} for $L_{\rm bol} \ > \ 10^{3} \ \rm L_{\odot}$, are shown with a dash-dotted line in the plot. 
The radio continuum flux can be inferred from the Lyman continuum flux, as the latter determines the amount of ionised material and thus the number of free electrons participating in the thermal bremsstrahlung process. 

The majority of CORNISH UCHIIs have radio luminosity that is between 1\% and 100\% of the theoretical value at the corresponding \hii region bolometric luminosity (i.e. below the model line). 
A mixture of UCHII angular sizes are found at equal fractions of the model luminosity, ruling out angular size as the culprit behind the large variation in observed luminosity. 
It is to be expected that a large portion of UCHIIs would be dimmer than predicted, as the stellar models do not account for the portion of Lyman continuum flux that is absorbed by dust.

 A significant number of sources (about a third) can be found above the line, with radio luminosities up to ten times or more than those predicted. 
The presence of \hii regions in the `forbidden' area above the model line is attributed to a Lyman excess. There are different explanations for such \hii regions, as discussed, for example, by \cite{sanchez:2013}, \cite{cesaroni:2015, cesaroni:2016}, and in the references therein. 
 One explanation is the assumption of spherical symmetry, whereas in reality Lyman photons could be leaking in directions tracing lower gas density, which could be away from the line of sight. 
It is unlikely for the contribution to be from overlapping 5$\sigma$ or 7$\sigma$ UCHIIs, as no systematic trends were found in their distribution in the plot. 
 \cite{cesaroni:2016} used molecular tracers to look for outflows and accretion shocks in the vicinity of 200 CORNISH \hii regions, and found no evidence to support any outflow-related phenomenon. Instead, they found that Lyman-excess sources are more associated with infall than non-excess sources, and propose ongoing accretion and accretion shocks as an explanation, but their HCO$^{+}$ measurements are only sensitive to large-scale infall ($\lesssim$ 1~pc), which is not direct evidence of accretion. 

Two sources, G065.2462\allowbreak+00.3505 and G011.0328\allowbreak+00.0274, were found to have computed bolometric luminosities below the lower limit for \hii region formation -- that is, for a B3 type star with $L_{\rm bol} \ \sim \ 10^{3} \ \rm L_{\odot}$ (\cite{boehm:1981}, \cite{meynet:2003}).  In the case of G011.0328\allowbreak+00.0274, this is likely due to an individual issue with the distance determination or the bolometric flux determination, as the radio-to-bolometric flux ratio is not extreme. 
G065.2462\allowbreak+00.3505, originally in the UCHII catalogue, was reclassified as a radio star. 
Although this unresolved radio source is embedded within an IR nebula $\sim$ 50$''$ in diameter, there is only a near-IR stellar counterpart but no trace of a compact counterpart at  8~$\upmu$m or 70~$\upmu$m.

It should be noted that 6 sources (G011.9786$-$00.0973, G014.5988\allowbreak+00.0198, G018.6654\allowbreak+00.0294, G031.1590\allowbreak+00.0465, G036.4062\allowbreak+00.0221, and G043.7960$-$00.1286) could not be shown in the plot. No bolometric luminosity was computed for them as the available supplementary data was not sufficient to build a good portion of their SEDs for a fit.  
Their $S_{\rm 5 GHz}D^{2}$ values  range from $\sim$ 34 to $\sim$ 828~$\rm mJy \ kpc^{2}$, and would thus still separate from the MYSO sample in the plot. 

\section{Comparison of UCHII search methods in blind surveys}\label{comparison_methods}

We argue that high resolution blind radio surveys are the most reliable way to obtain the UCHII population census of the Milky Way. 
With the CORNISH sample of genuine UCHII regions, our total estimate is $\sim$ 750 UCHIIs in the Galaxy. This was obtained by scaling up the sample size (239 sources) by a geometric correction factor of $\sim$ 3.1 -- the ratio between the detected number of RMS UCHII regions in the total RMS and CORNISH area -- 500 and 160, respectively (see \citealt{ urquhart:2008, lumsden:2013}). Colour-selected RMS sources with detectable radio emission were classified as UCHIIs due to their mid-IR morphologies. Practically the same scaling factor is obtained when taking the total RMS sample of 900 \hii regions and the 297 RMS \hii regions in the CORNISH area. The RMS survey encompasses $10\degree<l<350\degree$ and $|b|<5\degree$ and provides the best current map of the non-uniform distribution of massive star formation throughout the Galaxy. UCHII regions exhibit the same Galactic scale-height, $\sim$ 0.6\degree, in CORNISH and RMS. Since the RMS UCHII counts in the total and in the CORNISH area were obtained in the same manner, their ratio should not be greatly affected by the systematic limitations of the IR-selection.
We note that UCHIIs near the Galactic centre are missed in our total estimate, to avoid assumptions for the UCHII number density in this region not covered by RMS.

The sections below highlight limitations of the other UCHII search methods employed in Galactic plane surveys. 

\subsection{Sub-mm -- ATLASGAL}
\cite{urquhart:2013} used ATLASGAL-CORNISH associations to compute the surface density of UCHII regions as a function of Galactocentric distance. They estimated that the Galactic UCHII population comprises $\sim$ 400 sources around B0 and earlier type stars, out of which only $\sim$ 45 around O6 or earlier type stars are detectable.  
However, the depth of ATLASGAL falls short of detecting all UCHII regions within the common ATLASGAL-CORNISH area -- about 30 UCHII regions were missed as a result. The deeper SCUBA-2 survey should provide a higher detection certainty with the same search method. 

\subsection{Mid-IR -- RMS}
 The whole CORNISH region is covered in RMS; despite this, about half of the CORNISH UCHIIs have RMS counterparts. There are $\sim$ 40 further associations with the more diffuse CORNISH \hii sample (48 \hii regions). 
 Many of the  UCHIIs `missing' from RMS are, in fact, detected within big complexes but not listed as individual objects. \cite{urquhart:2013} discussed the larger total number of RMS versus ATLASGAL \hii regions -- the RMS \hii sample contains extended regions identified from their mid-IR morphology, with radio continuum emission lower than the CORNISH surface brightness sensitivity. The RMS sample thus includes fewer individually listed UCHII regions, together with a number of more extended \hii regions.
   
\subsection{Previous radio surveys of the Galactic plane}
\cite{giveon:2008} found 494 MSX matches to the \cite{white:2005} catalogue within the area shared with CORNISH ($\sim$~23\%). 
\cite{giveon:2005} believe their sample is dominated by UCHII regions.
As discussed in \S \ref{lowres}, through visual inspection we only found 162 CORNISH UCHII regions in common with the lower-resolution 6~cm data. A catalogue table cross-match found even fewer sources in common -- 111, with 20 cross-matches to 37 CORNISH diffuse \hii regions within the shared area. Thus we are finding a factor of 3-4 fewer UCHIIs than implied by \cite{giveon:2005, giveon:2008} -- the vast majority of their sample are not genuine ultra-compact \hii regions but rather represent more extended \hii region phases. 

\section{Summary}\label{summary}
The CORNISH UCHII sample is the largest complete and unbiased high-resolution collection of ultra-compact \hii regions to date. Within the mapped region ($10\degree<l<65\degree, |b|<1\degree$), 239 UCHIIs have been confirmed from 240 candidates visually identified at 5 GHz radio-continuum emission. 
In this work, we explored the observational properties, spectral indices, physical characteristics and spectral energy distributions of this early stage of massive star formation. 
In summary:
\begin{enumerate}
\item The selection procedure for the CORNISH UCHIIs is robust and the nature of the sample as a whole was reliably identified. 
\item The majority ($\sim$ 82\%) of UCHIIs have spectral indices that are consistent with the expected theoretical limits for thermal free-free emission. The instances of non-thermal spectral indices could be naturally resulting from the difference in VLA configuration between the higher- and lower-frequency datasets, or are the result of combining 6~cm and 20~cm fluxes of variable thermal sources from two epochs.  
\item We conclude that at least 5\% of UCHIIs have exhibited a significant flux increase (by $\sim$ 50\% or more) between two observational epochs separated by $\sim$ 15~years.
\item Distances were computed for 21 UCHIIs which had no literature distance (or their KDA had not been previously resolved) prior to this work. The derived physical properties of the UCHII sample agree well with theoretical expectations. 
\item We have presented results of extended source photometry of UCHII regions in the mid- and near-IR. The GLIMPSE and UKIDSS colours of the sample follow the expected trends set by results obtained from earlier, smaller samples. We expect the mid-IR results to be particularly reliable, as they combine the precise  knowledge of position, radio size, and shape provided by the CORNISH survey with the good correspondence (in the vast majority of cases) to the mid-IR counterparts. The results of the extended near-IR photometry (particularly the J and H bands) should be used with much more care, due to the difficulty in accurate subtraction of the stellar contamination in the busy, diffuse environments of star forming regions seen at these wavelengths. 
\item Extinctions towards the UCHII regions were computed using the intrinsic H- and K-band to radio flux ratio from \cite{willner:1972}, as well as from the $J-H$ and $H-K$ nebular colours. 
\item The average spectral energy distribution of the UCHII sample (from gathering available multi-wavelength data and combining them with the new near- and mid-IR results) is in excellent agreement with the expected shape (see e.g. \citealt{faison:1998}, Fig. 1 in \citealt{hoare:2012}), with a peak between 70 and 160~$\upmu$m. Bolometric luminosities were computed by fitting the individual SEDs. 
In a plot of radio luminosity against bolometric luminosity, the CORNISH UCHII sample is clearly a separate population to confirmed MYSO jets. About a third of the UCHIIs exhibit a Lyman excess. 
\item High resolution blind radio surveys are the best way to definitively find the UCHII population of the Galaxy. Radio selection provides a more reliable statistic than infrared and mm selection. We found a factor of 3-4 fewer genuine ultra-compact \hii regions than in previous lower resolution radio areal surveys, which, in conjunction with up-to-date models (see \citealt{davies:2011}), goes towards alleviating the \textit{lifetime problem} posed by \cite{churchwell:1989a}. 
\end{enumerate}

\begin{acknowledgements}
IEK acknowledges the support of the Science and Technology Facilities Council of the United Kingdom (STFC) through the award of a studentship. This publication has made use of data from the CORNISH survey database (\url{http://cornish.leeds.ac.uk/public/index.php}) and the RMS survey database (\url{http://rms.leeds.ac.uk/cgi-bin/public/RMS_DATABASE.cgi}).
This work has also made use of the SIMBAD database (CDS, Strasbourg, France). 
\end{acknowledgements}




\bibliographystyle{aa}
\bibliography{Breport.bib}

\begin{thebibliography}{87}
\expandafter\ifx\csname natexlab\endcsname\relax\def\natexlab#1{#1}\fi

\bibitem[{Acord {et~al.}(1998)Acord, Churchwell, \& Wood}]{acord:1998}
Acord, J.~M., Churchwell, E., \& Wood, D. O.~S. 1998, ApJ, 495, L107

\bibitem[{Anderson \& Bania(2009)}]{bania:2009}
Anderson, L.~D. \& Bania, T.~M. 2009, AJ, 690, 706

\bibitem[{Anderson {et~al.}(2009)}]{anderson:2009}
Anderson, L.~D. {et~al.} 2009, ApJSS, 181, 255

\bibitem[{Anglada(1995)}]{anglada:1995}
Anglada, G. 1995, RMxAASC, 1, 67

\bibitem[{Becker {et~al.}(1994)}]{becker:1994}
Becker, R.~H. {et~al.} 1994, Bull. Am. Astron. Soc., 26, 915

\bibitem[{Benjamin {et~al.}(2003)}]{benjamin:2003}
Benjamin, R.~A. {et~al.} 2003, PASP, 115, 953

\bibitem[{Boehm-Vitense(1981)}]{boehm:1981}
Boehm-Vitense, E. 1981, ARA\&A, 19, 295

\bibitem[{Boji{\v{c}}i{\'c} {et~al.}(2011)}]{bojicic:2011}
Boji{\v{c}}i{\'c}, I.~S. {et~al.} 2011, MNRAS, 412, 223

\bibitem[{Brand \& Blitz(1993)}]{brand:1993}
Brand, J. \& Blitz, L. 1993, A\&A, 275, 67

\bibitem[{Brussaard \& van~de Hulst(1962)}]{brussaard:1962}
Brussaard, P.~J. \& van~de Hulst, H.~C. 1962, RvMP, 34, 507

\bibitem[{Cardelli {et~al.}(1989)Cardelli, Clayton, \& Mathis}]{cardelli:1989}
Cardelli, J.~A., Clayton, G.~C., \& Mathis, J.~S. 1989, ApJ, 345, 245

\bibitem[{Carey {et~al.}(2009)}]{carey:2009}
Carey, S.~J. {et~al.} 2009, PASP, 121, 76

\bibitem[{Cesaroni {et~al.}(2015)}]{cesaroni:2015}
Cesaroni, R. {et~al.} 2015, A\&A, 579, A71

\bibitem[{Cesaroni {et~al.}(2016)}]{cesaroni:2016}
Cesaroni, R. {et~al.} 2016, A\&A, 588, L5

\bibitem[{Churchwell {et~al.}(2009)}]{churchwell:2009}
Churchwell, E. {et~al.} 2009, PASP, 121, 213

\bibitem[{Davies {et~al.}(2011)}]{davies:2011}
Davies, B. {et~al.} 2011, MNRAS, 416, 972

\bibitem[{de~la Fuente {et~al.}(2009)}]{fuente:2009}
de~la Fuente, E. {et~al.} 2009, RevMexAA, 37, 13

\bibitem[{{De Pree} {et~al.}(2005)}]{depree:2005}
{De Pree}, C.~G. {et~al.} 2005, ApJ, 624, L101

\bibitem[{Faison {et~al.}(1998)}]{faison:1998}
Faison, M. {et~al.} 1998, ApJ, 500, 280

\bibitem[{Feldt {et~al.}(1998)}]{feldt:1998}
Feldt, M. {et~al.} 1998, A\&A, 339, 759

\bibitem[{Franco-Hern{\'a}ndez \& Rodr{\'i}guez(2004)}]{hernandez:2004}
Franco-Hern{\'a}ndez, R. \& Rodr{\'i}guez, L.~F. 2004, ApJ, 604, L105

\bibitem[{Galv{\'a}n-Madrid {et~al.}(2008)}]{madrid:2008}
Galv{\'a}n-Madrid, R. {et~al.} 2008, ApJ, 674, L33

\bibitem[{Galv{\'a}n-Madrid {et~al.}(2011)}]{madrid:2011}
Galv{\'a}n-Madrid, R. {et~al.} 2011, MNRAS, 416, 1033

\bibitem[{Garay {et~al.}(1993)}]{garay:1993}
Garay, G. {et~al.} 1993, ApJ, 418, 368

\bibitem[{Gibb {et~al.}(2004)}]{gibb:2004}
Gibb, E.~L. {et~al.} 2004, ApJSS, 151, 35

\bibitem[{Giveon {et~al.}(2005)}]{giveon:2005}
Giveon, U. {et~al.} 2005, AJ, 129, 348

\bibitem[{Giveon {et~al.}(2008)}]{giveon:2008}
Giveon, U. {et~al.} 2008, AJ, 135, 1697

\bibitem[{Goldader \& Wynn-Williams(1994)}]{goldader:1994}
Goldader, J.~D. \& Wynn-Williams, C.~G. 1994, ApJ, 433, 164

\bibitem[{G{\'o}mez {et~al.}(2008)}]{gomez:2008}
G{\'o}mez, L. {et~al.} 2008, ApJ, 685, 333

\bibitem[{Hanson {et~al.}(2002)}]{hanson:2002}
Hanson, M.~M. {et~al.} 2002, ApJS, 138, 35

\bibitem[{Helfand {et~al.}(2006)}]{helfand:2006}
Helfand, D.~J. {et~al.} 2006, AJ, 131, 2525

\bibitem[{Hewett {et~al.}(2006)}]{hewett:2006}
Hewett, P.~C. {et~al.} 2006, MNRAS, 367, 454

\bibitem[{Hoare {et~al.}(1991)}]{hoare:1991}
Hoare, M.~G. {et~al.} 1991, MNRAS, 251, 584

\bibitem[{Hoare {et~al.}(2007)}]{hoare:2007}
Hoare, M.~G. {et~al.} 2007, Protostars and Planets, 5, 181

\bibitem[{Hoare {et~al.}(2012)}]{hoare:2012}
Hoare, M.~G. {et~al.} 2012, PASP, 124, 939

\bibitem[{Hummer \& Storey(1987)}]{hummer:1987}
Hummer, D.~G. \& Storey, P.~J. 1987, MNRAS, 224, 801

\bibitem[{Indebetouw {et~al.}(2005)}]{indebetouw:2005}
Indebetouw, R. {et~al.} 2005, ApJ, 619, 931

\bibitem[{Jackson {et~al.}(2006)}]{jackson:2006}
Jackson, J.~M. {et~al.} 2006, ApJSS, 163, 145

\bibitem[{Klassen {et~al.}(2012)}]{klassen:2012a}
Klassen, M. {et~al.} 2012, MNRAS, 421, 286

\bibitem[{Kurtz {et~al.}(1994)}]{kurtz:1994}
Kurtz, S. {et~al.} 1994, ApJS, 91, 659

\bibitem[{Lucas {et~al.}(2008)}]{lucas:2008}
Lucas, P.~W. {et~al.} 2008, MNRAS, 391, 136

\bibitem[{Lumsden {et~al.}(2013)}]{lumsden:2013}
Lumsden, S.~L. {et~al.} 2013, ApJSS, 208, 17

\bibitem[{Masci(2009)}]{masci:2009}
Masci, F. 2009, Version 2.0

\bibitem[{Meynet \& Maeder(2003)}]{meynet:2003}
Meynet, G. \& Maeder, A. 2003, A\&A, 404, 975

\bibitem[{Mezger \& Henderson(1967)}]{mezger:1967}
Mezger, P.~G. \& Henderson, A.~P. 1967, ApJ, 147, 471

\bibitem[{Miralles {et~al.}(1994)}]{miralles:1994}
Miralles, M.~P. {et~al.} 1994, ApJS, 92, 173

\bibitem[{Molinari {et~al.}(2014)}]{molinari:2014}
Molinari, S. {et~al.} 2014, Protostars and Planets, 6, 125

\bibitem[{Moore {et~al.}(2005)}]{moore:2005}
Moore, T. J.~T. {et~al.} 2005, MNRAS, 359, 589

\bibitem[{Mottram {et~al.}(2011)}]{mottram:2011}
Mottram, J.~C. {et~al.} 2011, A\&A, 525, A149

\bibitem[{Panagia \& Walmsley(1978)}]{panagia:1978}
Panagia, N. \& Walmsley, C.~M. 1978, A\&A, 70, 411

\bibitem[{Peters {et~al.}(2010{\natexlab{a}})}]{peters:2010a}
Peters, T. {et~al.} 2010{\natexlab{a}}, ApJ, 711, 1017

\bibitem[{Peters {et~al.}(2010{\natexlab{b}})}]{peters:2010b}
Peters, T. {et~al.} 2010{\natexlab{b}}, ApJ, 719, 831

\bibitem[{Peters {et~al.}(2010{\natexlab{c}})}]{peters:2010c}
Peters, T. {et~al.} 2010{\natexlab{c}}, ApJ, 725, 134

\bibitem[{Porter {et~al.}(1998)Porter, Drew, \& Lumsden}]{porter:1998}
Porter, J.~M., Drew, J.~E., \& Lumsden, S.~L. 1998, A\&A, 332, 999

\bibitem[{Price {et~al.}(2001)}]{price:2001}
Price, S.~D. {et~al.} 2001, AJ, 121, 2819

\bibitem[{Purcell {et~al.}(2013)}]{purcell:2013}
Purcell, C.~R. {et~al.} 2013, ApJSS, 205, 1

\bibitem[{Purser(2017)}]{purser:2017}
Purser, S. J.~D. 2017, PhD thesis (University of Leeds)

\bibitem[{Purser {et~al.}(2016)}]{purser:2016}
Purser, S. J.~D. {et~al.} 2016, MNRAS, 460, 1039

\bibitem[{Quireza {et~al.}(2006)}]{quireza:2006}
Quireza, C. {et~al.} 2006, ApJ, 653, 1226

\bibitem[{Ramos-Larios \& Phillips(2005)}]{larios:2005}
Ramos-Larios, G. \& Phillips, J.~P. 2005, MNRAS, 357, 732

\bibitem[{Reid {et~al.}(2009)}]{reid:2009}
Reid, M.~J. {et~al.} 2009, ApJ, 700, 137

\bibitem[{Robitaille {et~al.}(2007)}]{robitaille:2007}
Robitaille, T.~P. {et~al.} 2007, AJSS, 169, 328

\bibitem[{Rodr{\'i}guez {et~al.}(2007)}]{rodriguez:2007}
Rodr{\'i}guez, L.~F. {et~al.} 2007, ApJ, 663, 1083

\bibitem[{Rosolowsky {et~al.}(2010)}]{rosolowsky:2010}
Rosolowsky, E. {et~al.} 2010, ApJS, 188, 123

\bibitem[{Rubin(1968)}]{rubin:1968}
Rubin, R.~H. 1968, ApJ, 154, 391

\bibitem[{S{\'a}nchez-Monge {et~al.}(2013)}]{sanchez:2013}
S{\'a}nchez-Monge, {\'A}. {et~al.} 2013, A\&A, 550, A21

\bibitem[{Schuller {et~al.}(2009)}]{schuller:2009}
Schuller, F. {et~al.} 2009, A\&A, 504, 415

\bibitem[{Si{\'o}dmiak \& Tylenda(2001)}]{siodmiak:2001}
Si{\'o}dmiak, N. \& Tylenda, R. 2001, A\&A, 373, 1032

\bibitem[{Skrutskie {et~al.}(2006)}]{skrutskie:2006}
Skrutskie, M.~F. {et~al.} 2006, AJ, 131, 1163

\bibitem[{Stead \& Hoare(2009)}]{stead:2009}
Stead, J.~J. \& Hoare, M.~G. 2009, MNRAS, 400, 731

\bibitem[{Stil {et~al.}(2006)}]{stil:2006}
Stil, J.~M. {et~al.} 2006, ApJ, 132, 1158

\bibitem[{Thompson(1984)}]{thompson:1984}
Thompson, R.~I. 1984, ApJ, 283, 165

\bibitem[{Urquhart {et~al.}(2007)}]{urquhart:2007}
Urquhart, J.~S. {et~al.} 2007, A\&A, 461, 11

\bibitem[{Urquhart {et~al.}(2008)}]{urquhart:2008}
Urquhart, J.~S. {et~al.} 2008, ASP Conf. Ser., 387, 381

\bibitem[{Urquhart {et~al.}(2009)}]{urquhart:2009}
Urquhart, J.~S. {et~al.} 2009, A\&A, 501, 539

\bibitem[{Urquhart {et~al.}(2011)}]{urquhart:2011}
Urquhart, J.~S. {et~al.} 2011, MNRAS, 410, 1237

\bibitem[{Urquhart {et~al.}(2012)}]{urquhart:2012}
Urquhart, J.~S. {et~al.} 2012, MNRAS, 420, 1656

\bibitem[{Urquhart {et~al.}(2013)}]{urquhart:2013}
Urquhart, J.~S. {et~al.} 2013, MNRAS, 435, 400

\bibitem[{Watson {et~al.}(1997)}]{watson:1997}
Watson, A.~M. {et~al.} 1997, ApJ, 487, 818

\bibitem[{Watson {et~al.}(2008)}]{watson:2008}
Watson, C. {et~al.} 2008, ApJ, 681, 1341

\bibitem[{Weidmann {et~al.}(2013)}]{weidmann:2013}
Weidmann, W.~A. {et~al.} 2013, A\&A, 552, A74

\bibitem[{White {et~al.}(2005)White, Becker, \& Helfand}]{white:2005}
White, R.~L., Becker, R.~H., \& Helfand, D.~J. 2005, AJ, 130, 586

\bibitem[{Willner {et~al.}(1972)}]{willner:1972}
Willner, S.~P. {et~al.} 1972, ApJ, 175, 699

\bibitem[{Wood \& Churchwell(1989{\natexlab{a}})}]{churchwell:1989a}
Wood, D. O.~S. \& Churchwell, E. 1989{\natexlab{a}}, ApJS, 69, 831

\bibitem[{Wood \& Churchwell(1989{\natexlab{b}})}]{churchwell:1989b}
Wood, D. O.~S. \& Churchwell, E. 1989{\natexlab{b}}, ApJ, 340, 265

\bibitem[{Zinnecker \& Yorke(2007)}]{yorke:2007}
Zinnecker, H. \& Yorke, H.~W. 2007, ARA\&A, 481, 45

\bibitem[{Zoonematkermani {et~al.}(1990)}]{zoo:1990}
Zoonematkermani, S. {et~al.} 1990, ApJSS, 74, 181

\end{thebibliography}

\clearpage

\begin{appendices}


\section{KDA resolution -- plots}\label{appendixB}
\noindent\begin{minipage}{\textwidth}
    \centering
    \captionof{figure}{Continuum-subtracted VGPS spectra (black line) towards sources from the CORNISH UCHII region sample were used to solve their KDA. The red dashed line indicates the tangent point velocity. The blue spectrum shows the measured CO spectrum, with its corresponding y-axis also in blue. The green dot-dashed line represents the CO source velocity, with the region marked in grey on each side showing the expected uncertainty of  $ \pm $ 10$\kms$ due to streaming motions. The \hi spectrum of G024.4721+00.4877 is practically identical to G024.4698+00.4954 (top row, left) and therefore was omitted. Continues on next page.}
\includegraphics[width=0.45\columnwidth]{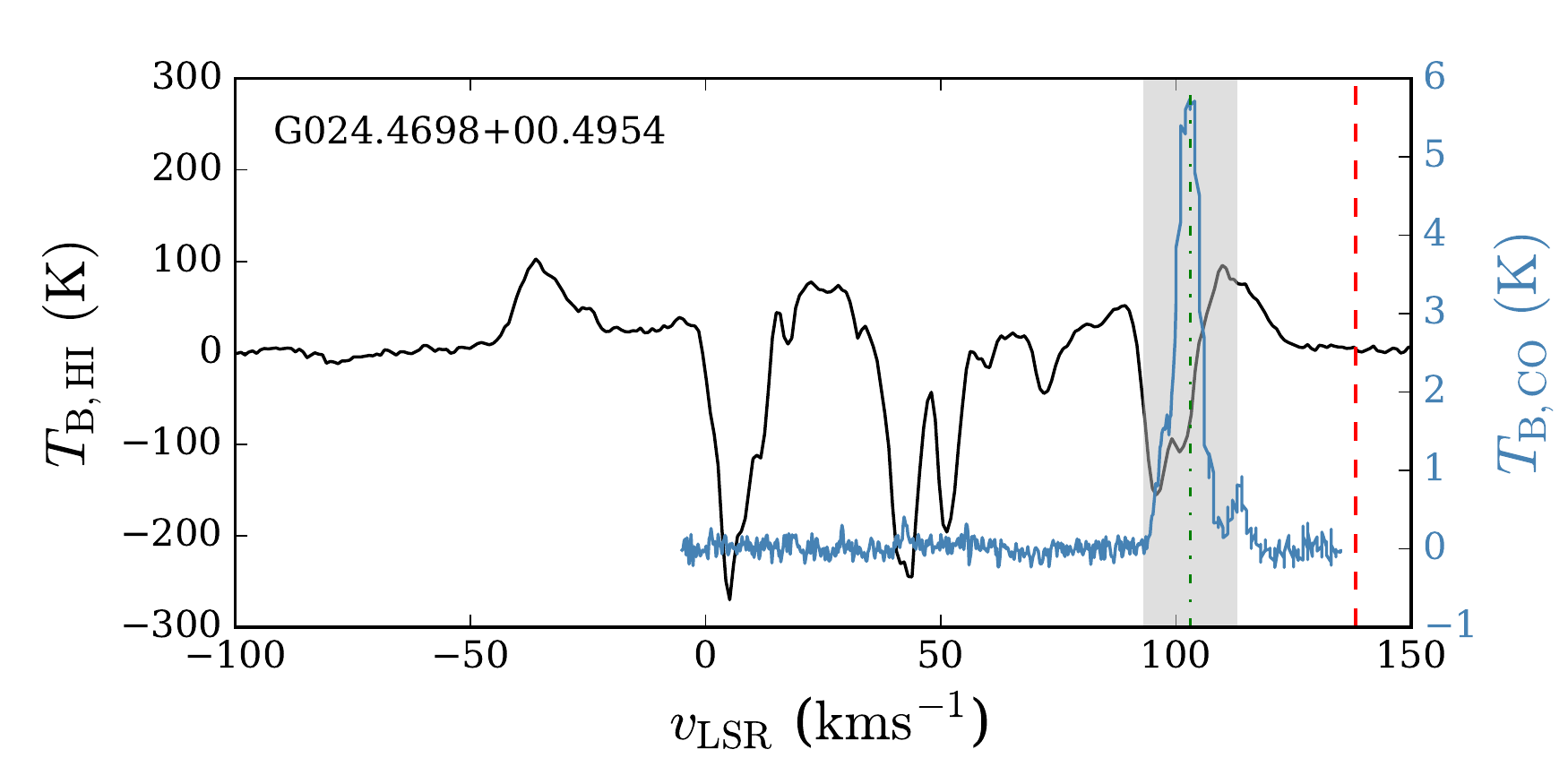}
\includegraphics[width=0.45\columnwidth]{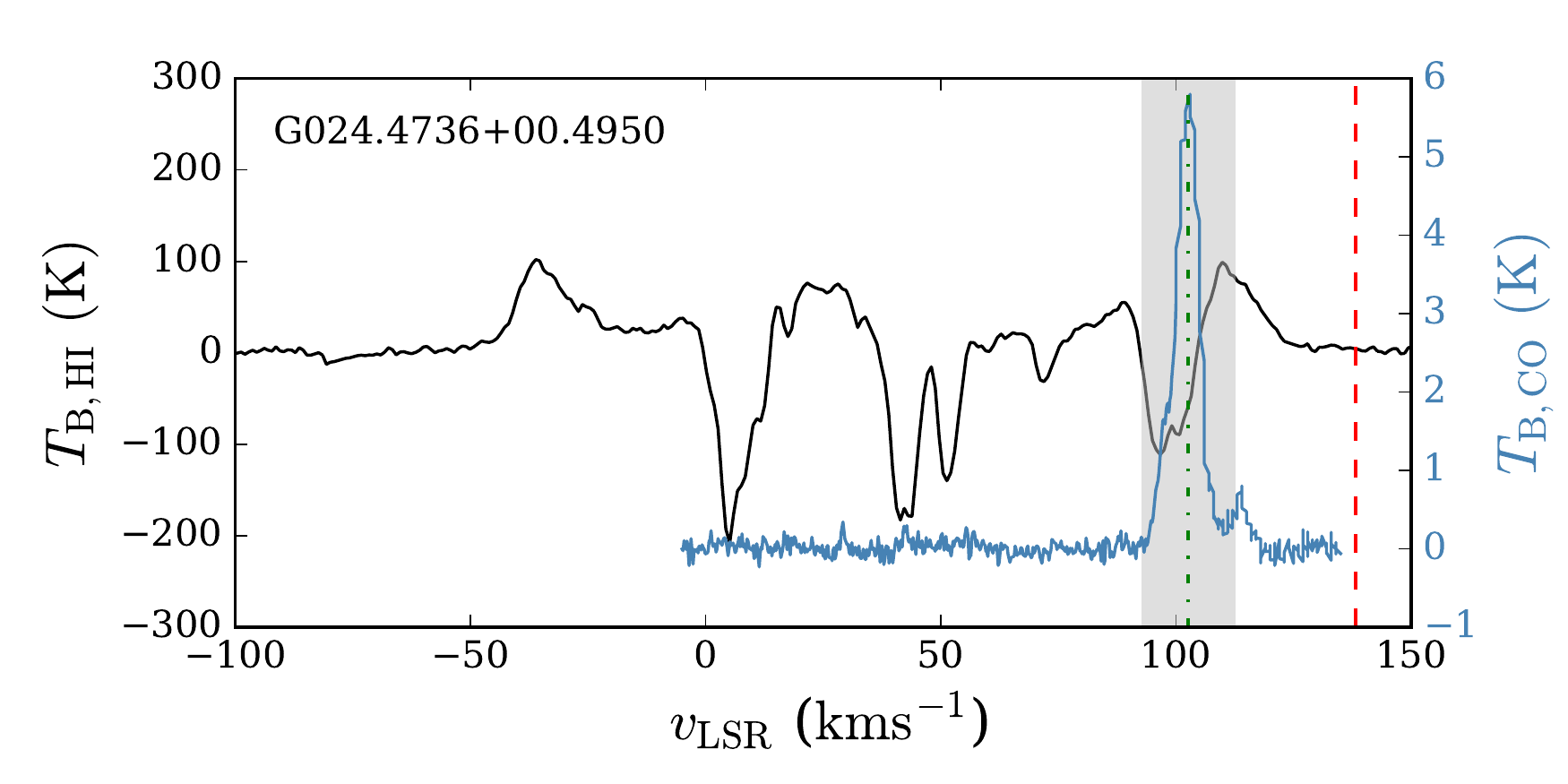}
\newline
\vspace{-1.8mm}
\includegraphics[width=0.45\columnwidth]{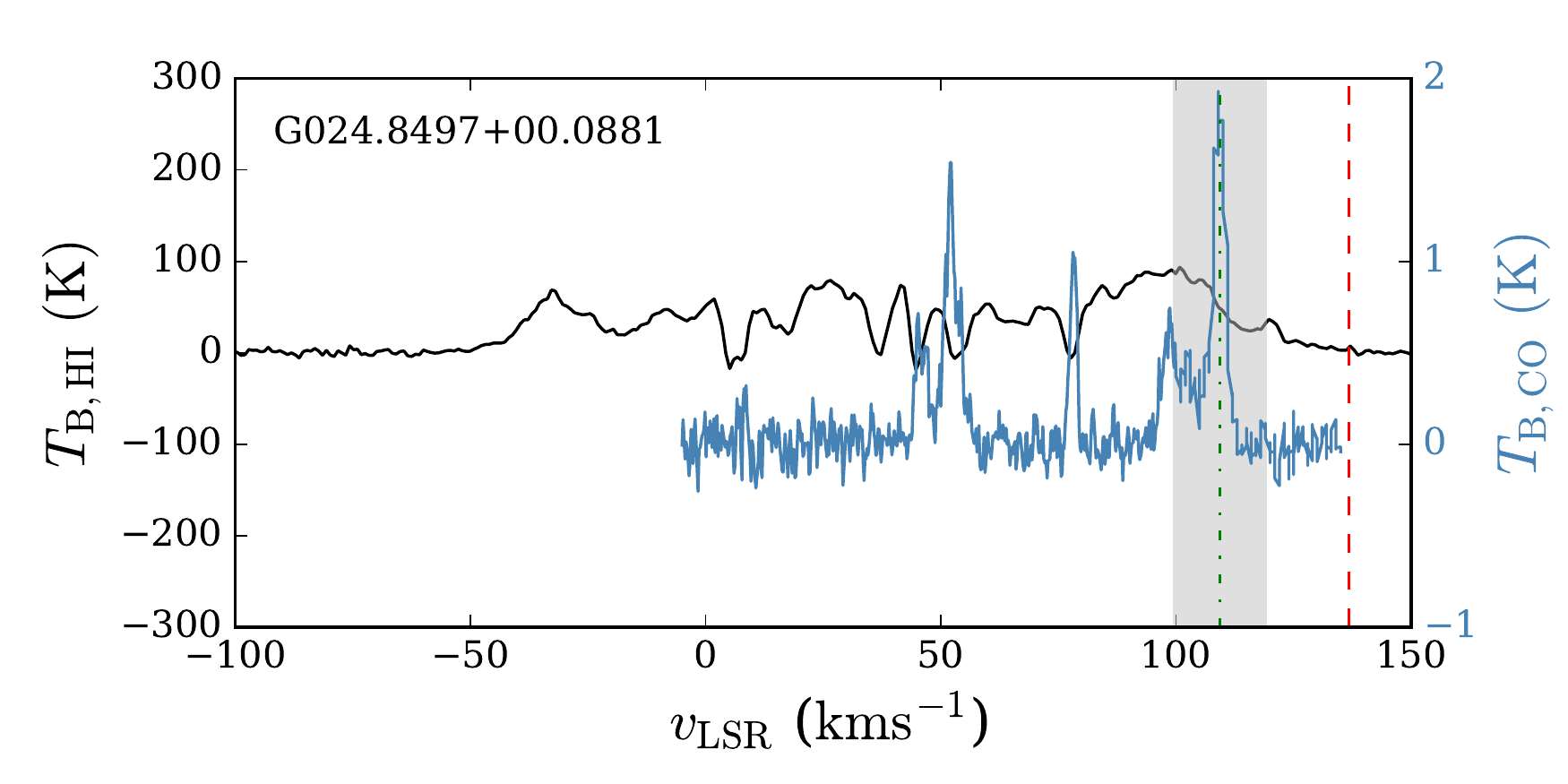}
\includegraphics[width=0.45\columnwidth]{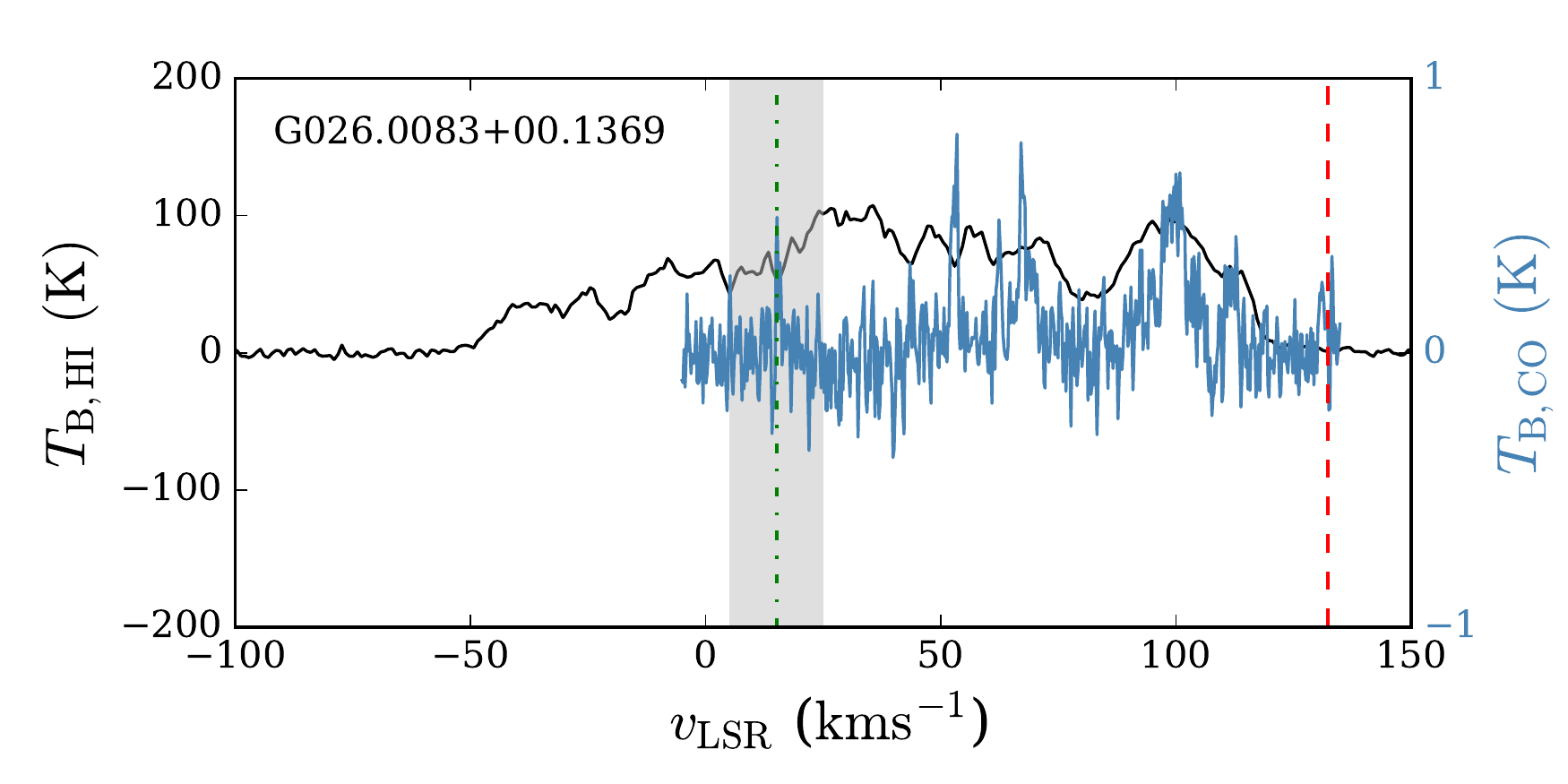}
\newline
\vspace{-1.8mm}
\includegraphics[width=0.45\columnwidth]{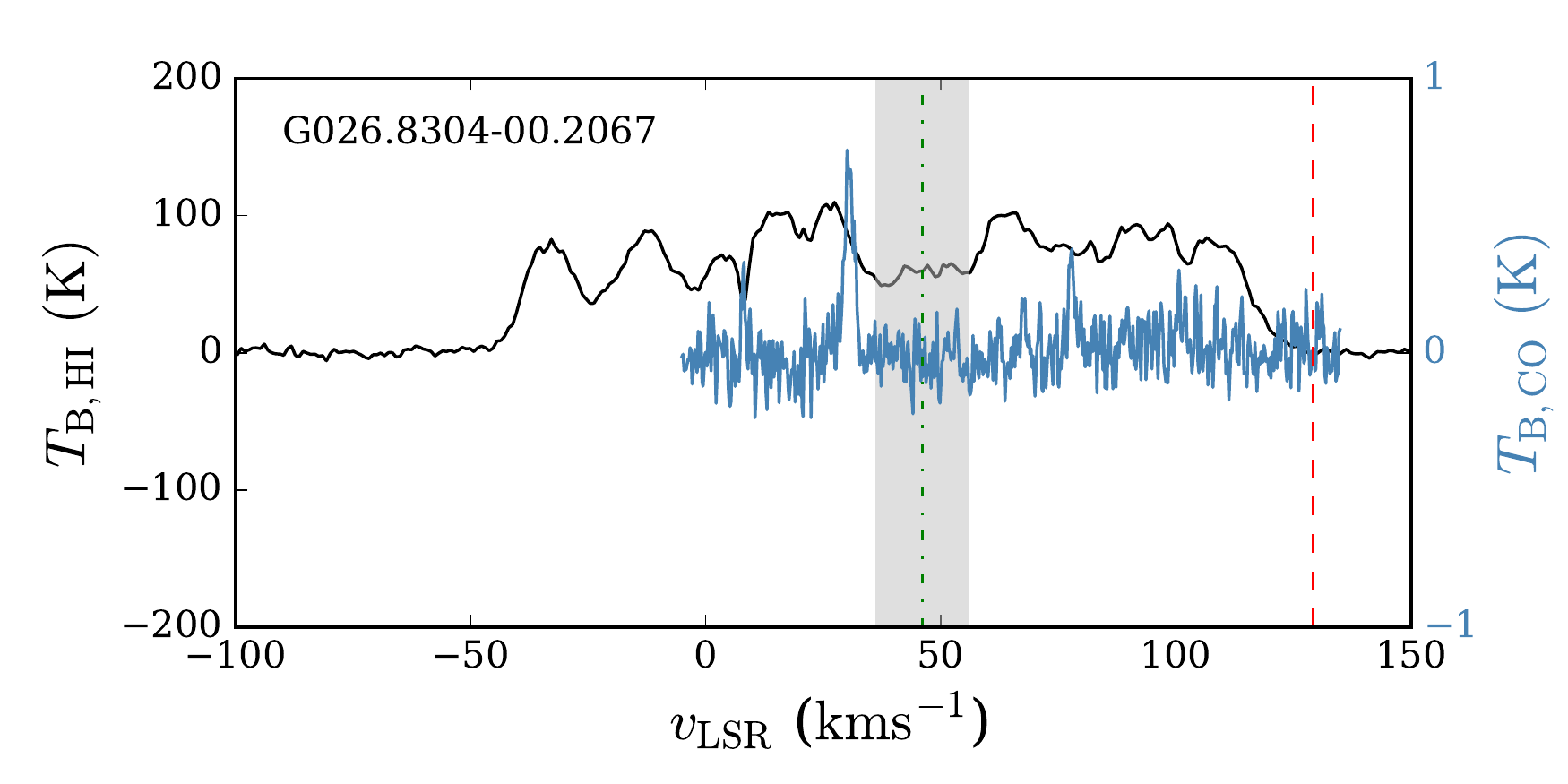}
\includegraphics[width=0.45\columnwidth]{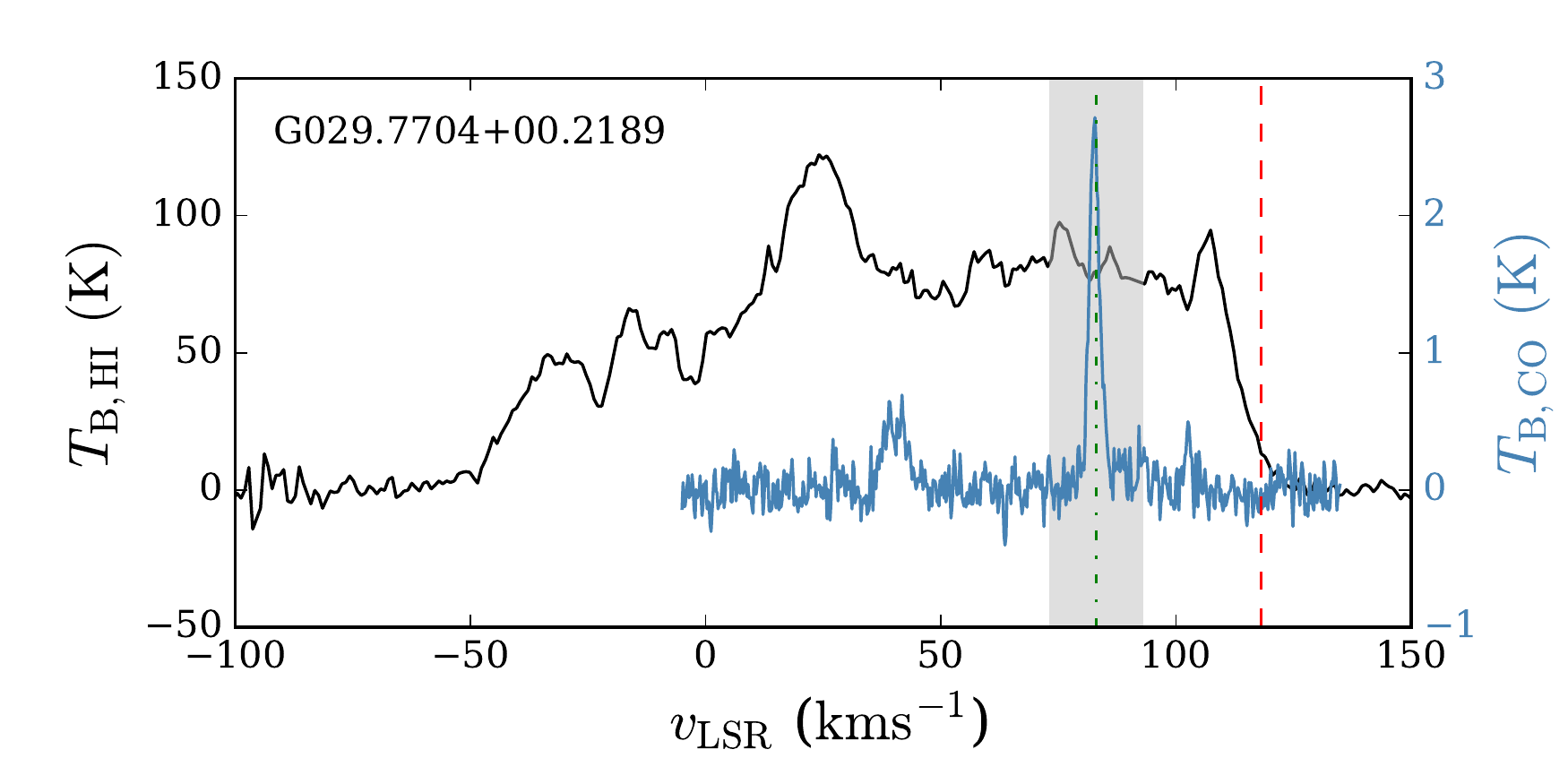}
\newline
\vspace{-1.8mm}
\includegraphics[width=0.45\columnwidth]{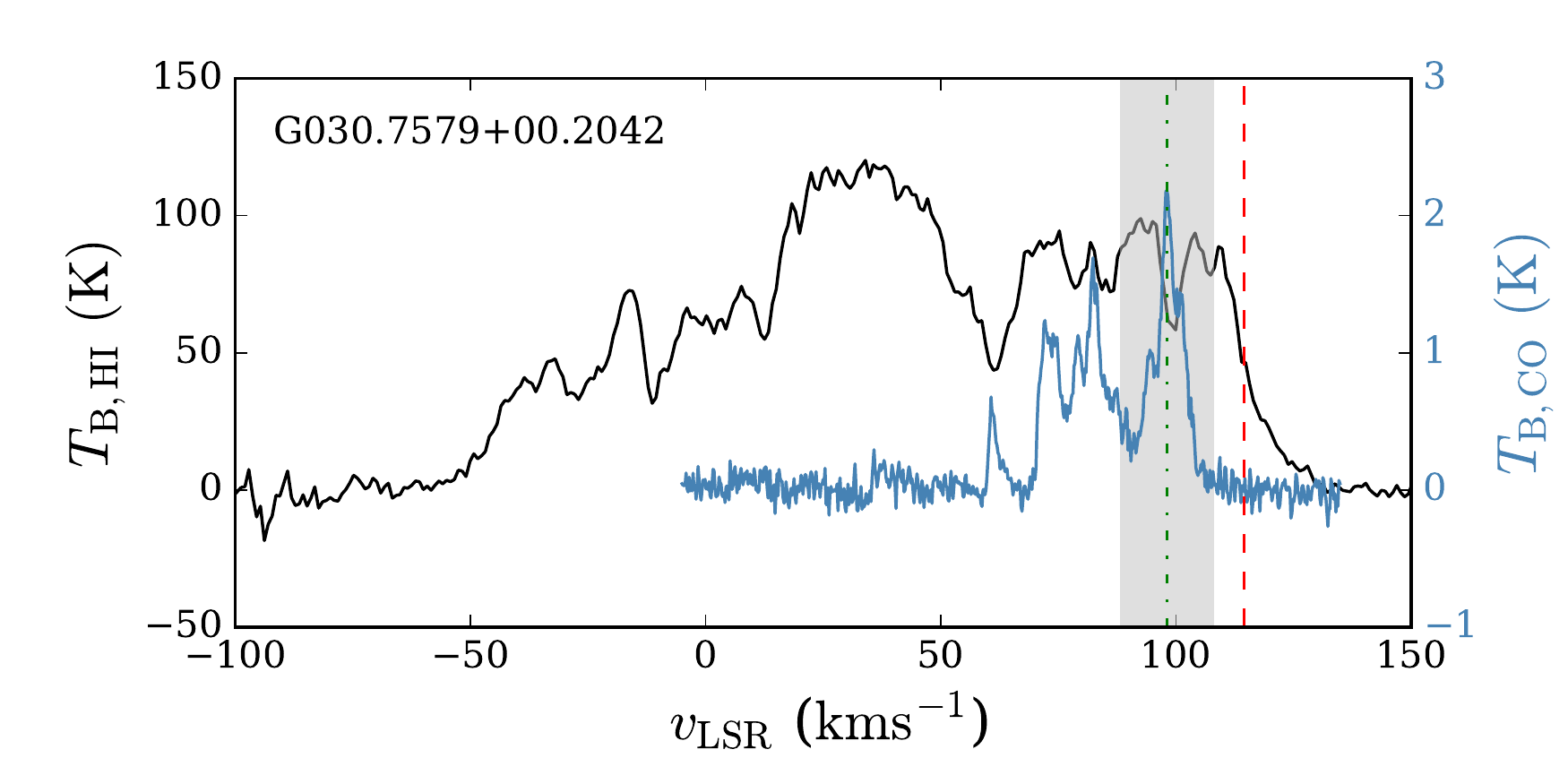}
\includegraphics[width=0.45\columnwidth]{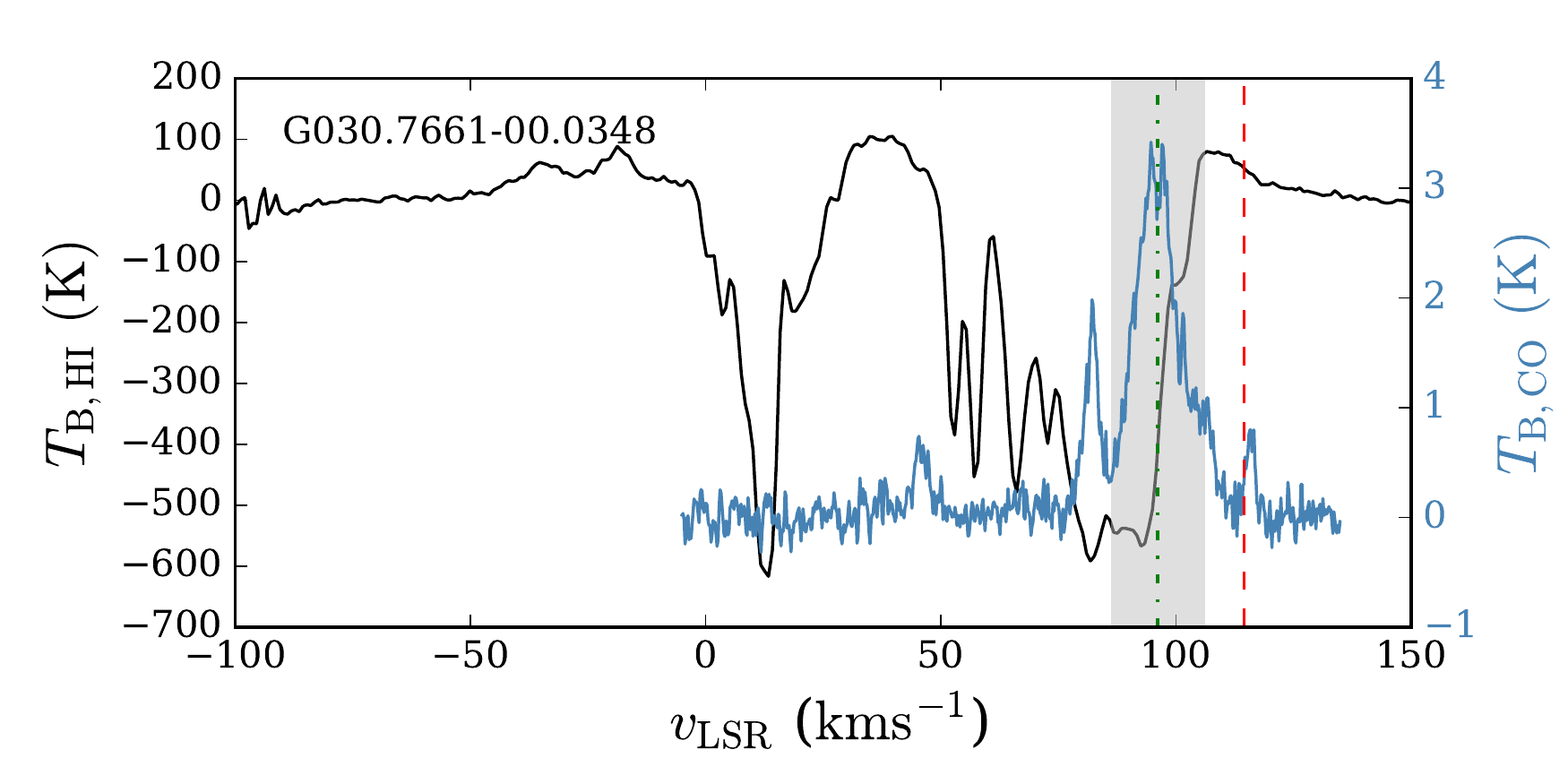}
\newline
\vspace{-1.8mm}
\includegraphics[width=0.45\columnwidth]{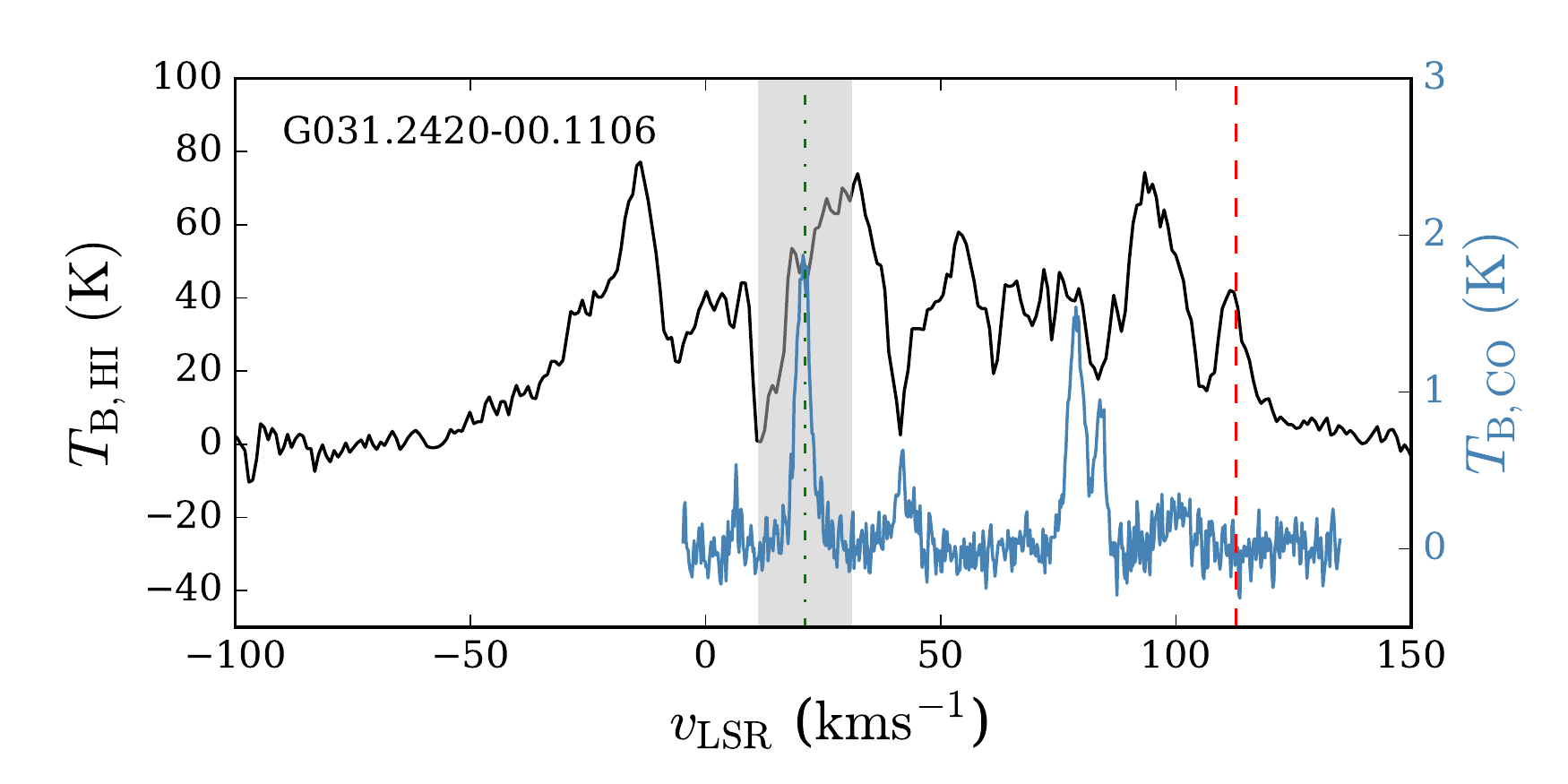}
\includegraphics[width=0.45\columnwidth]{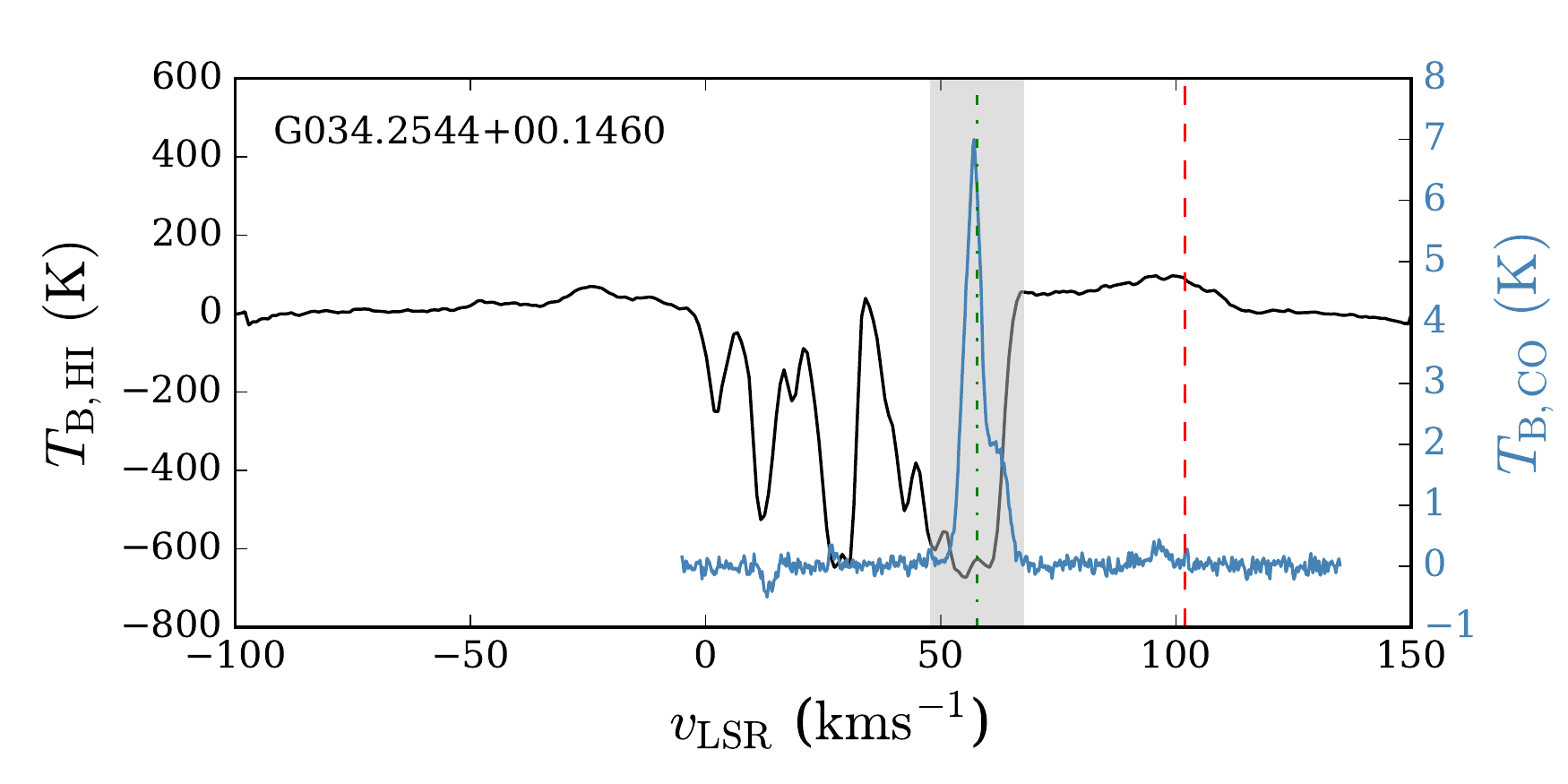}
\end{minipage}

\clearpage

\begin{minipage}{\textwidth}
    \centering
    \captionof{figure}{Continuum-subtracted VGPS spectra towards sources from the UCHII region sample, used to solve their KDA (continued).}
\includegraphics[width=0.45\columnwidth]{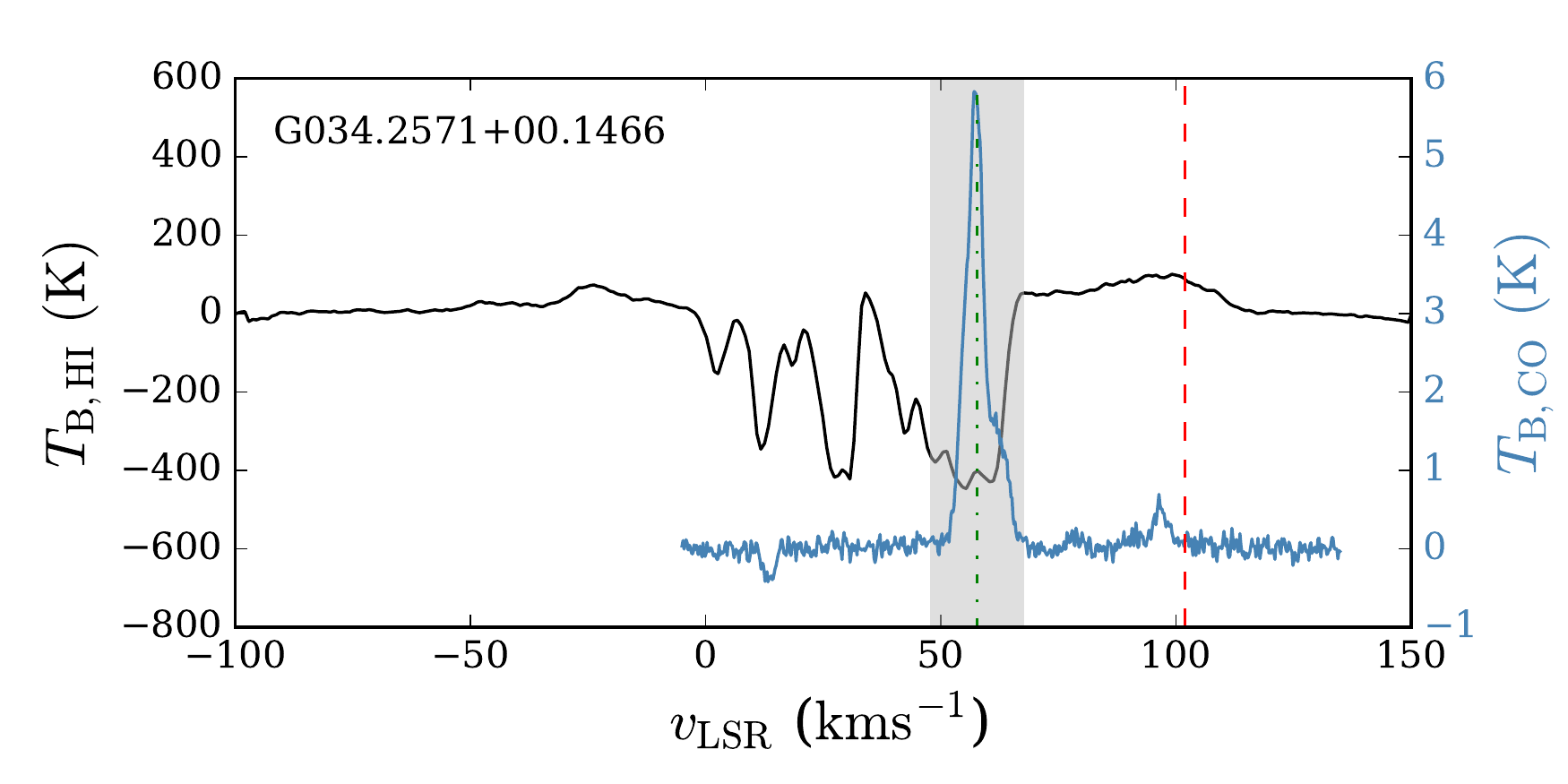}
\includegraphics[width=0.45\columnwidth]{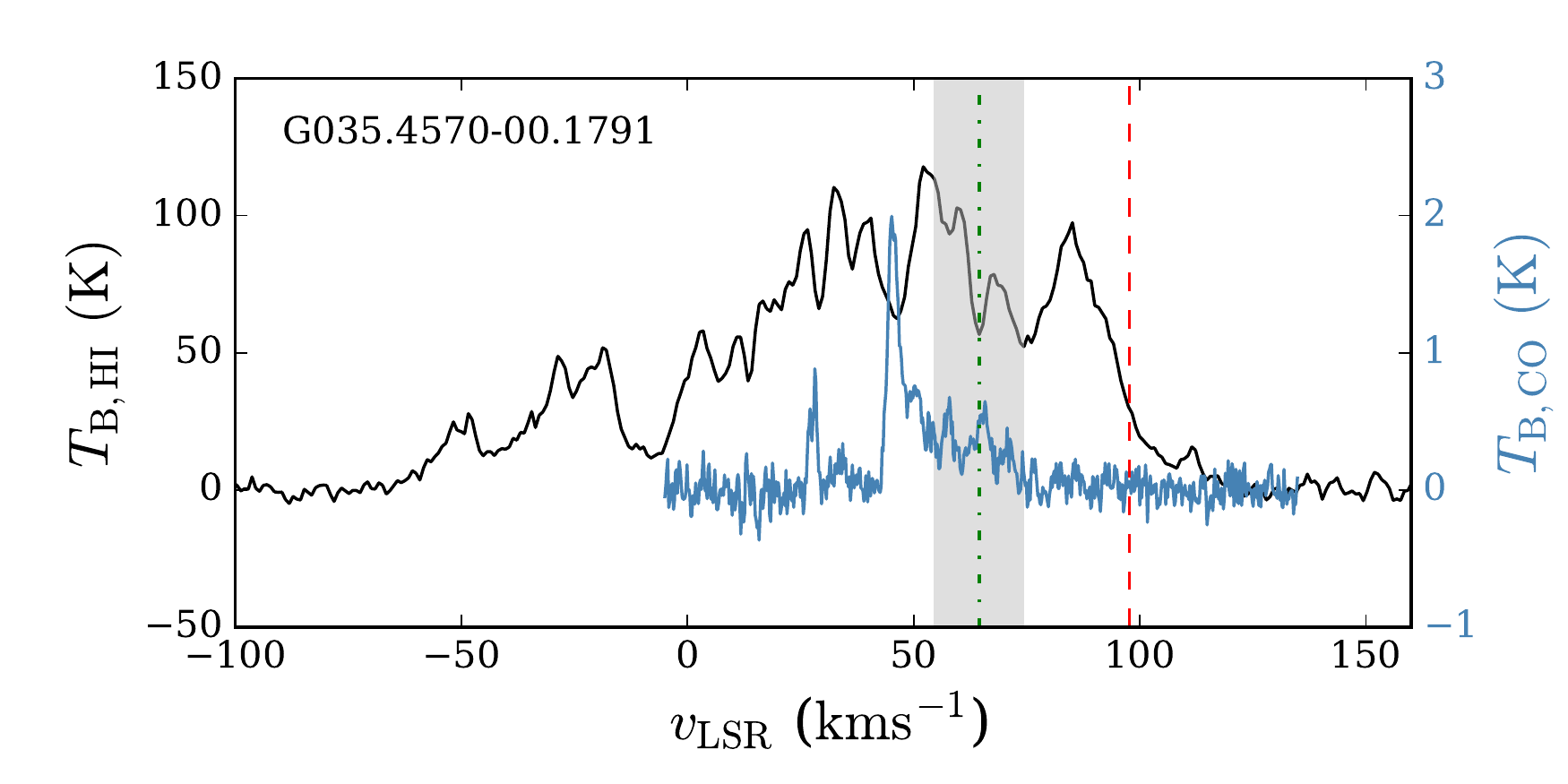}
\newline
\vspace{-1.8mm}
\includegraphics[width=0.45\columnwidth]{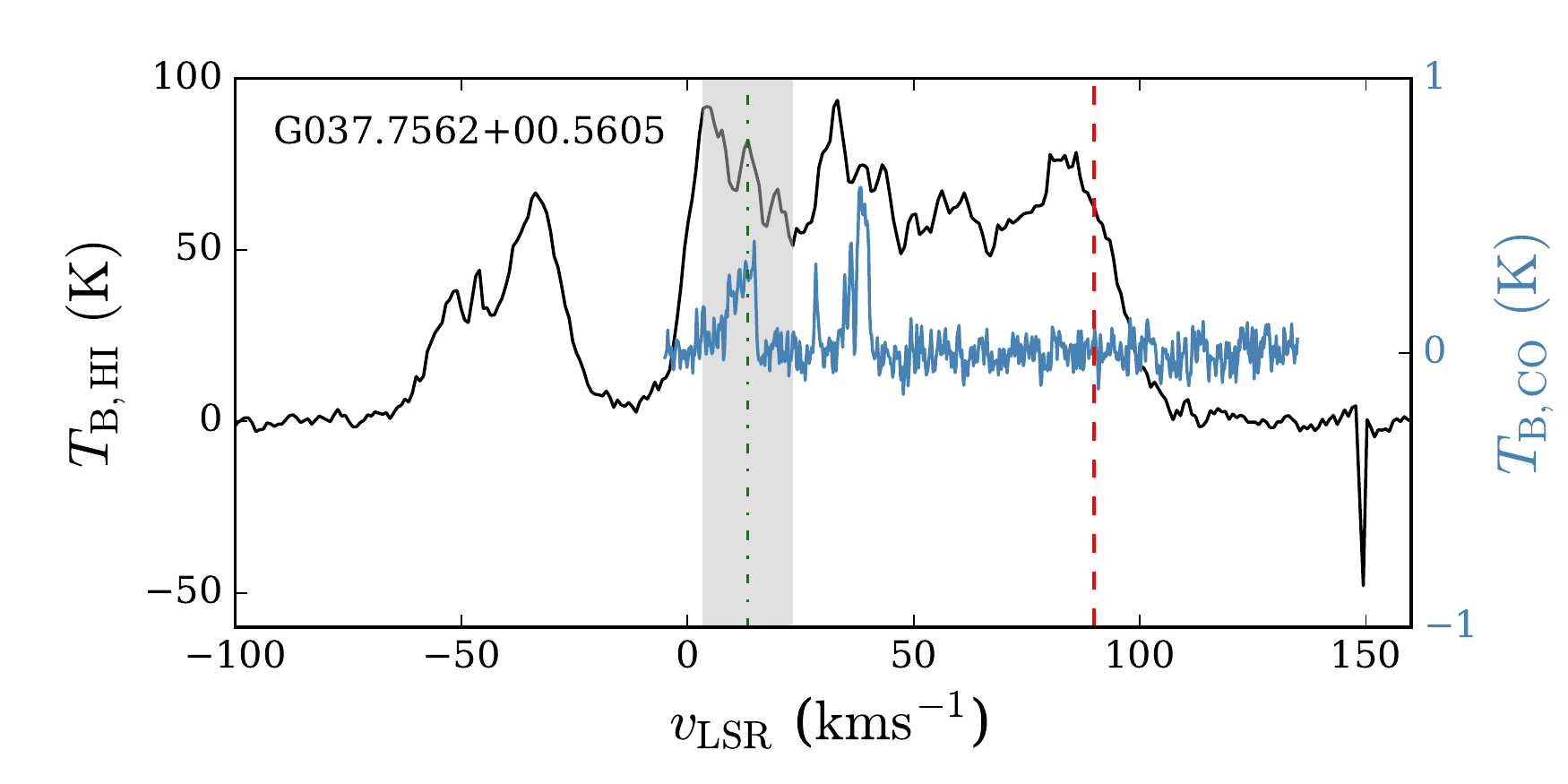}
\includegraphics[width=0.45\columnwidth]{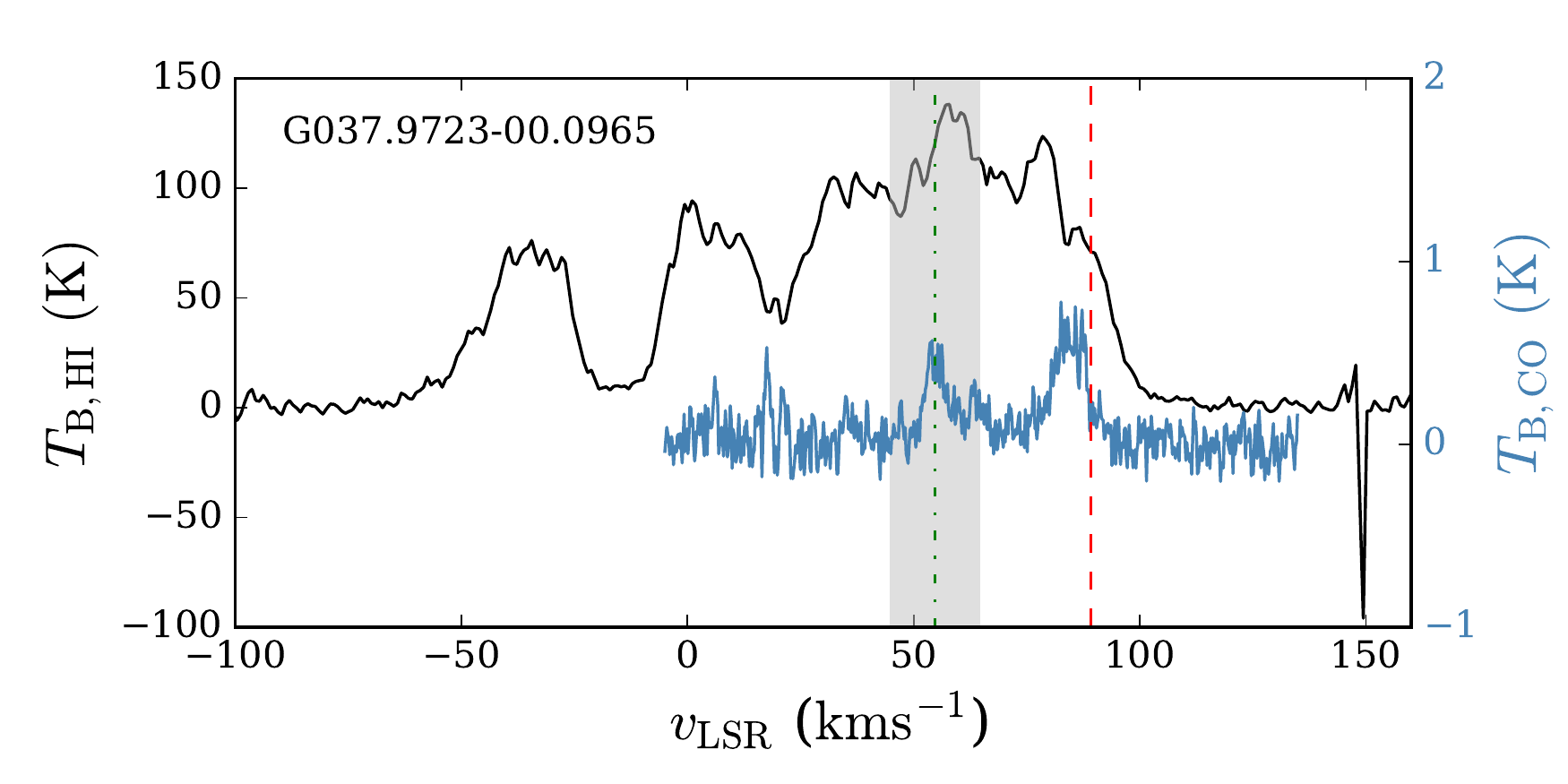}
\newline
\vspace{-1.8mm}
\includegraphics[width=0.45\columnwidth]{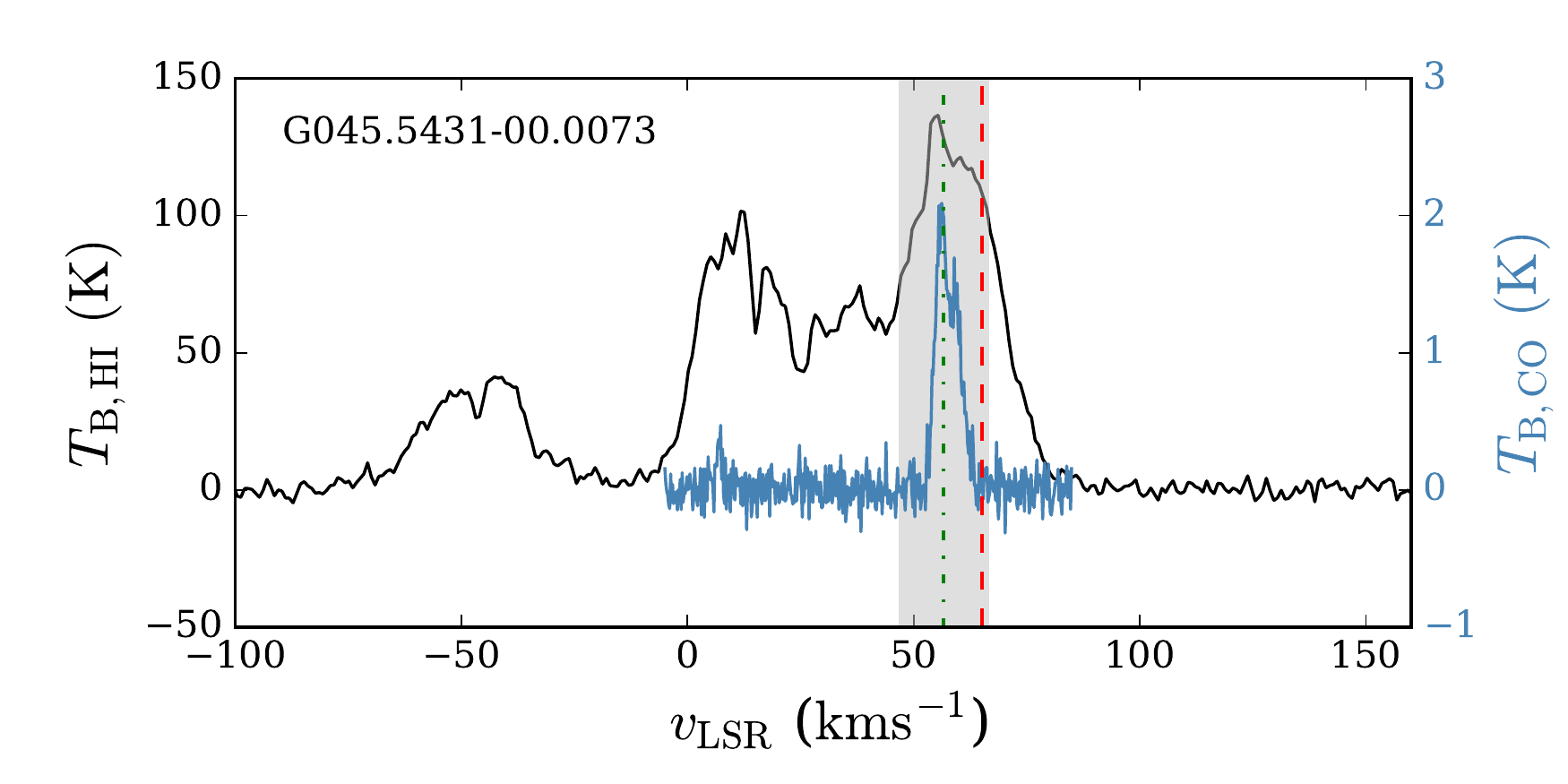}
\includegraphics[width=0.45\columnwidth]{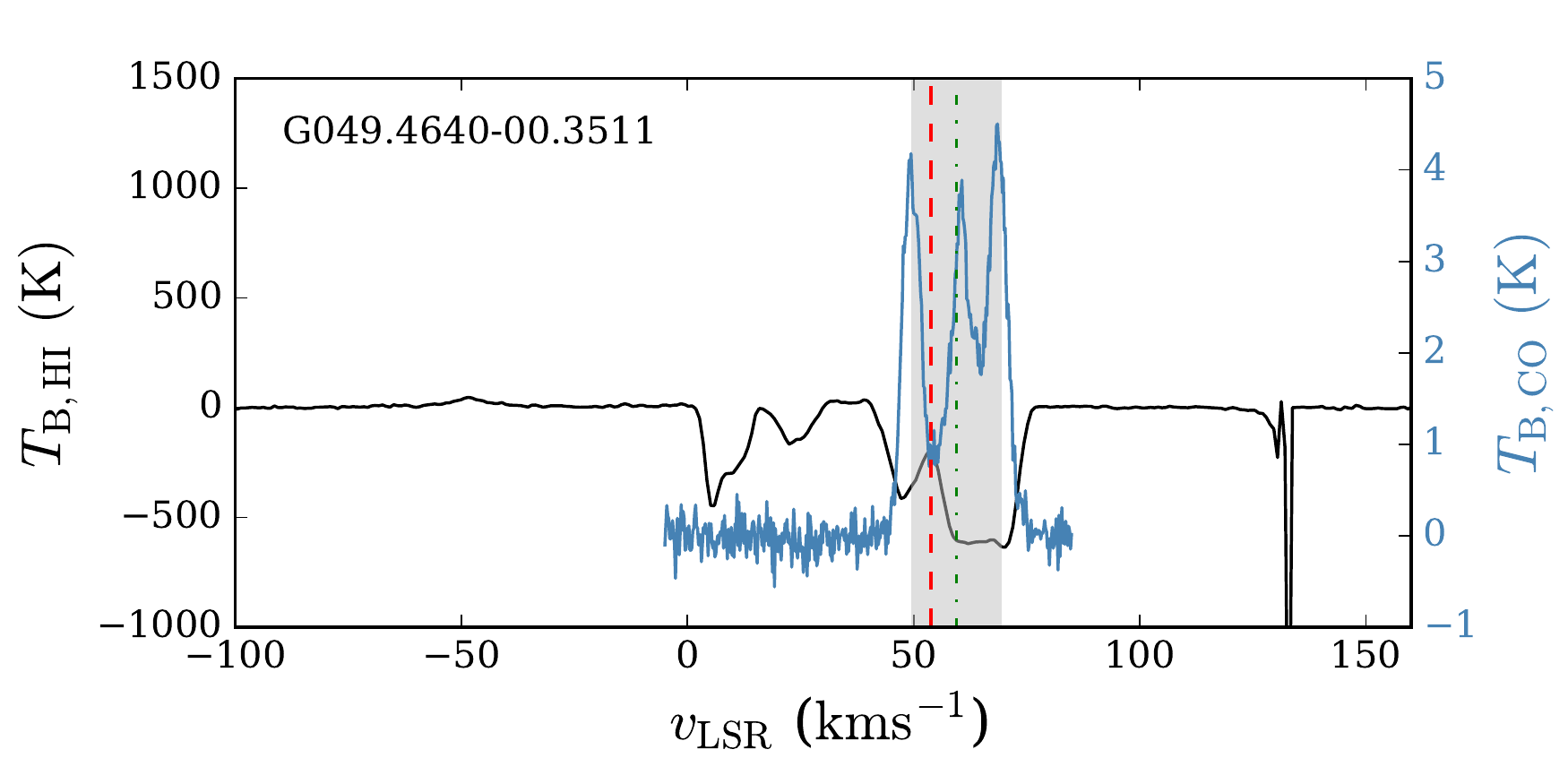}
\newline
\vspace{-1.8mm}
\includegraphics[width=0.45\columnwidth]{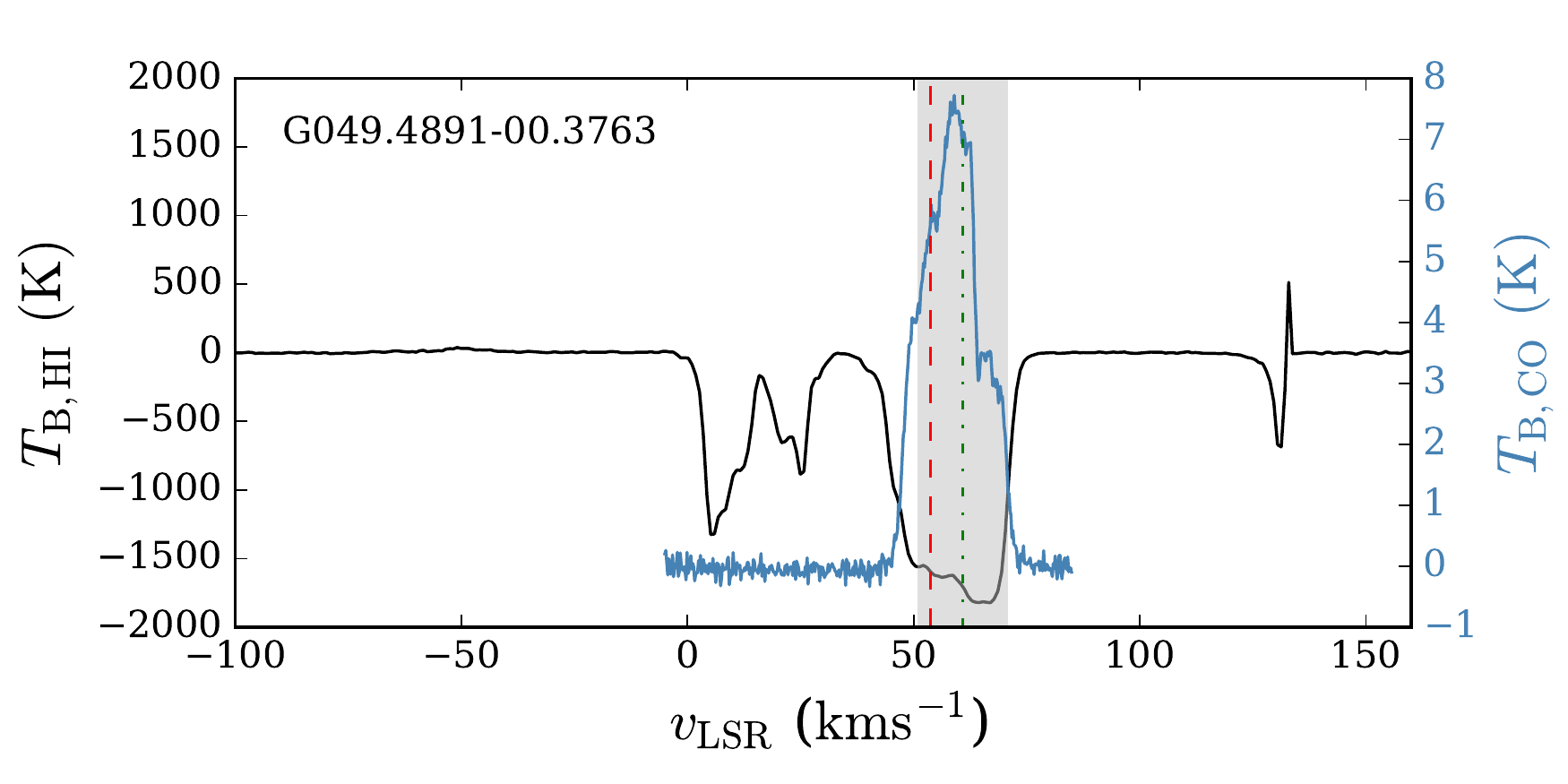}
\includegraphics[width=0.45\columnwidth]{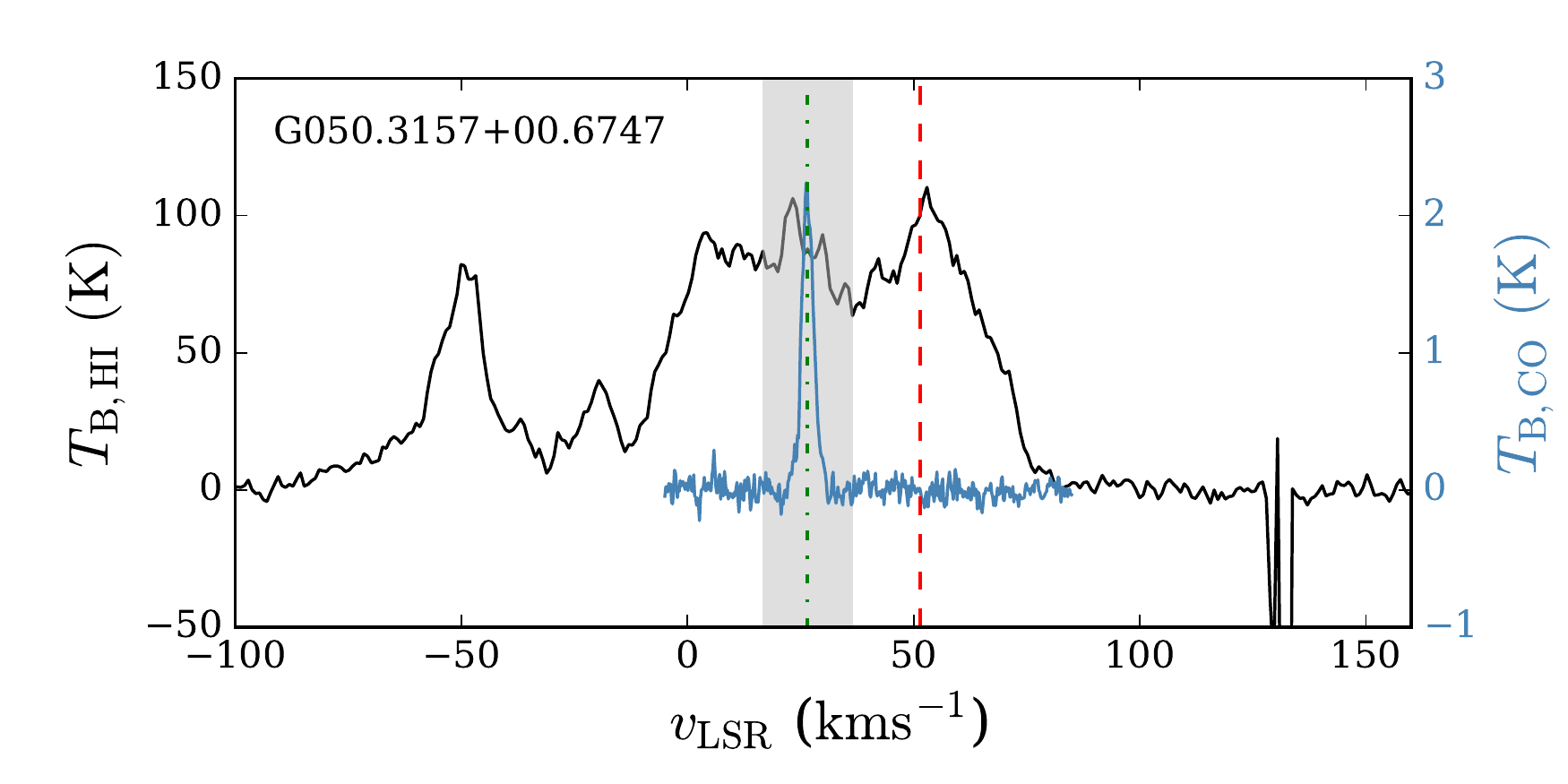}
\newline
\vspace{-1.8mm}
\includegraphics[width=0.45\columnwidth]{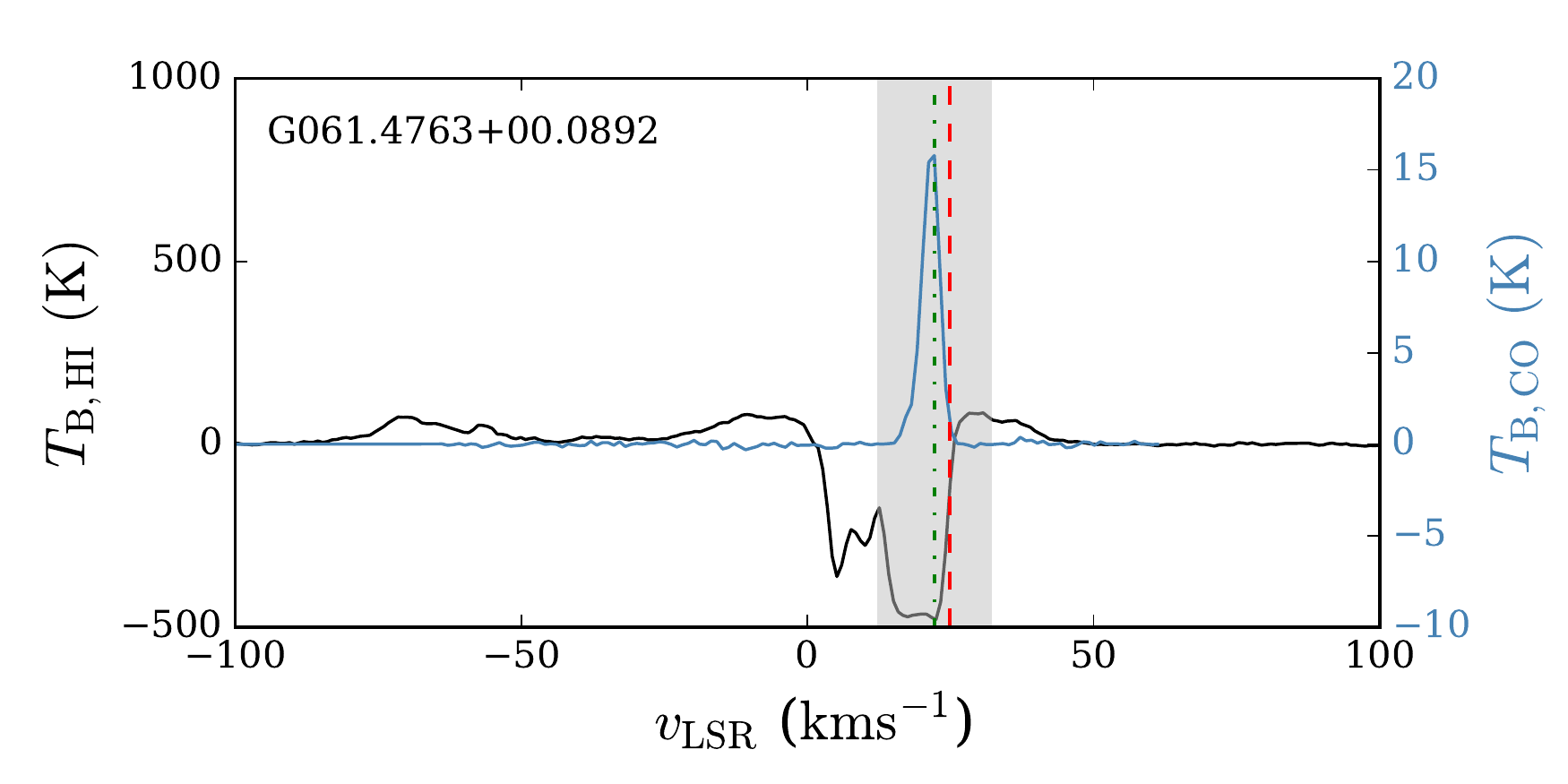}
\includegraphics[width=0.45\columnwidth]{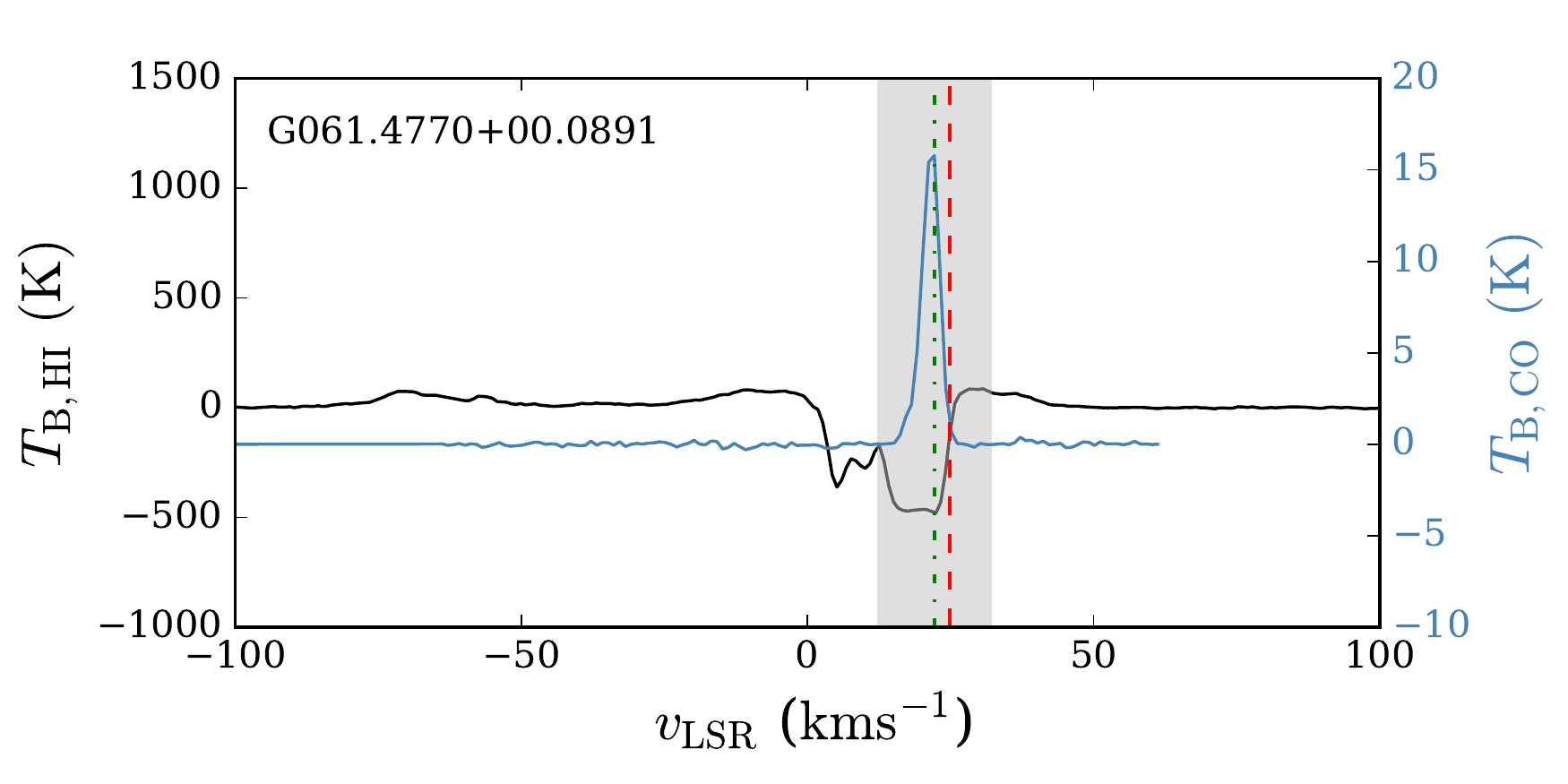}
\end{minipage}

\newpage
\onecolumn

\section{Radio fluxes and spectral indices}\label{appendixD}

\begin{scriptsize}
\setlength\LTcapwidth{\linewidth}
\begin{longtable}{rrrrr}
\caption{Fluxes of the MAGPIS counterparts to CORNISH UCHIIs at 6~cm and 20~cm, as remeasured in this work (described in \S \ref{lowres}), and computed spectral indices (CORNISH 6~cm -- MAGPIS 20~cm). Missing values are due to missing MAGPIS 6~cm images. In all cases of unreliable photometric results or non-detections, the 3$\sigma$ upper limits are included.} 
\label{table:fluxes_M} \\
\toprule
\toprule
  \multicolumn{1}{c}{Name} &
  \multicolumn{1}{r}{$S_{\rm C}$ (mJy)} &
  \multicolumn{1}{r}{$S_{\rm M, 6cm}$ (mJy)} &
  \multicolumn{1}{r}{$S_{\rm M, 20cm}$ (mJy)} &
  \multicolumn{1}{r}{$\alpha_{\rm C6cm-M20cm}$} \\
\hline
\endfirsthead

\multicolumn{5}{c}%
{{\bfseries \tablename\ \thetable{} -- continued from previous page}} \\
\hline
\hline
  \multicolumn{1}{c}{Name} &
  \multicolumn{1}{r}{$S_{\rm C}$ (mJy)} &
  \multicolumn{1}{r}{$S_{\rm M, 6cm}$ (mJy)} &
  \multicolumn{1}{r}{$S_{\rm M, 20cm}$ (mJy)} &
  \multicolumn{1}{r}{$\alpha_{\rm C6cm-M20cm}$} \\
\hline
\endhead

\hline \multicolumn{5}{r}{{Continued on next page}} \\ \hline
\endfoot

\hline 
\endlastfoot

  G010.3009$-$00.1477 & 631.39 $\pm$ 59.3 & 1169.19 $\pm$ 9.41 & 456.88 $\pm$ 74.73 & 0.25 $\pm$ 0.15\\
  G010.3204$-$00.2328 & 32.43 $\pm$ 5 & <28.78 & 48.94 $\pm$ 2.56 & $-$0.32 $\pm$ 0.13\\
  G010.3204$-$00.2586 & 18.2 $\pm$ 2.26 & 47.09 $\pm$ 3.28 & 38.2 $\pm$ 1.97 & $-$0.58 $\pm$ 0.11\\
  G010.4724+00.0275 & 38.43 $\pm$ 4.38 & 92.82 $\pm$ 1.35 & 20.69 $\pm$ 3.36 & 0.49 $\pm$ 0.16\\
  G010.4736+00.0274 & 19.3 $\pm$ 2.39 & 94.83 $\pm$ 1.01 & 17.68 $\pm$ 3.02 & 0.07 $\pm$ 0.17\\
  G010.6218$-$00.3848 & 37.06 $\pm$ 4.04 & 1238.9 $\pm$ 2.96 & 167.89 $\pm$ 4.76 & $-$1.19 $\pm$ 0.09\\
  G010.6223$-$00.3788 & 483.33 $\pm$ 49.87 & 1039.47 $\pm$ 7.42 & 518.5 $\pm$ 15.91 & $-$0.06 $\pm$ 0.08\\
  G010.6234$-$00.3837 & 1952.22 $\pm$ 176.18 & 3921.35 $\pm$ 6.97 & 747.85 $\pm$ 14.52 & 0.75 $\pm$ 0.07\\
\hline
\end{longtable}
\end{scriptsize}

\section{Extended IR source fluxes -- GLIMPSE and UKIDSS}\label{appendixC}

\begin{scriptsize}
\setlength\LTcapwidth{\linewidth}
\begin{longtable}{rrrrrrrrr}
\caption{GLIMPSE fluxes and magnitudes of the sample of CORNISH UCHII regions, computed in this work. All flux values and their associated errors are in mJy. Missing values are due to missing data, unreliable data, and the exclusion of non-detections. The  $\dagger$ symbol marks instances with possible saturation issues. The magnitudes presented here are shown in the AB system, for consistency with our UKIDSS results.} 
\label{glimpse_long} \\
\toprule
\toprule
  \multicolumn{1}{c}{Name} &
  \multicolumn{1}{c}{$S_{3.6 \rm \upmu m}$ (mJy)} &
  \multicolumn{1}{c}{$S_{4.5 \rm \upmu m}$ (mJy)} &
 \multicolumn{1}{c}{$S_{5.8 \rm \upmu m}$ (mJy)} &
  \multicolumn{1}{c}{$S_{8.0 \rm \upmu m}$ (mJy)} &
    \multicolumn{1}{c}{$[3.6]$} &
  \multicolumn{1}{c}{$[4.5]$} &
 \multicolumn{1}{c}{$[5.8]$} &
  \multicolumn{1}{c}{$[8.0]$} \\
\hline
\endfirsthead

\multicolumn{9}{c}%
{{\bfseries \tablename\ \thetable{} -- continued from previous page}} \\
\hline
\hline
  \multicolumn{1}{c}{Name} &
 \multicolumn{1}{c}{$S_{3.6 \rm \upmu m}$ (mJy)} &
  \multicolumn{1}{c}{$S_{4.5 \rm \upmu m}$ (mJy)} &
 \multicolumn{1}{c}{$S_{5.8 \rm \upmu m}$ (mJy)} &
  \multicolumn{1}{c}{$S_{8.0 \rm \upmu m}$ (mJy)} &
    \multicolumn{1}{c}{$[3.6]$} &
  \multicolumn{1}{c}{$[4.5]$} &
 \multicolumn{1}{c}{$[5.8]$} &
  \multicolumn{1}{c}{$[8.0]$} \\
\hline
\endhead

\hline \multicolumn{9}{r}{{Continued on next page}} \\ \hline
\endfoot

\hline
\endlastfoot
 
  G010.3009$-$00.1477 & 191.81 $\pm$ 80.80 & 316.14 $\pm$  99.64 & 1982.12 $\pm$  712.14 & 5117.70 $\pm$  2547.03 & 10.69 $\pm$  0.45 & 10.15 $\pm$  0.34 & 8.16 $\pm$  0.39 & 7.13 $\pm$  0.54\\
  G010.3204$-$00.2328 & 107.48 $\pm$  3.13 & 159.13 $\pm$  4.36 & 812.79 $\pm$  22.51 & 1646.41 $\pm$  53.34 & 11.32 $\pm$  0.03 & 10.90 $\pm$  0.03 & 9.13 $\pm$  0.03 & 8.36 $\pm$  0.03\\
  G010.3204$-$00.2586 & 8.68 $\pm$  2.98 & 57.50 $\pm$  4.64 & 213.39 $\pm$  41.53 & 495.73 $\pm$  136.43 & 14.05 $\pm$  0.37 & 12.00 $\pm$  0.09 & 10.58 $\pm$  0.21 & 9.66 $\pm$  0.30\\
  G010.6218$-$00.3848 & 5.80 $\pm$  3.48 & 26.34 $\pm$  4.77 &  &  & 14.49 $\pm$  0.65 & 12.85 $\pm$  0.20 &  & \\
  G010.6223$-$00.3788 &  & 53.54 $\pm$  23.54 &  &  &  & 12.08 $\pm$  0.48 &  & \\
  G010.6234$-$00.3837 & 36.98 $\pm$  15.23 & 147.58 $\pm$  21.27 & 498.96 $\pm$  259.22 & 1128.85 $\pm$  700.78 & 12.48 $\pm$  0.44 & 10.98 $\pm$  0.16 & 9.65 $\pm$  0.56 & 8.77 $\pm$  0.67\\
  G010.6240$-$00.3813 &  &  & 117.75 $\pm$  61.52 & 522.44 $\pm$  159.40 &  &  & 11.22 $\pm$  0.57 & 9.60 $\pm$  0.33\\
  G010.6297$-$00.3380 & 106.73 $\pm$  3.35 & 162.38 $\pm$  3.85 & 1325.45 $\pm$  29.06 & 3592.56 $\pm$  83.78 & 11.33 $\pm$  0.03 & 10.87 $\pm$  0.03 & 8.59 $\pm$  0.02 & 7.51 $\pm$  0.03\\
\hline
\end{longtable}
\end{scriptsize}

\begin{scriptsize}
\setlength\LTcapwidth{\linewidth}
\begin{longtable}{rrrrrrrrr}
\caption{UKIDSS fluxes, AB magnitudes, and colours of the CORNISH UCHII regions with a visible counterpart in at least the K band. All flux values and their associated errors are in mJy. Missing values are due to missing or unreliable data. All values without a quoted error are limits (upper flux limits and their corresponding lower magnitude limits).} 
 
\label{ukidss_long} \\
\toprule
\toprule
  \multicolumn{1}{c}{Name} &
  \multicolumn{1}{c}{$S_{\rm J}$ (mJy)} &
  \multicolumn{1}{c}{$S_{\rm H}$ (mJy)} &
 \multicolumn{1}{c}{$S_{\rm K}$ (mJy)} &
    \multicolumn{1}{c}{${J}$} &
  \multicolumn{1}{c}{${H}$} &
   \multicolumn{1}{c}{${K}$} &
  \multicolumn{1}{c}{$J-H$} &
  \multicolumn{1}{c}{$H-K$} \\
\hline
\endfirsthead

\multicolumn{9}{c}%
{{\bfseries \tablename\ \thetable{} -- continued from previous page}} \\
\hline
\hline
 \multicolumn{1}{c}{Name} &
  \multicolumn{1}{c}{$S_{\rm J}$ (mJy)} &
  \multicolumn{1}{c}{$S_{\rm H}$ (mJy)} &
 \multicolumn{1}{c}{$S_{\rm K}$ (mJy)} &
    \multicolumn{1}{c}{${J}$} &
  \multicolumn{1}{c}{${H}$} &
   \multicolumn{1}{c}{${K}$} &
  \multicolumn{1}{c}{$J-H$} &
  \multicolumn{1}{c}{$H-K$} \\
\hline
\endhead

\hline \multicolumn{9}{r}{{Continued on next page}} \\ \hline
\endfoot

\hline 
\endlastfoot

  G010.3009$-$00.1477 & 1.47 & 3.54 & 4.64 $\pm$ 1.81 & 15.05 & 13.65 & 12.83 $\pm$ 0.42 & 1.40 & 0.81\\
  G010.6297$-$00.3380 & 0.89 & 2.18 & 1.38 $\pm$ 0.58 & 15.59 & 14.17 & 14.15 $\pm$ 0.45 & 1.41 & 0.02\\
  G010.9584+00.0221 & 0.66 & 1.86 & 1.63 $\pm$ 0.77 & 15.92 & 14.35 & 13.97 $\pm$ 0.51 & 1.57 & 0.38\\
  G011.1104$-$00.3985 & 2.68 & 4.51 & 26.25 $\pm$ 1.53 & 14.39 & 13.39 & 10.95 $\pm$ 0.06 & 1.01 & 2.43\\
  G011.9368$-$00.6158 & 1.94 & 3.90 & 6.65 $\pm$ 1.79 & 14.74 & 13.54 & 12.44 $\pm$ 0.29 & 1.20 & 1.10\\
  G011.9446$-$00.0369 & 20.54 & 33.35 $\pm$ 18.56 & 43.24 $\pm$ 15.87 & 12.18 & 11.21 $\pm$ 0.60 & 10.41 $\pm$ 0.40 & 0.97 & 0.8 $\pm$ 0.72\\
  G012.4317$-$01.1112 &  & 4.48 & 5.24 $\pm$ 1.27 &  & 13.39 & 12.70 $\pm$ 0.26 &  & 0.69\\
  G012.8050$-$00.2007 & 5.61 & 7.15 $\pm$ 4.35 & 30.26 $\pm$ 5.86 & 13.59 & 12.88 $\pm$ 0.66 & 10.80 $\pm$ 0.21 & 0.70 & 2.09 $\pm$ 0.69\\
\hline
\end{longtable}
\end{scriptsize}

\begin{scriptsize}
\setlength\LTcapwidth{\linewidth}
\begin{longtable}{rrrrrrr}
\caption{Extinctions towards the CORNISH UCHII with visible near-IR nebulae (at least in the K band), computed from the measured near-IR fluxes using four different methods. $A_{\lambda}/A_{V}$ ratios were computed following \protect \cite{cardelli:1989}: those are $\sim$ 0.28, 0.18, and 0.11, for UKIDSS $\lambda$ = $J$, $H$, $K$, respectively, assuming $R_{\rm V}$ = 3.1. Missing values in the table are due to missing or unreliable data. All values without a quoted error are lower limits.} 
 \label{ukidss_Av} \\
\toprule
\toprule

  \multicolumn{1}{c}{Name} &
  \multicolumn{1}{c}{$A_{\rm H}$} &
  \multicolumn{1}{c}{$A_{\rm K}$} &
 \multicolumn{1}{c}{$A_{\rm V} (H)$} &
    \multicolumn{1}{c}{$A_{\rm V} (K)$} &
  \multicolumn{1}{c}{$A_{\rm V} (J-H)$} &
   \multicolumn{1}{c}{$A_{\rm V} (H-K)$} \\
\hline
\endfirsthead

\multicolumn{7}{c}%
{{\bfseries \tablename\ \thetable{} -- continued from previous page}} \\
\hline
\hline
  \multicolumn{1}{c}{Name} &
  \multicolumn{1}{c}{$A_{\rm H}$} &
  \multicolumn{1}{c}{$A_{\rm K}$} &
 \multicolumn{1}{c}{$A_{\rm V} (H)$} &
    \multicolumn{1}{c}{$A_{\rm V} (K)$} &
  \multicolumn{1}{c}{$A_{\rm V} (J-H)$} &
   \multicolumn{1}{c}{$A_{\rm V} (H-K)$} \\
\hline
\endhead

\hline \multicolumn{7}{r}{{Continued on next page}} \\ \hline
\endfoot

\hline 
\endlastfoot

  G010.3009$-$00.1477 & 4.16 & 4.03 $\pm$ 0.44 & 36.71 & 35.49 $\pm$ 3.85 & 19.60 & 9.37\\
  G010.6297$-$00.3380 & 1.24 & 1.9 $\pm$ 0.46 & 10.97 & 16.74 $\pm$ 4.08 & 19.74 & $-$1.89\\
  G010.9584+00.0221 & 3.59 & 3.89 $\pm$ 0.52 & 31.69 & 34.31 $\pm$ 4.6 & 21.31 & 3.17\\
  G011.1104$-$00.3985 & 3.11 & 1.36 $\pm$ 0.07 & 27.45 & 11.96 $\pm$ 0.65 & 15.63 & 32.41\\
  G011.9368$-$00.6158 & 4.72 & 4.29 $\pm$ 0.31 & 41.57 & 37.84 $\pm$ 2.77 & 17.57 & 13.43\\
  G011.9446$-$00.0369 & 2.17 $\pm$ 0.60 & 2.04 $\pm$ 0.40 & 19.1 $\pm$ 5.33 & 17.98 $\pm$ 3.54 & 15.24 $\pm$ 6.07 & 9.21 $\pm$ 10.24\\
  G012.4317$-$01.1112 & 1.51 & 1.49 $\pm$ 0.27 & 13.29 & 13.15 $\pm$ 2.38 &  & 7.64\\
  G012.8050$-$00.2007 & 6.65 $\pm$ 0.68 & 5.24 $\pm$ 0.25 & 58.65 $\pm$ 6.00 & 46.22 $\pm$ 2.22 & 12.58 $\pm$ 6.65 & 27.48 $\pm$ 9.81\\
\hline
\end{longtable}
\end{scriptsize}

\section{Distances and physical properties}\label{appendixE}

\begin{scriptsize}
\setlength\LTcapwidth{\linewidth}
\begin{longtable}{rrrrrrrrrr}
\caption{Heliocentric and galactocentric distances and physical properties of the 239 CORNISH UCHIIs. Heliocentric distances adopted from \protect \cite{cesaroni:2015} are marked with $\dagger$ (the KDA is not resolved and the far distance is quoted); those from the RMS database are denoted by $\ddagger$, and the ones computed in this work -- with $\star$. All remaining heliocentric distances (and the corresponding Lyman continuum flux values) were adopted from \protect \cite{urquhart:2013}. Upper limits on the physical sizes are given for unresolved sources of known distance.}
\label{phys_prop} \\
\toprule
\toprule
  \multicolumn{1}{c}{Name} &
  \multicolumn{1}{r}{$d_{h}$ (kpc)} &
  \multicolumn{1}{r}{$d_{G}$ (kpc)} &
  \multicolumn{1}{r}{$D$ (pc)} & 
  \multicolumn{1}{r}{log$N_{\rm i}$ $(\rm s^{-1})$} &
  \multicolumn{1}{r}{log$EM$ $(\rm cm^{-6}pc)$} &
  \multicolumn{1}{r}{log$n_{\rm e}$ $(\rm cm^{-3})$} &
  \multicolumn{1}{r}{log$T_{\rm b}$ (K)} &
  \multicolumn{1}{r}{$\tau_{\rm 5GHz, 10^{4} \rm K}$ } & 
   \multicolumn{1}{r}{log$L_{\rm bol}$ ($\rm L_{\odot}$)}\\
\hline
\endfirsthead

\multicolumn{10}{c}%
{{\bfseries \tablename\ \thetable{} -- continued from previous page}} \\
\hline
\hline
  \multicolumn{1}{c}{Name} &
  \multicolumn{1}{r}{$d_{h}$ (kpc)} &
  \multicolumn{1}{r}{$d_{G}$ (kpc)} &
  \multicolumn{1}{r}{$D$ (pc)} & 
  \multicolumn{1}{r}{log$N_{\rm i}$ $(\rm s^{-1})$} &
  \multicolumn{1}{r}{log$EM$ $(\rm cm^{-6}pc)$} &
  \multicolumn{1}{r}{log$n_{\rm e}$ $(\rm cm^{-3})$} &
  \multicolumn{1}{r}{log$T_{\rm b}$ (K)} &
  \multicolumn{1}{r}{$\tau_{\rm 5GHz, 10^{4} \rm K}$ } & 
   \multicolumn{1}{r}{log$L_{\rm bol}$ ($\rm L_{\odot}$)}\\
  
\hline
\endhead

\hline \multicolumn{10}{r}{{Continued on next page}} \\ \hline
\endfoot

\hline 
\endlastfoot

  G010.3009$-$00.1477 & 2.4 & 6.15 & 0.061 & 47.5 & 7.03 & 4.2 & 3.02 $\pm$ 0.04 & 0.11 & \\
  G010.3204$-$00.2328 & 12.6$^\dagger$ & 4.5 & 0.18 & 47.7 & 6.19 & 3.54 & 2.18 $\pm$ 0.07 & 0.015 & 4.51\\
  G010.3204$-$00.2586 & 3.5 & 5.1 & < 0.028 & 46.3 & 6.51 & 4.12 & 2.5 $\pm$ 0.06 & 0.032 & 4.18\\
  G010.4724+00.0275 & 11.0 & 3.06 & 0.091 & 47.6 & 6.58 & 3.84 & 2.57 $\pm$ 0.05 & 0.038 & 5.84\\
  G010.4736+00.0274 & 11.0 & 3.06 & 0.075 & 47.3 & 6.37 & 3.76 & 2.36 $\pm$ 0.05 & 0.023 & \\
  G010.6218$-$00.3848 & 2.4 & 6.16 & 0.02 & 46.3 & 6.55 & 4.15 & 2.54 $\pm$ 0.05 & 0.035 & \\
  G010.6223$-$00.3788 & 2.4 & 6.16 & 0.065 & 47.4 & 6.86 & 4.11 & 2.85 $\pm$ 0.04 & 0.074 & \\
  G010.6234$-$00.3837 & 2.4 & 6.16 & 0.051 & 48.0 & 7.66 & 4.55 & 3.65 $\pm$ 0.04 & 0.59 & 4.95\\
\bottomrule
\end{longtable}

\end{scriptsize}

\end{appendices}

\end{document}